\documentclass[preprint,3p]{elsarticle}

\usepackage{graphicx}
\usepackage{amsmath}
\usepackage{amssymb}
\usepackage{xcolor}
\def\mat#1{{\boldsymbol #1}}
\def\vec#1{{\boldsymbol #1}}

\def\r{\mat r}

\def\x{\mat x}

\def\C{{\cal C}}

\def\mylabel#1{\label{#1}}
\def\revisited#1{\tag{\ref{#1} revisited} }
\def\myslabel#1{\label{#1}}

\def\<{\langle}
\def\>{\rangle}
\def\[{\left[}
\def\]{\right]}
\def\({\left(}
\def\){\right)}

\def\sign#1{{\rm sign}(#1)}

\def\Si{{\rm Si}}

\def\re{ {\rm Re} }
\def\im{ {\rm Im} }

\def\a0{~$a_0$}

\def\balpha{{\bar\alpha}}

\def\tt{0}

\def\EE{{\cal E}}

\def\head#1{{\it #1}.}
\def\binf{\exp\[ -\frac{i}{\hbar}\int\limits_{0}^{\infty} dt'\, \epsilon_+(t')\]}
\def\rinf{\exp\[ \frac{i}{\hbar}\int\limits_{0}^{-\infty} dt'\, \epsilon_-(t')\]}

\def\E#1{\vec \EE #1\hbar\omega}
\def\M{\frac{\varepsilon_0}{2}\vec\mu}
\def\O{\vec 0}

\def\hh{\frac{1}{\hbar\sqrt{2\pi}}}
\def\hH{\frac{2}{\hbar\sqrt{2\pi}}}

\def\bv{\bar v}

\begin{document}
\title{
Complex time method for quantum dynamics when an exceptional point is encircled in the parameter space
}
\author[1]{Petra Ruth Kapr\'alov\'a-\v Z\v d\'ansk\'a}
\ead{kapralova@fzu.cz}

\affiliation[1]{ organization={Department of Radiation and Chemical Physics, Institute of Physics, Academy of Sciences of the Czech Republic},
addressline={Na Slovance 2}, 
postcode={182 21},
city={Prague 8}, 
country={Czech Republic} }

\begin{abstract}
	We revisit the complex time method for
	the application to quantum dynamics
	as an exceptional point is encircled in the 
	parameter space of the Hamiltonian.
	The basic idea of the complex time method is using complex
	contour integration to perform the first-order 
	adiabatic perturbation integral.
	In this way, the quantum dynamical problem is transformed to
	a study of singularities in the complex time plane -- transition points -- which
	represent complex degeneracies of the adiabatic Hamiltonian as 
	the time-dependent parameters defining the encircling contour
	are analytically continued to complex plane.
	As an underlying illustration of the approach we discuss a switch between
	Rabi oscillations and rapid adiabatic passage which occurs
	upon the encircling of an exceptional point in a special time-symmetric
	case.
\end{abstract}

\maketitle

%\setcounter{tocdepth}{2}
%\tableofcontents

\newpage

\section{Introduction}

\noindent
Dynamics of a quantum system near exceptional points
(EPs) is of a fundamental importance in many areas of atomic, molecular, and
optical physics~\cite{Rotter:2009,Heiss:2012,Rotter:2015}.
EPs are degeneracies that appear under
certain conditions
at {\it complex}-defined potential energy surfaces (PESs) 
where unstable (non-Hermitian) quantum states are involved.
They are found
in optical devices~\cite{Klaiman:2008,Doppler:2016,Miri:2019}, 
laser cavities~\cite{Liertzer:2012,Peng:2014,Feng:2014,Ozdemir:2019}, 
laser driven atoms or atoms perturbed by fields~\cite{Kapralova-Zdanska:2014,Peng:2016,Oberreiter:2018,Li:2019}, 
molecular vibrations~\cite{Estrada:1986,Lefebvre:2009,Cederbaum:2011,Leclerc:2017,Benda:2018}, etc..
In this paper we focus on the non-adiabatic dynamics
which takes place as an EP is {\it dynamically encircled}
in the parameter space of the PESs. 
%Depending on a problem, such a space is 
%represented by the frequency and strength of a laser,
%index of refraction defining waveguides, 
%or nuclear coordinates for vibrational dynamics in molecules, etc..

It has been established so far that EP encircling 
is manifested via the {\it time-asymmetric mode switching (TAMS)}~\cite{Uzdin:2011,Gilary:2013,Kapralova-Zdanska:2014,Feng:2014,Doppler:2016,Xu:2016,Zhong:2018,Oberreiter:2018,Li:2019,Fernandez:2020,Feilhauer:2020}. 
Mathematically,
the TAMS is caused by the presence of imaginary components of
the energies of the coupled non-Hermitian states (which
are unstable resonances), as they imply exponential
suppression or enhancement of the evolution operator.

Another phenomenon 
is represented by a {\it behavior switch between Rabi oscillations and
rapid adiabatic passage (Rabi-to-RAP switch)} 
introduced in Ref.~\cite{LETTER}. This
phenomenon occurs for {\it time-symmetric EP encircling},
in contrast to TAMS.
It has been shown that the Rabi-to-RAP switch can
be used to directly localize the encircled EP.
It has been discussed to this end that
under specified conditions, zeros of Rabi oscillations
converge to the same point 
which is also identical with the Rabi-to-RAP change of behavior.
This intriguing fact, it has been proposed, could be a basis
of a new spectroscopic experiment
to localize the Rabi-to-RAP switch, which is hard to find by other means
due to low resolution of shallow oscillations near the sought critical point.
The weight of theoretical explanation of the Rabi-to-RAP switch in Ref.~\cite{LETTER} 
relies on a complex time method, the mathematical basis of which is
provided in the present paper.

This paper represents a detailed introduction
to the {\it complex time method} as a novel
 theoretical tool to study different quantum
phenomena associated with the dynamical encircling of EP.
 From the historical point of view,
the complex time method has been introduced
 in the scattering theory by Dykhne, Davis, and Pechukas~\cite{Dykhne:1962,Davis:1975} 
who studied non-adiabatic jumps in potential curve crossings
for a quadratic coupling model.
Our current work presents a new
generalization of the complex time method {\it to the non-Hermitian (dissipative) cases}.

The underlying idea of the complex time method is given by using
the {\it complex contour integration} to perform
the first-order perturbation integral over the {\it adiabatic time}
(conditions for a convergence of the adiabatic perturbation theory in
our case are discussed in the Sections below, see also Ref.~\cite{Dridi:2010}).
The integrand includes {\it poles} in the complex time plane due to 
the non-adiabatic coupling element. At the same time, however,
the poles of the coupling element represent 
{\it branchpoint singularities} of the adiabatic Hamiltonian,
which must be taken into account when the complex contour is
defined. 
We note that these singularities are known in the literature
as {\it transition points} (TPs)~\cite{Child:1978}.

The solution within the complex time plane method 
relies on the proper choice of the complex integration contour.
In particular, the integration contour used by Dykhne, Davis, and Pechukas
is suitable only for Hermitian systems which are
characterized by a time-symmetric adiabatic Hamiltonian.
However, {\it to describe the EP encircling dynamics, a different 
choice of the complex integration contour is required}.
The proposal of such a general contour (together with the associated
solution which includes contributions of
individual singularities to the survival amplitude) 
represents the main achievement of the present work.
The novel integration contour presented here can be used to describe
both time-symmetric and 
time-asymmetric dissipative dynamics (such as the one associated
with TAMS).

The presented complex contour method is developed on the
background of the problem where bound and resonance states
are coupled via a linearly chirped Gaussian laser pulse.
Importantly, the dipole coupling element between the states
is real valued. This physical problem results in time-symmetric
dynamics which is manifested as the behavior switch between
Rabi oscillations and rapid adiabatic passage.
Therefore, in this paper we also discuss particular
aspects of this problem, such as laser-atom interaction,
we derive new effective laser parameters for coherent control,
and based on the complex contour integration,
we provide an analytical fit to the solution for the
survival probability.

The present paper is organized as follows:
In Section~II we introduce the quantum dynamical
description of an atom driven by a chirped laser
pulse as a physical system for which the theory
will be practically demonstrated.

In Section~III we introduce the exceptional point, EP,
as a branchpoint singularity which arises when bound and
resonance states are coupled by a continuous wave laser.
We discuss the phenomenon of the
time-asymmetric mode switching, TAMS,
when the EP is dynamically encircled.
Then we go on to prove that the TAMS does not take place
upon the specific assumption of the {\it time-symmetry} of the adiabatic Hamiltonian.

In Section IV, we derive expressions for
the adiabatic amplitudes and survival probability
based on the {\it adiabatic perturbation theory, APT}.

In Section V, we introduce new {\it effective parameters for
Gaussian linearly chirped laser pulses} which greatly
simplify the calculations making them fully independent on
a particular atomic system. 
The effective parameters include the laser pulse area
which has been known in 
laser physics (see pulse area theorem~\cite{McCall:1967,McCall:1969,AllenEberly} 
or laser control~\cite{Vitanov:2001}).

In Section VI, we introduce {\it transition points}, TPs,
which pop up if Hamiltonian is analytically continued to the
complex time plane (through time-dependent parameters).
For the system under the study we discuss possible configurations
of paired TPs near the axis origin as well as asymptotic series
of TPs which evolve in distant regions of the plane.

In Section VII, we discuss {\it Puiseux series based on the branch
points (TPs) on the complex time plane}. 
The knowledge of the Puiseux series, apart from being necessary to
perform the
complex contour integration, 
represents a nice tool to derive non-adiabatic coupling
or study what happens when  two TPs coalesce.

In Section VIII, we study {\it residua} at TPs given by the
integration over the quasi-energy split (defined for
the adiabatic Hamiltonian at a given
complex time).

In Section~IX we introduce the so called {\it equivalue lines},
which are curves starting at the TPs and connecting points
 with the same imaginary component of the integral over the
 quasi-energy split.
These lines, essential in defining the complex integration
contour, have been introduced
earlier in the literature on the complex time method~\cite{Dykhne:1962,Davis:1975}.
We show explicitly how these lines are associated with
the first order local Puiseux expansion coefficient and
we discuss their asymptotic behavior, which is specifically
different in the Hermitian vs. non-Hermitian
cases. 

In Section~X, we define the {\it new complex contour} and
compare it with the contour used by Dykhne, Davis, and Pechukas (DDP).
It is demonstrated that while solutions based on the 
so called DDP formula, which has been applied at times also
to dissipative dynamics~\cite{Schilling:2006,Dridi:2010,Dridi:2012},
are approximate, the new integration contour represents a correct solution.

In Section~XI we present a derivation of analytical formulas for
the {\it contributions of the individual TPs to the survival amplitude}.
Two different types of contributions of each TP exist,
namely the residual and the branch cut contributions.

In Section~XII, we apply the results of the previous
Section~XI to the {\it concrete physical problem of Rabi-to-RAP behavior
switch} to the point of deriving formulas for the complex survival amplitude,
whereas in Section~XIII, we find analytical expressions for the survival probability.
The practical application of the obtained expressions 
relies on the knowledge of the residua at the TPs.
The residua must be obtained using a numerical procedure,
however, in order to make the formulas usable straight away, 
we provide an analytical fit for the residua. 

In Section~XIV we present our conclusions.

After the Section of conclusions we add a number of Appendixes
which include more details on the derivations included
in the main body of the paper, and/or numerical
illustrations associated with the derivations.

\section{Laser driven two-level atom}

\subsection{\label{system}Atomic transition}
\noindent
We assume an atom, which is temporarily driven by a chirped linearly polarized
electromagnetic pulse
which couples two atomic levels.
The atom is initially in one of its bound states, $|1\>\equiv|E_1\>$, 
and the field frequency 
is approximately tuned near a selected transition to another state,
which is a metastable resonance, 
$|2\>\equiv|E_2-i\Gamma/2\>$. 
Since the final state is a resonance, the state has a nonzero energy width $\Gamma$,
which is reflected in the fact that the Hamiltonian eigenvalue $(E_{2}-i\Gamma/2)$ is complex, 
where the real part $E_2$ indicates the mean energy value and the imaginary part is the
half-width of the energy uncertainty, see Ref.~\cite{Moiseyev} for a general
discussion of non-Hermitian
quantum mechanics.

We are interested in the level occupation of the initial state $|1\>$
as the pulse is over (i.e. the survival probability).
Such a quantity is relatively feasible in spectroscopic measurements and
as such it may be well thought of as an indicator of effects associated with
the exceptional point (EP) in different experimental setups.

\subsection{Rotating wave approximation}

\noindent
Since the laser pulses considered here include many
oscillations of field, it is sensible to give a separate
consideration to the fast motion represented by the radiative
field oscillations on one hand and
slow motion represented by the relatively slow change of the
pulse envelope and frequency.

By putting an atom to a laser field with a constant 
frequency $\omega$ and amplitude $\varepsilon_0$, the atomic levels are changed and shifted. This quantum
system is referred to as a {\it laser driven atom} and its levels
as {\it Floquet states}, solutions of the Floquet Hamiltonian
\begin{align}
\hat H_F = \hat H_0 + \hat V_{int-cw}(t';\omega,\varepsilon_0)
- i\hbar \frac{\partial}{\partial t'} ,
\end{align}
where $\hat H_0$ is the Hamiltonian for the field-free atom
and $\hat V_{int-cw}$ represents the interaction Hamiltonian
of the atom and field.  
The latter operator is sometimes
referred to as the ``photon operator'' as it allows to
include general multi-photon interaction in the classical
quantum dynamics limit.
Note that the time variable ($t'$) is here part of the
phase space, not a parameter.

We apply 
the rotating wave approximation (RWA) to the laser driven atom 
where only two Floquet states are taken into account (Appendix ~\ref{AP1}).
The RWA approximation is accurate if the driving frequency is tuned such that
it couples two particular levels of the field-free atom
supposed that the field strength is rather low.
The corresponding two-level Floquet Hamiltonian is represented as
\begin{align}
\hat H_F \approx \hbar\, |\mat \Psi\>
 \[ 
\begin{matrix}
0 & \frac{1}{2}\Omega \\
 \frac{1}{2}\Omega & \Delta
\end{matrix}
\]  \< \mat \Psi |,
\mylabel{H0}
\end{align}
in the matrix form, where the basis set is given by the atomic
field-free states with the added phase oscillations
\begin{align}
|\mat \Psi\> =
\[\begin{matrix}e^{-iE_2t'/\hbar+i\omega t'}\, |1\>,
\mylabel{EQbas}
&
e^{-iE_2t'/\hbar}\, |2\> \end{matrix}\] .
\end{align}
$\Omega$ represents  the Rabi frequency,
\begin{align}
	\Omega=\mu\varepsilon_0/\hbar ,
\mylabel{rab}
\end{align}
where $\mu$ is the
transition dipole element between field-free atomic states $|1\>$ and $|2\>$ (Appendix~\ref{mu12}). 
Further,
\begin{align}
\Delta=\omega - (\EE_2-\EE_1)/\hbar
\mylabel{detun0}
\end{align}
 is the frequency detuning, where $\EE_1$, $\EE_2$
represent the eigen-energies
of the field-free atomics states $|1\>$, $|2\>$, respectively. 

In the non-Hermitian case,
$\EE_1$, $\EE_2$ represent the {\it complex} eigen-energies
of the field-free atomics states in contrast to $E_1$, $E_2$
which represent only their respective real components.
In our case, where the state $|2\>$ is a metastable resonance,
\begin{align}
	&\EE_1 = E_1,
	&\EE_2 = E_2 - i\Gamma/2 ,
	\mylabel{EQcalE}
\end{align}
the detuning is complex, given by 
\begin{align}
	\Delta=\omega - [(E_2-i\Gamma/2)-E_1]/\hbar 
	=\omega - \omega_r + \frac{i\Gamma}{2\hbar} ,
\mylabel{detun}
\end{align}
where we used the resonance frequency $\omega_r$ defined as
\begin{align}
	\omega_r = \frac{1}{\hbar}(E_{2} - E_{1}) .
	\mylabel{omegar}
\end{align}
We
note that $bra$ of $\<2|$ represents the so called {\it left}
vector of the excited resonance state which satisfies
the special closure defined in the non-Hermitian quantum mechanics~\cite{Moiseyev}.

The Floquet states $\Phi_\pm$ obtained using the approximate Floquet Hamiltonian Eq.~\ref{H0},
\begin{align}
\hat H_F \Phi_\pm(\vec r,t';\omega,\varepsilon_0) =
\epsilon_\pm \Phi_\pm(\vec r,t';\omega,\varepsilon_0),
\mylabel{EQepm}
\end{align}
are given by,
\begin{align}
	&\[\begin{matrix}|\Phi_+\> &
|\Phi_-\> \end{matrix}\]
= |\mat \Psi\> \[ \begin{matrix} \sin \Theta & \cos \Theta \\ -\cos \Theta & \sin \Theta \end{matrix} \]
\mylabel{EQ2}
\end{align}
where $\Theta$ is defined as 
\begin{align}
\Theta = 1/2 \arctan {\Omega \over \Delta}
\mylabel{EQTh}
\end{align}
 and the associated adiabatic energies $\epsilon_\pm$ are given by
\begin{align}
\epsilon_\pm = \frac{\hbar}{2}\(\Delta\pm\delta\),
\mylabel{EQ3}
\end{align}
where
\begin{align}
\delta =  \sqrt{\Delta^2+\Omega^2} .
\mylabel{delta}
\end{align}

\subsection{\label{SectionCC}Close-coupled equations for laser pulse driven atom}

\noindent
As we mention above, we assume an atom driven by
 a long laser pulse including many optical cycles.
Such a laser pulse is defined by its envelope given by varying
amplitude $\varepsilon_0(t)$ and varying frequency $\omega(t)$
which defines the frequency chirp.
The Floquet wavefunctions defined above
represent a reasonable time-dependent adiabatic basis set
\begin{align}
\psi_{ad,\pm} =
e^{-\frac{i}{\hbar}\int\limits_0^t dt' \epsilon_\pm(t)}\,\Phi_\pm[\r,t;\omega(t),\varepsilon_0(t)] 
\end{align}
for the actual time-dependent wavefunction.
Note that the included exponential factor pops up when
the Floquet state is substituted into time-dependent Sch\"odinger
equation as shown in  Appendix~\ref{APADIAB}.

In many cases, a single adiabatic
basis function would be sufficient to describe the laser dynamics.
However, in our case a strong non-adiabatic coupling 
between
the Floquet states $|\Phi_\pm\>$ 
defined by the element 
\begin{align}
N(t) = \(\psi_+^{(l)}|\partial_t \psi_-\)
\end{align}
takes place along the adiabatic path, which is defined by
$\omega(t)$, $\varepsilon_0(t)$.
This situation happens due to the fact that two field-free states are
strongly mixed by the field.
Using the definition of the Floquet states Eq.~\ref{EQ2}
one can show that the non-adiabatic coupling $N(t)$ is given by 
\begin{align}
N(t) = \frac{d\Theta(t)}{dt} .
\mylabel{EQ7a}
\end{align}
The variable $\Theta$ represents a measure of mixing of the field-free
states within the Floquet states. As we will show below, the two field
free states are switched along the encircling contour, thus $\Theta$
and its time-derivative is not negligible.

The exact time-dependent wavefunction beyond the adiabatic approach is
defined in terms of the adiabatic basis set $\psi_{ad,\pm}(t)$ such that
\begin{align}
\psi(\vec r,t) = e^{-\frac{i}{\hbar}\int\limits_{\tt}^t dt' \epsilon_-(t')} a_-(t) \Phi_{-}[\vec r,t;\omega(t),\varepsilon_0(t)] \cr
\quad + e^{-\frac{i}{\hbar}\int\limits_{\tt}^t dt' \epsilon_+(t')} a_+(t) \Phi_{+}[\vec r,t;\omega(t),\varepsilon_0(t)] ,
	\mylabel{adiabexp}
\end{align}
Then we substitute this ansatz into the time-depenedent Schr\"odinger equation, 
\begin{align}
i\hbar \frac{\partial}{\partial t} \psi(\vec r,t) =
\left \{ \hat H_0 + \hat V_{int-cw}[t;\omega(t),\varepsilon_0(t)]  \right\} \psi(\vec r,t) ,
\mylabel{EQTDSE}
\end{align}
from where we get the close-coupled equations:

\begin{align}
\dot a_+(t) = - a_-(t) N(t) e^{i\int\limits_{\tt}^t dt' \delta(t')} \cr
\dot a_-(t) = a_+(t) N(t) e^{-i\int\limits_{\tt}^t dt' \delta(t')}  .
\mylabel{CC}
\end{align}
where $a_\pm$ are non-adiabatic amplitudes of the adiabatic states $\psi_{ad,\pm}$;

\section{Contour encircling of exceptional point}

\subsection{\label{SEP1}Exceptional point in the frequency-amplitude plane}

\noindent
Let us investigate, how the complex energies $\epsilon_\pm$ 
vary as functions of the continuous wave (CW)
laser parameters -- frequency $\omega$ and laser strength $\varepsilon_0$,
see Fig.~\ref{FigSurf}. The complex energy surfaces consist the real parts $\re\, \epsilon_\pm$,
and also the widths $\Gamma_\pm = -2\,\im\,\epsilon_\pm$. 
The surfaces include an {\it exceptional point -- EP}, where a complex degeneracy is found,
\begin{align}
	\epsilon_+ = \epsilon_- 
	\quad
	\Leftrightarrow
	\quad
	\delta = \sqrt{\Delta^2+\Omega^2} = 0.
\end{align}
If we substitute for the detuning $\Delta$ and Rabi frequency $\Omega$ (Eqs.~\ref{detun} and
\ref{rab}), we get the critical laser parameters for the EP,
\begin{align}
	& \varepsilon_0^{EP} = \frac{\Gamma}{2\re \mu}, 
\end{align}
and
\begin{align}
	& \omega^{EP} = \omega_r - \frac{\Gamma}{2\hbar}\frac{\im\mu}{\re\mu}  ,
	\mylabel{EQomegaEP}
\end{align}
where $\omega_r$ is the resonance frequency between the two field-free states
(Eq.~\ref{omegar}).
Note that in this paper we will address the special case where 
the transition dipole moment $\mu$ is real defined 
therefore the frequency position of the EP ($\varepsilon_0^{EP}$, $\omega^{EP}$) is given by
\begin{align}
&\varepsilon_0^{EP} = \frac{\Gamma}{2 \mu}, 
	&\omega^{EP} \equiv \omega_r .
\mylabel{epsEP}
\end{align}
This assumption will take effect only
few paragraphs below when dealing with time-symmetry of quantum dynamics.
\begin{figure}[!h]
\includegraphics[width= 3 in
]{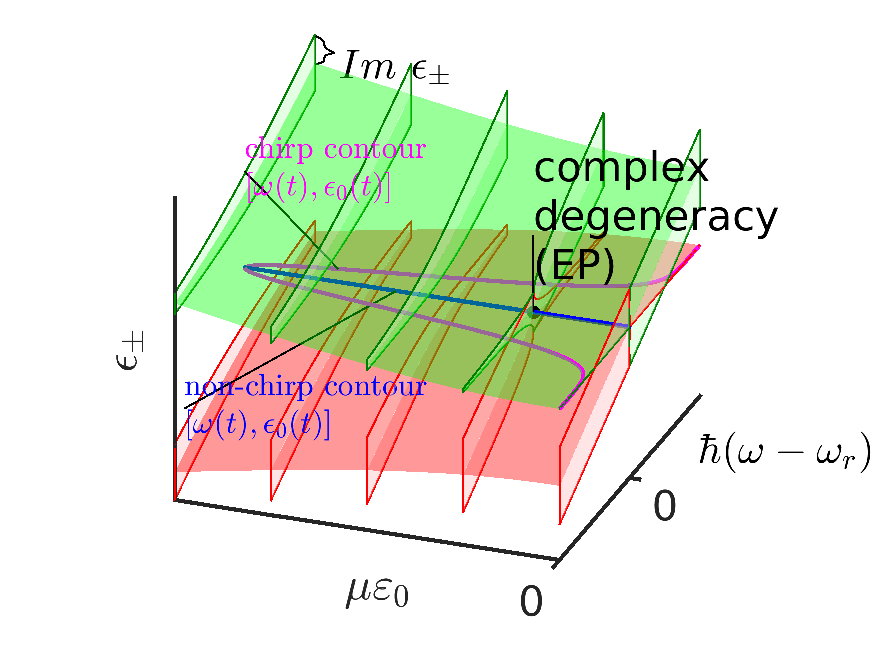}
\caption{Adiabatic energy surfaces obtained as solutions of a 2x2 Hamiltonian
	for two coupled states by a CW laser of frequency $\omega$ and
	laser strength $\varepsilon_0$. The energies are complex, since
	the excited state is a resonance. They consist of the real parts,
	which are demonstrated by the horizontal surfaces, and the imaginary parts,
	which are responsible for finite widths $\Gamma_\pm=-2\im\epsilon_\pm$ 
	of the surfaces. The widths are displayed using the vertical ``cuts''.
	We mark a point of complex degeneracy (exceptional point, EP).
	We also draw possible types of the laser ``contours'', which
	correspond to the definition of the studied Gaussian linearly
	chirped laser pulses, Eq.~\ref{S2}, \ref{S3}.}
\mylabel{FigSurf}
\end{figure}

\subsection{\label{SEP2}Quasi-energy split near exceptional point}

\noindent
The quasi-energy split is defined in Eq.~\ref{delta} which can
be also written as,
\begin{align}
& \delta = \Delta\sqrt{1 + \lambda^2 } = \Delta\sqrt{\lambda + i}\cdot\sqrt{\lambda-i}, 
\end{align}
where
\begin{align}
& \lambda \equiv \frac{\Omega}{\Delta} .
\mylabel{EQlambda}
\end{align}
It features two distinct EPs given by
\begin{align}
\lambda_{EP} = \pm i .
\end{align}
It is known the neigborhood of the EP can be expressed using the
Puiseux expansion~\cite{Moiseyev},
\begin{align}
\delta(\lambda) = \sum\limits_{k=1}^\infty a_k (\lambda - \lambda_{EP})^{k/2}.
\end{align}
Since we have two EPs, we define two neighborhood expansions such that
\begin{align}
\delta_\pm(\omega,\varepsilon_0) =
\sqrt{\Delta}\,\sqrt{\lambda\pm i} =
=\sqrt{\omega-\omega_r +\frac{i\Gamma}{2\hbar} \pm i\frac{\mu}{\hbar}\varepsilon_0} ,
\end{align}
such that apparently
\begin{align}
\delta(\omega,\varepsilon_0) = \delta_+ \cdot \delta_- .
\end{align}
We assume that the quasi-energy split is a product function of the 
Poiseux series which are associated with $\delta_+$ and $\delta_-$,
respectively, i.e. with the individual EPs.

Persuing this goal we replace $\omega_r$ with
$\omega_{EP}$ as the reference using Eq.~\ref{EQomegaEP},
\begin{align}
\delta_\pm
=\sqrt{\omega-\omega_{EP} 
-\frac{\Gamma}{2\hbar}\frac{\im \mu}{\re \mu}
+\frac{i\Gamma}{2\hbar} \pm i\frac{\mu}{\hbar}\varepsilon_0} ,
\end{align}
and then we combine the terms including $\Gamma$ such that
% \begin{align}
% \delta_\pm
% =\sqrt{\omega-\omega_{EP} 
% -\frac{\Gamma}{2\hbar}\frac{(\im \mu - i\re\mu)}{\re \mu}
 % \pm i\frac{\mu}{\hbar}\varepsilon_0}
% \end{align}
%which is further simplified as
\begin{align}
\delta_\pm
=\sqrt{\omega-\omega_{EP} 
+i\frac{\Gamma}{2\hbar}\frac{\mu}{\re \mu}
 \pm i\frac{\mu}{\hbar}\varepsilon_0} .
\end{align}
Now using the definition of $\varepsilon_0^{EP}$ we get
(the sign of $\varepsilon_0^{EP}$ must correlate with the definition
of $\lambda_{EP}$)
% \begin{align}
% \delta_\pm
% =\sqrt{\omega-\omega_{EP} 
% +i\frac{\mu}{\hbar}(\pm\varepsilon_0 \mp \varepsilon_{0\pm}^{EP})}
% \end{align}
\begin{align}
\delta_\pm
=\sqrt{(\omega-\omega_{EP})
\pm i\frac{\mu}{\hbar}(\varepsilon_0 - \varepsilon_{0\pm}^{EP})} .
\mylabel{EQ32}
\end{align}

\subsection{State exchange along contours encircling EP\myslabel{Sencirc}}

\noindent
It has been shown before that when EP is stroboscopically encircled,
the Floquet states are exchanged~\cite{Moiseyev,Lefebvre:2009}.
Let us show here another short proof to this.
We define a general encircling contour such that
\begin{align}
&\omega(\varphi) = \omega_{EP} + \rho(\varphi)\, \cos\(\varphi\) , \cr
&\varepsilon_0(\varphi) = \varepsilon_0^{EP} + \rho(\varphi)\frac{\re\mu}{\hbar}\,\sin\(\varphi\)
, \cr
&\quad
0<\varphi<2\pi , \cr
&\quad \rho(\varphi) > 0 , \quad \rho(\varphi) = \rho(\varphi+k\cdot 2\pi).
\end{align}
where $\phi$ represents the angular variable of the encircling. 
$\rho$ represents
the varying distance from the encircled EP, always a positive real defined number.
The quasi-energy split along this encircling contour is given by,
\begin{align}
\delta_\pm(\varphi)
=\sqrt{\rho(\varphi)}\ \sqrt{\cos(\varphi)
\pm i\frac{\mu}{\hbar}\sin(\varphi)} .
\end{align}
We intentionally simplify this expression such that the
phase due to the encircling appears before the square root such that
\begin{align}
&\delta_\pm(\varphi)
=\cr
%& \sqrt{\rho(\varphi)}\ \sqrt{\cos(\varphi) \mp \frac{\im\mu}{\re\mu}\sin(\varphi)
%\pm i \sin(\varphi)}
%=\cr
& \sqrt{\rho(\varphi)}\ 
e^{\pm i\varphi/2} \, \sqrt{1 \mp \frac{\im\mu}{\re\mu}\sin(\varphi) e^{\mp i \varphi}
}
\cr
\end{align}
Assuming $|\im\mu| < |\re\mu|$, one can use the Taylor expansion for
the square root which is periodic in $\phi$
with the period $2\pi$. The exponential prefactor shows that
as $\varphi=2\pi$ we get the sign change.
The sign change applies for the encircling of
one single EP, either $\lambda_{EP}=+i$ or $\lambda_{EP}=-i$.
Note that in the physical sense, it is impossible to encircle
the EP $\lambda_{EP}=-i$, which corresponds to a negative
field amplitude $(-\varepsilon_0^{EP})$, encircling of the
physical EP therefore is always associated with the sign change
of the quasi-energy split.

The sign change of the quasi-energy split $\delta$ upon the encircling implies
 an exchange between the Floquet states, see the definition 
 of the state energies, Eq.~\ref{EQ3}.
We will assume that the initially occupied state is bound,
and it is coupled with a resonance.
The quasi-energy split before the interaction is
given by the diagonal Floquet Hamiltonian, which is given by Eq.~\ref{H0}
for $\Omega=0$.
Using Eqs.~\ref{detun0} and \ref{EQcalE} we get,
\begin{align}
& \delta(t_i) = \[\omega(t_i) -\omega_r \] + i\frac{\Gamma}{\hbar} .
\mylabel{deltaini}
\end{align}
For simplicity, let us first assume a contour defined by a the same 
initial and final frequencies,
$\omega(t_i) = \omega(t_f)$, and of course it is assumed
that the laser amplitude is again zero
in the end of such a process.
Then in the end of the process, assuming the sign change proved
above,
\begin{align}
& \delta(t_f) = -\[\omega(t_f) -\omega_r \] - i\frac{\Gamma}{\hbar} .
\mylabel{deltafin}
\end{align}

Now, in the most simple case, the pulse chirp is linear as we show in Fig.~\ref{EQTDSE}, 
where the final frequency is {\it not} the same as the initial one. 
Yet, even such a process represents an EP encircling where the
contour is closed hypothetically
through the zero amplitude axis, $\varepsilon_0(t>t_f)\to +0$.
Importantly, the Floquet states, being completely decoupled
from each other along the gedanken 
closing contour where $\Omega\to+0$ (Eq.~\ref{H0}),
are given by the same field-free solutions
from the actual end of the laser interaction up to the end
of the hypothetical closing contour.
Eq.~\ref{deltafin} remains formally the same, but now
$\omega(t_f)\ne \omega(t_i)$.

In the case of the laser pulse specifically, not only the Floquet states
are exchanged, but the field-free states are exchanged as well~\cite{Lefebvre:2009}.
This fact is related to the mixing angle $\Theta(t)$ which is 
defined in Eqs.~\ref{EQ2} and \ref{EQTh} and now it depends on 
the time parameter $t$ through the varying $\varepsilon_0(t)$ and $\omega(t)$
in the pulse.
$\Omega(t)/\Delta(t)$ has a zero
value both on the start and end of the pulse, as the Rabi frequency $\Omega$ 
is proportional to 
the field amplitude $\varepsilon_0(t)$ (Eq.~\ref{rab}).
This implies that $\Theta(t\to\pm \infty)$
 is either equal to 0 or $\pi$.
$\Theta$ determines the Floquet states as
linear combinations of the field free states (the states $|1\>$ and
$|2\>$, optionally including the additional phase factors, Eq.~\ref{EQbas}).
Bearing in mind that the Floquet states
are not the same as the EP is stroboscopically encircled ($\varphi=0$ $\to$
$\varphi=2\pi$, Eq.~\ref{EQbas}), we assess that
$\Theta$ is initially given by 0 and finally by $\pi$.
The radical change of $\Theta$ is the reason for the necessity of
including the non-adiabatic coupling as we discussed in Section~\ref{SectionCC}.
Combining the asymptotic values of $\Theta$ and the definition of
Floquet states in Eq.~\ref{EQ2} also represents a proof
that the field-free states are 
always exchanged along the contours
defined by finite laser pulses which encircle the EP.

\subsection{Dynamical encircling of exceptional point}

\noindent
The state exchange that appears for stroboscopic encircling
led to designing and studying realistic systems where 
the EP was {\it dynamically encircled}.
The encircling dynamics occurs when the interaction term which takes
place in the time-dependent Schr\"odinger equation Eq.~\ref{EQTDSE}
(the interaction term is defined explicitely in Appendix~\ref{AP1})
is constructed
based on the stroboscopic encircling contour $[\omega(t),\varepsilon_0(t)]$
defined above.
{\it Time-asymmetric state switch} represents a characteristic behavior
in the systems where EP is dynamically encircled~\cite{Uzdin:2011,Gilary:2013,Kapralova-Zdanska:2014,Feng:2014,Doppler:2016,Xu:2016,Zhong:2018,Oberreiter:2018,Li:2019,Fernandez:2020,Feilhauer:2020}.
This phenomenon  means that the quantum dynamics is different for the two possible opposite
encircling directions, and in particular, the population switch, which
we saw for the stroboscopic encircling, is open in
one direction but closed in the other.

Let us explain here shortly the reason why this phenomenon takes place (see also Refs.~\cite{Uzdin:2011,Gilary:2013,Kapralova-Zdanska:2014}).
The quasi-energy split $\delta(t)$ which is in the exponents of
the close coupled equations (Eq.~\ref{CC}) includes a non-zero imaginary component
due to complex-defined Floquet energies $\epsilon_\pm$ (compare Eqs.~\ref{EQepm}, \ref{EQ3}).
It is illustrative to split between these components such that,
\begin{align}
& \dot a_\pm(t) = \mp a_\mp(t)\  N_{\mp\im\delta}(t)\ 
e^{\pm\int\limits_0^t dt' \re \delta(t')} , \cr
& N_{\mp\im\delta}(t) = N(t)\  e^{\mp\int\limits_0^t dt' \im \delta(t')} .
\mylabel{EQtimas}
\end{align}
Although it is not usually done this way, we have intentionally
assigned the contribution of the imaginary energy components 
to the non-adiabatic coupling term to show it
functioning as a dumping factor that may dump/promote
non-adiabatic jumps between the Floquet states,
depending on the {\it sign} of the exponent.
As the sign is opposite for each one of the amplitudes $a_+$
and $a_-$, respectively,
the time-derivative of one of the amplitudes, either $\dot a_+$ or
$\dot a_-$ on the left hand side of Eq.~\ref{EQtimas},
is always dumped with respect to the other one.
Importantly, the effect of damping/promotion is switched with the sign of the propagation time $t$.

\subsection{Time-symmetric encircling}

\noindent
The time-asymmetric state switch has been approved in different cases, including an exprimental verification~\cite{Doppler:2016},
however, the explanation of the time-asymmetry itself~\cite{Uzdin:2011,Gilary:2013}
 suggests that there
is a condition which has to be fulfilled should the time-asymmetric switch take
place, which is the time-asymmetry of the quasi-energy split $\delta(t)$.
Let us suggest the contrary, namely,
\begin{align}
	\delta(-t^*) = \[\delta(t)\]^* .
	\mylabel{EQ8d10}
\end{align}
It is the time-integral over quasi-energy split that figures in the exponents
in equations, Eqs.~\ref{EQtimas} or~\ref{CC}.
It is suggested that if Eq.~\ref{EQ8d10} is applicable,
a {\it time-symmetrical EP encircling dynamics} takes place
and no time-asymmetric state switch can be observed although EP apparently 
{\it is encircled}.

In order to put our argument on solid grounds, we 
prove that Eq.~\ref{EQ8d10} implies proper time-symmetry relations
 for the integral over the quasi-energy split, which is what indeed
figures in the exponents of the close-coupled equations Eq.~\ref{EQtimas}.
% (the same relation will be used also below within a solution using the complex time-plane method).
Let us start the derivation by defining
the integral over $\delta(t)$ to $(-t^*)$.
As the limit is taken in analogy to the left hand side of Eq.~\ref{EQ8d10}, we denote it as $L$, 
\begin{align}
	L=\int\limits_0^{-t^*} dt' \,\delta(t') .
\end{align}
By substituting $t' = -t''^*$ for the integration variable we get
\begin{align}
	L= 
	-\int\limits_0^{t} dt''^* \,\delta(-t''^*)
\end{align}
and then by using the symmetry of the quasi-energy split defined in Eq.~\ref{EQ8d10}
we get
\begin{align}
	L= 
	-\int\limits_0^{s} dt''^* \,\delta^*(t'')
\end{align}
which yields the final time-symmetry relation for the integral over quasi-energy split
given by
\begin{align}
	\int\limits_0^{-t^*} dt' \,\delta(t') =
	-\(\int\limits_0^{t} dt' \,\delta(t') \)^* .
\end{align}

Now as we divide the real and imaginary components of the quasi-energy split 
assuming integration on the real axis we get
\begin{align}
	& \int\limits_0^{t} dt' \,\re \delta(t') = 	-\int\limits_0^{-t} dt' \,\re \delta(t')  ,	\cr
	& \int\limits_0^{t} dt' \,\im \delta(t') = 	 \int\limits_0^{-t} dt' \,\im \delta(t')  .
	\mylabel{EQsymenc}
\end{align}
When the second equation is substituted to the close coupled 
equations Eq.~\ref{EQtimas}, namely to the term defining the non-adiabatic term $N_{\mp\im\delta}(t)$,
we prove that the obtained exponential damping of the non-adiabatic coupling element
is exactly the same upon the exchanged direction of encircling.
As the ultimate reason for the time-asymmetric atomic switch is not present,
the dynamics follows other rules that are yet to be studied.

To be clear, Eqs.~\ref{EQsymenc} do not prove that the dynamics will be the same when the
direction of the encircling is switched because the non-adiabatic
coupling element $N(t)$	does not necessarily posses any time-symmetry based on Eq.~\ref{EQ8d10}
alone.
In some cases, however, it is possible to achieve a fully symmetrical dynamics
where the strict condition
 \begin{align}
	& \Delta(-t^*) = - [\Delta(t)]^* , \cr
	& \Omega(-t^*) = [\Omega(t)]^* ,
	\mylabel{EQ5d5}
\end{align}
is applicable.
Eqs.~\ref{EQ5d5} assure the time-symmetry of the non-adiabatic coupling
\begin{align}
N(-t^*) = [N(t)]^* ,
\end{align}
as one can show by using Eqs.~\ref{EQ7a} and \ref{EQTh}.
The conditions Eqs.~\ref{EQ5d5} imply also the validity of Eq.~\ref{EQ8d10} as one can show 
by expressing the square of the energy split (Eqs.~\ref{delta} and \ref{EQ5d5}) 
\begin{align}
	& \delta^2(-t^*) = \Delta^{2}(-t^*) + \Omega^2(-t^*) \cr
	& \quad = \Delta^{*2}(t) + \Omega^{*2}(t) = \[ \delta^2(t)\]^*  .
\end{align}
As the square is a complex conjugated value, the same relation
is applicable for the quasi-energy split itself,
Eq.~\ref{EQ8d10}.

\section{Adiabatic perturbation theory}

\subsection{Boundary conditions for time-dependent wavefunction}

\noindent
Let us consider the initial and final conditions of the time-dependent wavefunction
$\psi(\r, t)$ defined in Eq.~\ref{adiabexp}.
It is assumed that the initial and final wavefunctions,
$\psi(\r, t\to \pm\infty)$, are associated with the field-free states $|1\>$, $|2\>$,
or more precisely, with the basis set states given in Eq.~\ref{EQbas}.
Namely we set,
\begin{align}
& \Phi_-(t\to -\infty) =  e^{-\frac{i}{\hbar}E_2 t + i\omega(t) t}\  |1\> , \cr
& \Phi_+(t\to -\infty) = e^{-\frac{i}{\hbar}E_2 t} \ |2\>, \cr
& \Phi_-(t\to \infty) =  e^{-\frac{i}{\hbar}E_2 t}\  |2\> , \cr
& \Phi_+(t\to \infty) = e^{-\frac{i}{\hbar}E_2 t + i\omega(t) t} \ |1\>, \
\mylabel{EQfloq0}
\end{align}
This setting reflects the fact that the adiabatic Floquet states
are switched as the EP is encircled.
When the definitions of the Floquet states given in Eqs.~\ref{EQfloq0} 
are substituted to the definition of $\psi(\r, t)$ (Eq.~\ref{adiabexp}),
for  $t\to -\infty$ first, we get,
\begin{align}
&\psi(\r,t\to-\infty) = \cr
&\quad
e^{-\frac{i}{\hbar}\int\limits_0^{-\infty} dt' \epsilon_-(t')} 
 a_-(t\to-\infty) \ 
e^{-\frac{i}{\hbar}E_2 t + i\omega(t) t}\  |1\> \cr & \quad
+
e^{-\frac{i}{\hbar}\int\limits_0^{-\infty} dt' \epsilon_+(t')} 
 a_+(t\to-\infty) \ 
e^{-\frac{i}{\hbar}E_2 t}\  |2\> .\cr
\end{align}
This expression must be compared with 
the actual initial condition following from the time-evolution of the initial field-free state,
\begin{align}
\psi(\r, t\to -\infty) = e^{-\frac{i}{\hbar}E_1 t}\  |1\> .
\end{align}
This implies the initial values of the non-adiabatic amplitudes
\begin{align}
	& a_-(t\to-\infty)=e^{ \frac{i}{\hbar}\int\limits_{0}^{-\infty} dt'\, \epsilon_-(t')} \  e^{i[\omega_r - \omega(t)]t}, \cr
	& a_+(t\to-\infty)=0 .
	\mylabel{EQini}
\end{align}
Note that we used the definition of the resonance frequency $\omega_r$ in Eq.~\ref{omegar}.
The same is done for the final boundary condition where
\begin{align}
&\psi(\r,t\to\infty) = \cr
&\quad e^{-\frac{i}{\hbar}\int\limits_0^{\infty} dt' \epsilon_-(t')} 
 a_-(t\to\infty) \ 
e^{-\frac{i}{\hbar}E_2 t}\  |2\> \cr & \quad
+
e^{-\frac{i}{\hbar}\int\limits_0^{\infty} dt' \epsilon_+(t')} 
 a_+(t\to\infty) \ 
e^{-\frac{i}{\hbar}E_2 t + i\omega(t) t}\  |1\> \cr
\end{align}
is obtained using Eqs.~\ref{adiabexp} and \ref{EQfloq0}.
The actual wavefunction after the pulse 
may be written as a linear combination
of the two possibly occupied non-interacting field free states
such that
\begin{align}
\psi(\r, t\to \infty) = a_1\ e^{-\frac{i}{\hbar}E_1 t}\  |1\>  + a_2 \ e^{-\frac{i}{\hbar}{\EE_2} t}\  |2\> ,
\mylabel{Psifin}
\end{align}
where $a_{1,2}$ represent time-independent complex survival and excitation amplitudes, respectively.
Note that $\EE_2$ is the complex energy of the excited resonance state $|2\>$, Eq.~\ref{EQcalE}.
From here we get the relation between the non-adiabatic amplitude $a_-(t\to\infty)$
to the survival amplitude given by
\begin{align}
a_+(t\to\infty) = a_1\ e^{\frac{i}{\hbar}\int\limits_0^{\infty} dt' \epsilon_+(t')} 
e^{i[\omega_r-\omega(t)] t} .
\mylabel{EQfin}
\end{align}

\subsection{Adiabatic perturbation theory for non-adiabatic amplitudes}

\noindent
By integrating Eq.~\ref{CC} we obtain
the amplitude of the coupled adiabatic state $\Phi_+$ , such that
\begin{align}
a_+(t) & = a_+(t\to-\infty) \cr
&- \int\limits_{-\infty}^t dt'\, a_-(t') N(t) e^{i\int\limits_{\tt}^{t'} dt'' \delta(t'') } ,
\end{align}
which is simplified using the initial conditions
Eq.~\ref{EQini} such that
\begin{align}
a_+(t) = 
- \int\limits_{-\infty}^t dt'\, a_-(t') N(t) e^{i\int\limits_{\tt}^{t'} dt'' \delta(t'') } ,
\mylabel{EQ6}
\end{align}
Similarly, $a_-(t)$ is given by
\begin{align}
a_-(t) & = a_-(t\to-\infty) \cr
&+  \int\limits_{-\infty}^t dt'\, a_+(t') N(t') e^{-i\int\limits_{\tt}^{t'} dt'' \delta(t'') } .
\mylabel{EQ6b}
\end{align}
By subsequently substituting Eqs.~\ref{EQ6} and \ref{EQ6b}, we obtain
the {\it perturbation series},
\begin{align}
&a_-(t) = a_-(t\to-\infty)\cdot \sum\limits_{j=0,2,...}^\infty v^{(j)} (t),\cr
&a_+(t) = a_-(t\to-\infty)\cdot \sum\limits_{j=1,3,...}^\infty v^{(j)} (t),
\mylabel{EQperts}
\end{align}
where
\begin{align}
	&
	v^{(j+1)}(t) = (-)^{j+1} \cr
	& \times
	\int_{-\infty}^t dt' \,
v^{(j)}(t')\, N(t')\,
e^{i (-1)^{j} 
\int_{\tt}^{t'} dt'' \delta(t'') }
\mylabel{V1}
\end{align}
and
\begin{equation}
v^{(0)}(t) = 1.
\end{equation}
The prefactor $a_-(t\to-\infty)$ has been defined in Eq.~\ref{EQini}.

A convergence of the perturbation series is the key
assumption here.
The adiabatic perturbation series has been proven generally convergent 
if the {\it less dissipative state is initially occupied}, see Refs.~\cite{Dridi:2010}. 
This condition holds in the present case 
where the initial state is represented by the
{\it bound} state $|1\>$ while the excited
state is a {\it resonance}, $|2\>$.
We provide an additional numerical verification of the convergency
applicable to the studied case described below based
on Gaussian chirped pulses in Appendix~\ref{AP:APTnum}.

\subsection{Survival probability\myslabel{Ssurv}}

\noindent
The survival probability $p_1$ is defined
as the square of the absolute value of the complex
survival amplitude $a_1$, Eq.~\ref{Psifin},
\begin{align}
p_1 = |a_1|^2 .
\end{align}
The survival amplitude $a_1$ is related to the final
non-adiabatic amplitude $a_+(t\to\infty)$ through
Eq.~\ref{EQfin}.
Using the adiabatic perturbation theory to
obtain $a_+(t\to\infty)$, Eqs.~\ref{EQperts},
we get the expression,
\begin{align}
a_1 & = a_-(t\to -\infty) 
\lim\limits_{t\to-\infty}
e^{-i[\omega_r-\omega(t)] t} \cr
& \times 
e^{-\frac{i}{\hbar}\int\limits_0^{\infty} dt' \epsilon_+(t')}  
 \sum\limits_{j=1,3,...}^\infty v^{(j)} (t\to\infty) .
\end{align}
Using the initial condition Eq.~\ref{EQini}
we obtain
\begin{align}
a_1 & = f\cdot
 \sum\limits_{j=1,3,...}^\infty v^{(j)} (t\to\infty) ,
\mylabel{EQa10}
\end{align}
where
$f$ represents an important {\it normalization} and phase factor given by
\begin{align}
	f = \binf \cdot \rinf .
\mylabel{EQfdef}
\end{align}
The survival probability within the adiabatic perturbation
theory is defined as
\begin{align}
p_1 = |f|^2 \cdot \left |
\sum\limits_{j=1,3,...}^\infty v^{(j)} (t\to\infty)
\right |^2 .
\mylabel{psurv2}
\end{align}
The normalization and phase factor $f$ can be further simplified
for the case of {\it time-symmetric EP encircling} defined above
in Eq.~\ref{EQ5d5}, taking the following steps.
First,  the exponents of the two terms are merged into one integral such that
\begin{align}
	f=\exp\[ -\frac{i}{\hbar} \int\limits_0^{\infty}
	dt\,
	\epsilon_+(t) + \epsilon_-(-t)\] .
\end{align}
Now, we substitute for $\epsilon_\pm$ using Eq.~\ref{EQ3}:
\begin{align}
	f=\exp\[ -\frac{i}{2} \int\limits_0^{\infty}
	dt\, \{
	\Delta(t) + \Delta(-t) 
	+ \delta(t) - \delta(-t)\} \] .
\end{align}
For the time-symmetric EP encircling defined in Eq.~\ref{EQ8d10} we get
\begin{align}
	f=\exp\[ \int\limits_0^{\infty}
	dt\,
	\{ -i \frac{\Delta(t) + \Delta(-t) }{2}
	+\im \delta(t)\} \]  .
\end{align}
The dynamical detuning $\Delta(t)$ has the constant imaginary part given by the
resonance width $\Gamma$, Eq.~\ref{detun}, thus
\begin{align}
	f=f_0 \cdot \lim\limits_{T\to\infty}
	\exp\[ \frac{\Gamma\,T}{2\hbar} + \int\limits_0^{T}
	dt\, \im \delta(t)\]  ,
\mylabel{EQfdefsym}
\end{align}
where $f_0$ represents a mere phase factor given by
\begin{align}
f_0 & = \exp\[ -i \int\limits_0^{\infty}
	dt\,
	\re \left\{ \frac{\Delta(t) + \Delta(-t) }{2}
	\right\} \]  \cr
& = \exp\[ -i \int\limits_0^{\infty}
	dt\,
	\left\{ \frac{\omega(t) + \omega(-t) }{2}
	\right\} \] 
,
\end{align}
with no effect on the calculated survival probability $p_1$, Eq.~\ref{psurv2}.
Note that in the case of the fully time-symmetric dynamics
which fulfills the more strict conditions Eq.~\ref{EQ5d5},
\begin{align}
f_0 = 1.
\end{align}

	% \begin{align}
% \tau \gg \frac {2\pi}{\omega_r} ,
% \mylabel{S1}
% \end{align}

\section{Gaussian laser pulse}

\subsection{\label{Gauss}Gaussian encircling contour}
\noindent
Let us consider a Gaussian encircling contour defined in the
frequency-amplitude plane such that
\begin{align}
\varepsilon_0(t) = \varepsilon_0^{max} e^{-t^2/2\tau^2}
\mylabel{S2}
\end{align}
and 
\begin{align}
\omega(t) = \omega_r + \alpha t .
\mylabel{S3}
\end{align}

This contour leads to the time-symmetric encircling
supposed that the transition dipole moment $\mu$ (which codefines
the Rabi frequency Eq.~\ref{rab}) is {\it real defined}.
By substituting definitions Eqs.~\ref{S2} and \ref{S3} 
to Eqs.~\ref{rab} and \ref{detun} such that we obtain
\begin{align}
& \Omega (t) = \frac{\mu\varepsilon_0^{max}}{\hbar} e^{-t^2/2\tau^2} , \cr
& \Delta (t) = \alpha \, t + \frac{i\Gamma}{2\hbar} .
\mylabel{EQOD}
\end{align}
Using these particular definition and assuming
$\mu\in\Re$ we see that the conditions given by Eqs.~\ref{EQ5d5} are satisfied.

\subsection{Relative Gaussian pulse parameters}

\noindent
In this Section we derive effective pulse parameters for two-level atoms in linear Gaussian chirps.
To this point, the laser pulse is defined via its length $\tau$,
chirp $\alpha$, carrier frequency $\omega_r$, and
maximum peak strength $\varepsilon_0^{max}$, as given by Eqs.~\ref{S2} and \ref{S3}.
These parameters lead to expressions for the quantum dynamics
including atomic parameters such as the transition dipole moment $\mu$
and resonance width $\Gamma$.

It is known however that the quantum dynamics can be reduced to a problem which is
independent on particular atomic parameters in some cases. Such is 
the case of non-dissipative two-level atoms in non-chirped pulses where
the result of the dynamics is defined by a single
effective parameter of a laser pulse, namely the {\it pulse area}, 
\begin{align}
\theta = \int\limits_{-\infty}^\infty dt\, \Omega(t) ,
\mylabel{EQarea}
\end{align}
see the pulse area theorem~\cite{McCall:1967,McCall:1969,AllenEberly}
and $\pi$-pulse method in laser control~\cite{Vitanov:2001}.

Let us start our considerations 
by replacing the physical time $t$
with the effective relative time $s$ such that
\begin{align}
s = t/\tau ,
	\mylabel{sdef}
\end{align}
using the pulse length $\tau$ defined generally as
\begin{align}
\tau = \frac{\int\limits_{-\infty}^\infty dt\, t\, \Omega(t)}{\theta} .
\end{align}
This definition of 
$\tau$ coincides with the one given in Eq.~\ref{S2}.
We redefine the key quantities for the dynamics
as functions of $s$.
The quasi-energy split and non-adiabatic amplitudes
are defined such that
\begin{align}
	&\bar\delta(s) \equiv \delta(s\tau), \cr
	&\bar a_\pm(s) \equiv a_\pm(s\tau) .
	\mylabel{deltaGL1}
\end{align}
The non-adiabatic coupling will be defined as
\begin{equation}
\bar N(s) = \tau N(s\tau) ,
\mylabel{NA2}
\end{equation}
where the prefactor $\tau$ is added due to the derivative
(Eq.~\ref{NA}).
The non-adiabatic amplitudes satisfy
 the close-coupled equations obtained from
Eqs.~\ref{CC} which read
\begin{align}
& \dot{\bar a}_\pm(s) = \mp {\bar a}_\pm(s) \ \bar N(s) \  e^{\pm i\tau \int\limits_0^s ds'\, \bar \delta(s')} .
\mylabel{CC2}
\end{align}

Next follows an {\it ad hoc} step where we
deliberately introduce the pulse area into the
evolution equations Eq.~\ref{CC2} as a substitute for $\tau$.
We start by rewritting the dynamical Rabi frequency and detuning
using the relative time $s$ such that
\begin{align}
& \bar\Omega(s) = \frac{\mu\varepsilon_0^{max}}{\hbar} \, e^{-s^2/2}, \cr
& \bar\Delta(s) = \alpha \tau\cdot s + \frac{i\Gamma}{2\hbar} .
\mylabel{EQOD2}
\end{align}
As the pulse shape is not changed with the pulse length $\tau$, the 
dynamical Rabi frequency $\bar \Omega(s)$ is now independent on $\tau$.
The time-integral in Eq.~\ref{EQarea}
is redefined using $s$ instead of the physical time such that
\begin{align}
\theta = \tau \cdot \int\limits_{-\infty}^\infty ds \, \bar\Omega(s) ,
\end{align}
showing that $\theta$ is linearly proportional to the pulse length.
In particular, it is given by
\begin{align}
\theta = \frac{\mu\varepsilon_0^{max}}{\hbar}\,\tau\,g,
\quad
g = \int\limits_{-\infty}^\infty ds \, e^{-s^2/2} = \sqrt{2\pi} ,
\mylabel{EQth}
\end{align}
where $g$ is defined by the particular shape of the pulse,
here specified by the Gaussian,
see Eq.~\ref{EQOD2}.
\begin{align}
& \dot{\bar a}_\pm(s) = \mp {\bar a}_\pm(s) \ \bar N(s) \  e^{\pm \frac{i\theta}{\sqrt{2\pi}} \int\limits_0^s ds'\, \tilde \delta(s')} ,
\mylabel{CC3}
\end{align}
where $\tilde \delta(s)$ is related to $\bar \delta(s)$ through the factor,
\begin{align}
\tilde \delta(s) = \frac{\hbar}{\mu\varepsilon_0^{max}} \, \bar \delta(s) .
\mylabel{EQdeltil}
\end{align}

By assuming that the pulse area $\theta$ is responsible for the exponential
factor rather the the pulse length $\tau$, we obtained the factor needed to
get the appropriate reduced quasi-energy split $\tilde \delta(s)$.
From its explicit form given by
\begin{align}
\tilde \delta(s) =
\sqrt{
e^{-s^2} + 
\(
\frac{\hbar}{\mu\varepsilon_0^{max}} \alpha\tau \cdot s + \frac{i\Gamma}{2\mu\varepsilon_0^{max}}
\)^2
}
\mylabel{EQdelred}
\end{align}
it is clear that $\tilde \delta(s)$ is independent on any atomic
parameters in the case of unchirped pulses ($\alpha=0$) and bound-to-bound transitions ($\Gamma=0$).
This is in agreement with what is known and has beed stated above that
in such a case final populations of the laser driven quantum dynamics only depend
on the pulse area.
Of course, $\bar N(s)$ must prove independent on other laser or atomic parameters as well,
which we will show now.
$\bar N(s)$ is defined by Eq.~\ref{NA2} and also by the definitions of $N(t)$ (Eq.~\ref{EQ7a})
and $\Theta(t)$ (Eq.~\ref{EQTh}). When we put them together we get
\begin{align}
\bar N(s) = \frac{1}{2} \frac{d \bar\lambda(s)}{ds} \frac{1}{1 + \bar\lambda^2(s)} ,
\quad 
\bar\lambda(s) = \frac{\bar \Delta(s)}{\bar \Omega(s)} ,
\mylabel{EQNAf}
\end{align}
which shows that the non-adiabatic coupling element is a functional 
of the ratio between the dynamical detuning and Rabi frequency $\bar\lambda(s)$.
By substituting from Eqs.~\ref{EQOD2} to the definition of $\bar\lambda(s)$ we get
\begin{align}
\bar\lambda(s) = 
\(\frac{\hbar}{\mu\varepsilon_0^{max}}\alpha\tau\,s
+ \frac{i\Gamma}{2\mu\varepsilon_0^{max}}\)\, e^{s^2/2}.
\mylabel{EQf}
\end{align}
We can see that $\bar\lambda(s)=0$ for the examined case ($\alpha=0$, $\Gamma=0$)
which proves the fact $\bar N(s)$ is independent of any atomic
and laser parameters in that particular case.

Let us define other reduced laser parameters (as if added to a set with the pulse area) for
the more general laser-atom two-level dynamics based on
the definitions of $\tilde \delta(s)$ and $\bar\lambda(s)$ which
clearly define the system dynamics.
We define the {\it relative laser strength} $x$ such that
\begin{align}
x = \frac{2\mu\varepsilon_0^{max}}{\Gamma} \equiv \frac{\varepsilon_0^{max}}{\varepsilon_0^{EP}},
\mylabel{EQdefx}
\end{align}
where we used the position of the EP given above in Eq.~\ref{epsEP},
and
the {\it effective chirp} $\bar \alpha$,
\begin{align}
\bar \alpha = \frac{2\hbar}{\mu} \cdot
\frac{\alpha\tau}{\varepsilon_0^{max}} .
\mylabel{EQdefalpb}
\end{align}
The definitions of the basic dynamical quantities from Eqs.~\ref{CC3} now read
\begin{align}
& \tilde \delta(s) =
\sqrt{
e^{-s^2} + 
\(
\frac{\bar\alpha}{2} \cdot s + \frac{i}{x}
\)^2
}, \cr
& \bar\lambda(s) = 
e^{s^2/2} \cdot 
\(
\frac{\bar\alpha}{2}\,s  + \frac{i}{x}
\) ,
\mylabel{EQdelred2}
\end{align}
where $\bar\lambda(s)$ relates to $\bar N(s)$ through Eq.~\ref{EQNAf}.

It is clear that $x$ represents a measure of
non-Hermiticity of the quantum dynamics as it reduces
the imaginary components in Eqs.~\ref{EQdelred2}.
This simple analysis alone shows that
the general quantum dynamics of the bound-to-resonance
transitions coincides with the bound-to-bound systems
in the $x\to\infty$ limit. Hermitian regime of the
EP encircling {\it can} be achieved (for the presently studied
fully time-symmetric cases, Eqs.~\ref{EQ5d5}), namely by
seting large laser intensity $\varepsilon_0^{max} \gg \varepsilon_0^{EP}$.

\section{\label{TP}Singularities in complex time plane}

\subsection{Effective equations for survival amplitude}
\noindent
The  transformation 
using the effective time $s$ defined in Eq.~\ref{sdef} (above applied only to
 the evolution equations Eqs.~\ref{CC} leading to Eqs.~\ref{CC2} and \ref{CC3})
is now applied to the equations for the survival amplitude $a_1$
as based on the adiabatic perturbation approach.
Eqs.~\ref{V1} are rewritten as
\begin{align}
&\bv^{(j)}(s) = \cr
&
(-)^j 
\int\limits_{-\infty}^{s} ds' e^{-i\tau (-1)^{j} \int\limits_0^{s'} ds''\, \bar \delta(s'')}
\,  \bv^{(j-1)}(s') \bar N(s')  \cr
\mylabel{V02}
\end{align}
where $\bv^{(j)}$ correspond with $v^{(j)}$ through
\begin{align}
\bv^{(j)}(s) = v^{(j)}(s\tau).
\end{align}
and further using the effective quasi-energy split $\tilde \delta(s)$ defined
above in Eq.~\ref{EQdeltil}
\begin{align}
&\bv^{(j)}(s) = \cr
&
(-)^j 
\int\limits_{-\infty}^{s} ds' e^{-i\frac{\theta }{\sqrt{2\pi}}  \, (-1)^{j} \int\limits_0^{s'} ds''\, \tilde \delta(s'')}
\,  \bv^{(j-1)}(s') \bar N(s')  ,\cr
\mylabel{V03}
\end{align}
where we have also used Eq.~\ref{EQth} to include the pulse area $\theta$ instead of pulse length $\tau$.
Eqs.~\ref{V03} clearly correspond to Eqs.~\ref{CC3}.

The survival amplitude $a_1$ given in Eq.~\ref{EQa10}
is rewritten as
\begin{align}
a_1 = f \cdot 
\sum\limits_{j=1,3,\dots} \bv^{(j)}(s\to\infty),
\mylabel{EQa1}
\end{align}
where $f$ has been defined in Eq.~\ref{EQfdef} and simplified for the
time-symmetric case in Eq.~\ref{EQfdefsym}.

\subsection{Quasi-energy split in the real time axis\myslabel{Sdelre}}

\noindent
Let us show how the exponential in Eq.~\ref{V03} behaves for
the odd corrections $\bar v^{(j)}(s)$ which sum up to the
survival amplitude, Eq.~\ref{EQa1}.
The exponent is given by the integral of the quasi-energy
split $\tilde \delta(s)$ (defined in Eq.~\ref{EQdelred2}).
The asymptotic behavior of $\tilde \delta(s)$ for $s\to\pm\infty$
is given by 
\begin{align}
& \tilde \delta(s) =
sign(s)\cdot \(
\frac{\bar\alpha}{2} \cdot s + \frac{i}{x}
\),
\end{align}
which we illustrate in Fig.~\ref{FIGreals1} for two typical
values of the parameters $\bar \alpha$ and $x$.
\def\balp{\bar \alpha}
\begin{figure}[!hb]
(a)\includegraphics[width= 2 in ]{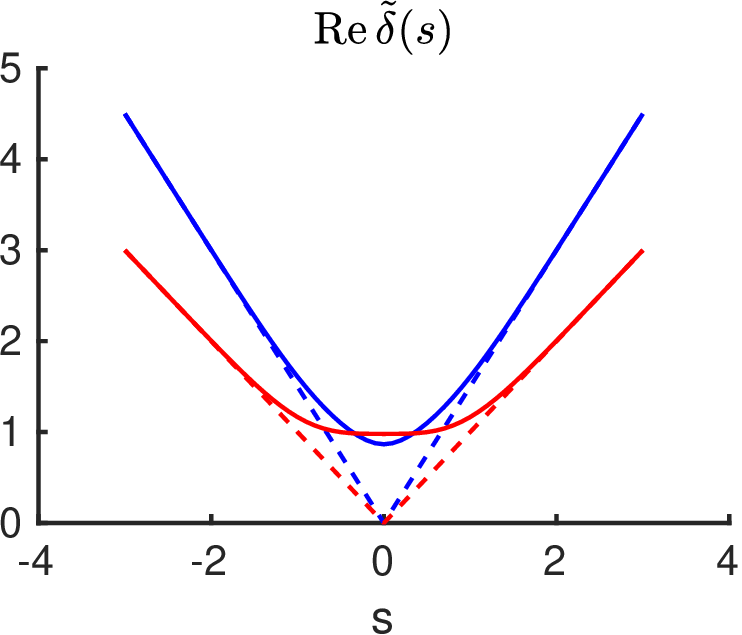}
(b)\includegraphics[width= 2 in ]{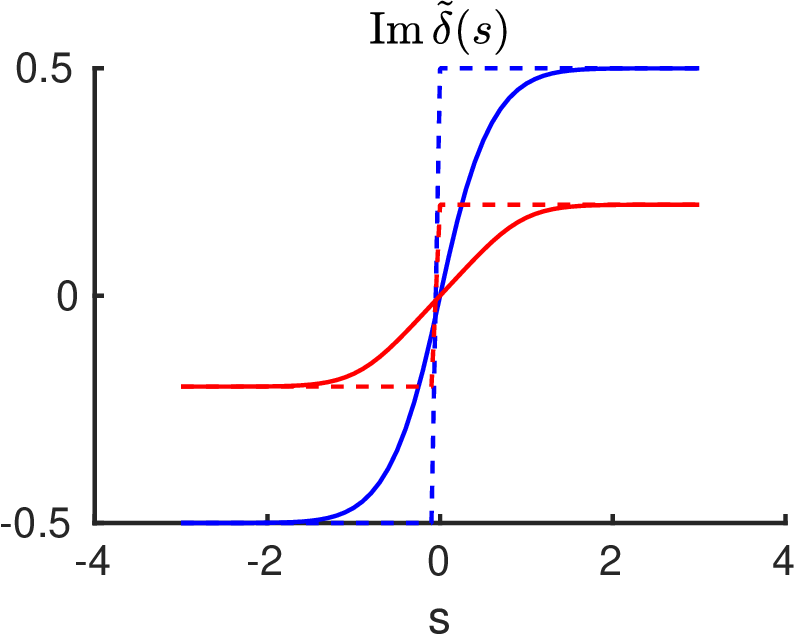}
\caption{Quasi-energy split for two typical values of the parameters
$[x,\balp]$. The red curves correspond to $[5,2]$
(two minima are found on the real part of the split
and the limits of the imaginary part are give by $\pm 1/5$)/
The blue curves correspond to $[2,3]$
(single minimum on the real part of the split whereas
the imaginary part has the limits given by $\pm 1/3$.
}
\mylabel{FIGreals1}
\end{figure}
After integrating such a function over the time $s$ we obtain the asymptotic
behavior of the real part which is quadratic but it has
an inflex point at $s=0$, Fig.~\ref{FIGreals2}a,
and linear asymptotic behavior of the imaginary part
with the minimum at $s=0$, Fig.~\ref{FIGreals2}b.
\begin{figure}[!hb]
(a)\includegraphics[width= 2 in ]{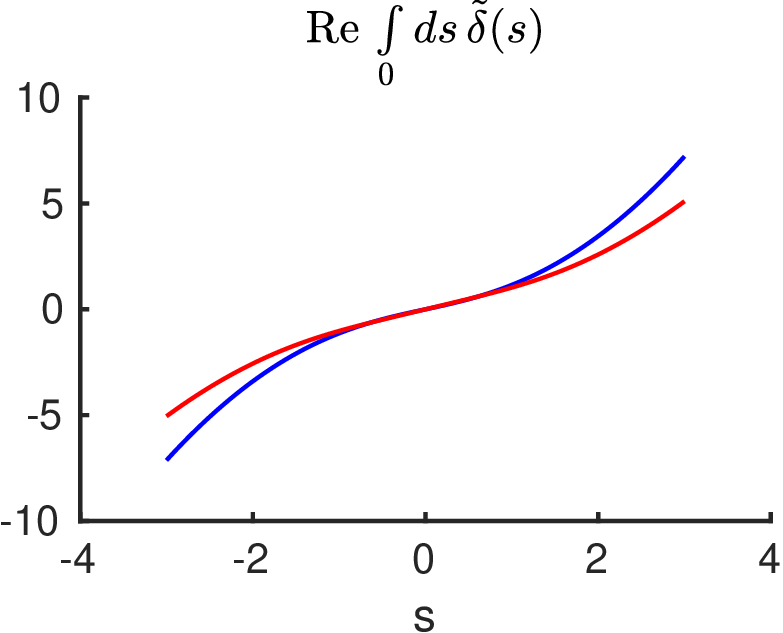}
(b)\includegraphics[width= 2 in ]{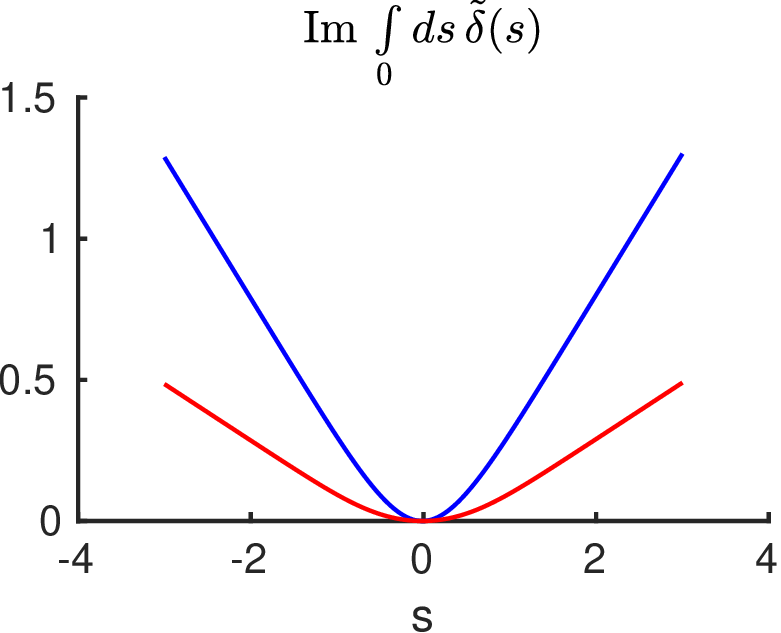}
\caption{Integral over
quasi-energy split for two typical values of the parameters
$[x,\balp]$ corresponding to Fig.~\ref{FIGreals1}.
Note that the imaginary part has a clear maximum,
while the real part has a clear inflexion point.
}
\mylabel{FIGreals2}
\end{figure}

When we multiply the integral by the imaginary unit to get the
exponent, and additionally multiply this by the pulse area $\theta$,
see Eq.~\ref{V03}, we obtain a bell-type complex function,
Fig.~\ref{FIGreals3}, where the bell envelope is due to the 
fact that the imaginary part of the integral goes to infinity
in the asymptotic limits where its minimum corresponds to
the maximum of the bell function.
As $\theta$ is increased, the bell function would become
infinitely narrow. This trend is illustrated by increasing 
$\theta$ from $\theta=4\pi$ to $\theta=50\pi$ when comparing
the left and right columns in Fig.~\ref{FIGreals3}.

From this point of view, the integrals in Eq.~\ref{V03} for
odd corrections $\bar v^{(j)}$ seem to reduce to simple
integrations over the Dirac $\delta$-function for very large
pulse areas $\theta$.
However, the complex phase of the exponential function
must neither be neglected. The inflexion point shown in Fig.~\ref{FIGreals2}a
indicates that such a $\delta$-function 
would occur in the complex plane, $\delta(s-s_0)$, $s_0\in \C$,
where
$s_0$ would represent a sort of extreme of both the imaginary and real
parts of the integral over the quasi-energy split and thus
a ``zero'' of the quasi-energy split.
\begin{figure}[!hb]
\includegraphics[width= 2.8 in ]{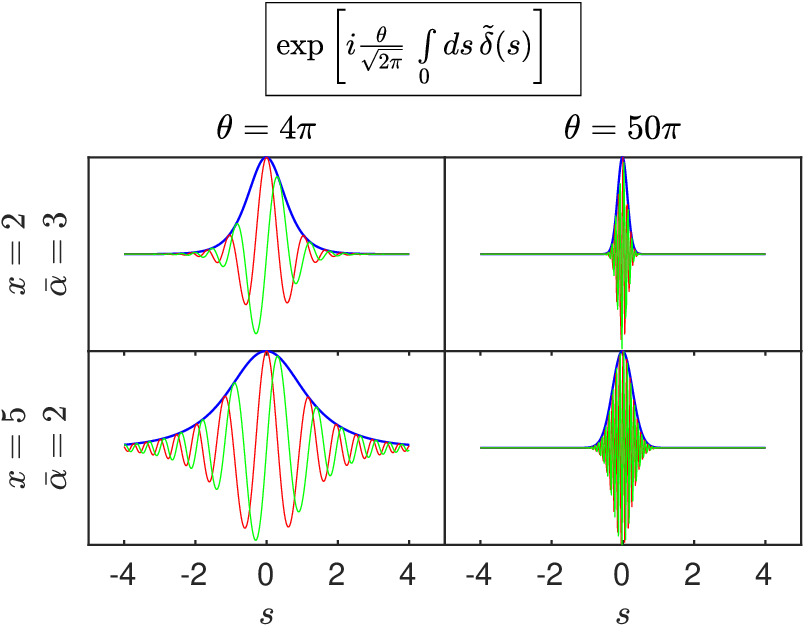}
\caption{Exponent of the integral Eq.~\ref{V03} for different
pulse parameters. This exponent is always a bell-function
with maximum at $s=0$. Note that the frequency of oscillations
of the complex phase  
is not constant but it is notably decreased near $s=0$.
}
\mylabel{FIGreals3}
\end{figure}

\subsection{Definition of transition points}

\noindent
Above we suggested that there is likely a ``zero'' of the quasi-energy
split that occurs in the complex time-plane ($s_0\in\C$) 
\begin{align}
\tilde\delta(s_0) = 0,
\end{align}
which is reflected
as the inflection point and the minimum on the real and imaginary
components of the integrated quasi-energy split, respectively (Fig.~\ref{FIGreals2}).
Such a zero is not a stationary point as one could expect
(either a maximum, a minimum, or a saddle point)
but rather a {\it branchpoint}.
While the stationary points are
characterized by a finite second derivative of the studied 
function (i.e. here the first derivative of the quasi-energy split),
the first derivative of the quasi-energy split given by
\begin{align}
\frac{ d \tilde\delta}{d s} = \(\frac{\hbar}{\mu \varepsilon_0^{max}}\)^2 \, 
\frac{1}{\tilde\delta} \, 
\( \frac{d\bar\Delta}{ds}\bar\Delta + \frac{d\bar\Omega}{ds}\bar\Omega\) 
\end{align}
is rather infinite at the zero of the quasi-energy split $\tilde \delta$ which
occurs in the denominator.

Importantly, the non-adiabatic element $\bar N(s)$
has a pole in the branchpoint $s_0$, as follows from
Eq.~\ref{EQNAf} which can be written as
\begin{equation}
\bar N(s) = 
\frac{1}{4i} \frac{d (\bar \lambda)}{ds} 
\(
\frac{1}{i + \bar \lambda} +
\frac{1}{i - \bar \lambda} 
\) ,
\mylabel{EQ7s}
\end{equation}
realizing that
\begin{align}
\bar\lambda(s_0) = \pm i
\end{align}
at the branchpoints due to the definition
\begin{align}
\tilde\delta(s) = \frac{2\hbar}{x\Gamma} \, \bar\Delta(s)\, \sqrt{1 + \bar\lambda^2(s)} .
\end{align}
Namely, the integrand of Eq.~\ref{V03} is singular for $s_0$, which
makes it impossible to use the previous idea of using Dirac $\delta$-function
to simplify the integral over time for large pulse areas $\theta\to\infty$.
Rather, a proper application of the residuum theorem 
represents a possible way to solve this problem, still using the 
zeros in the complex time plane.

We shall note that the branchpoints in the complex time plane
are long known of, being referred to as the {\it transition points (TPs)}.
Although having the same mathematical nature as the exceptional points (EPs), we will
distinguish between the EPs as branchpoints defined in the laser parameter plane $[\omega,\varepsilon_0]$ ,
and  the TPs  in contrast as branchpoints defined in the complex time plane.

\subsection{\label{coal}Transition points on the imaginary time axis}

\noindent
Let us search for the TP in the complex time plane in the
concrete example given by the linearly chirped Gaussian pulse,
where we particularly use Eq.~\ref{EQdelred2} for the quasi-energy
split definition.
Let us denote the TP by $s_k$ as we will soon see that there are
more than one TPs in the complex time plane.
The TP satisfies the equation,
\begin{align}
& e^{-s_k^2} + \(\frac{\balpha}{2} s_k + \frac{i}{x}\)^2 = 0 .
\mylabel{e1}
\end{align}
%Eq.~\ref{e1} has an apparent symmetry.
%Note that
%if $\balpha < 0$ then the situation is just reversed in the sign, therefore
%we will not treat it explicitly and confine the further analysis to
%\begin{align}
%\balpha > 0 .
%\end{align}

It is reasonable to assume that at least for some parameters
$[x,\bar\alpha]$ a TP occurs for $\re \,s_k = 0$. Such is the case
of $[x=2,\bar\alpha=3]$ for which one can see a single
minimum on the imaginary part of the quasi-energy split, see
the blue curve in Fig.~\ref{FIGreals2}b. On the other hand,
the second example given by sthe red curve on the same figure
indicates two minima, which may possibly correspond to two
distinct TPs with different non-zero real parts, $\re\, s_k \ne 0$.

Let us study for what laser parameters ($x$ and $\balpha$)
 Eq.~\ref{e1} has {\it purely imaginary roots}.
It is instructive to change the variable $s$ such that
\begin{align}
s_k = i \xi_k
\mylabel{zdef}
\end{align}
and study the real defined roots $\xi_k$ instead.
Eq.~\ref{zdef} is substituted to Eq.~\ref{e1}:
\begin{align}
& e^{\xi_k^2} = \(\frac{\balpha}{2} \xi_k + \frac{1}{x}\)^2 .
\mylabel{e3}
\end{align}
Apparently, the two sides of Eq.~\ref{e3} include even, parabolic or
parabolic-like ($\exp(z^2)$), functions, respectively, see Fig.~\ref{FIGcoal}.
If these parabolas cross each other for real $-i s_{0i} \in \Re$,
we find the real roots, if they avoid each other, the roots are
found in the complex plane. 
\begin{figure}[!hb]
\includegraphics[width= 2.8 in ]{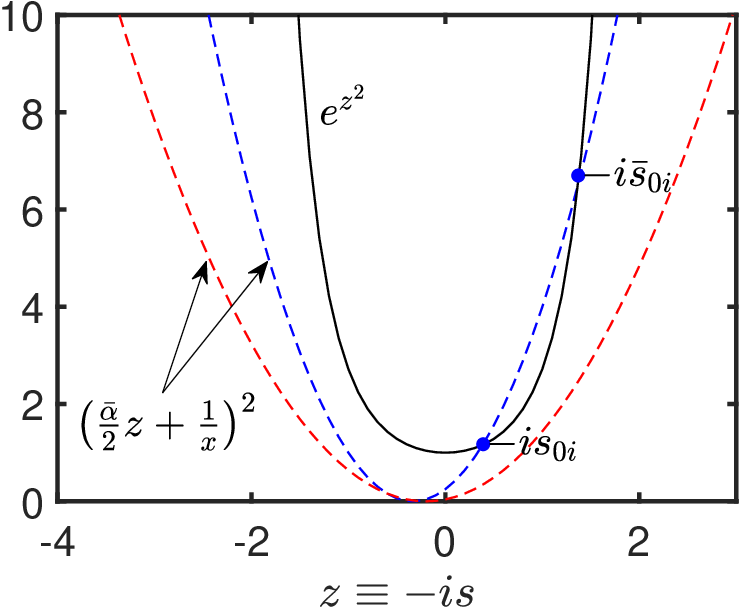}
\caption{Left and right hand sides of Eq.~\ref{e3} are compared
for the two sets of laser pulse parameters described in Fig.~\ref{FIGreals1}.
Red color stands for $[x=5,\balp=2]$, while blue color stands for 
$[x=2,\balp=3]$. 
}
\mylabel{FIGcoal}
\end{figure}

Let us determine the pulse parameters for the
situation, where the two parabolas touch each other.
At this critical point, the two purely imaginary
TPs happen to go to the complex plane.
To distinguish the imaginary and complex roots we
use different notations for them, $s_{0i}$, $\bar s_{0i}$ and
$s_0$, $\bar s_0$, respectively. 
We will obtain the critical laser parameters by requiring that the first derivatives of
the left and right hand sides of Eq.~\ref{e3} are equal:
\begin{align}
e^{{\(\xi_k^{coal}\)}^2} \xi_k = \frac{\balpha}{2} \(\frac{\balpha}{2} \xi_k^{coal} + \frac{1}{x}\) .
\mylabel{e4}
\end{align}
By combining Eqs.~\ref{e3} and \ref{e4} we obtain 
\begin{align}
	&\bigg[\(\xi_k^{coal} + \frac{2}{\balpha x}\) 
	\( \xi_k^{coal} + \frac{1}{\balpha x} + \sqrt{\frac{1}{(\balpha x)^2} + 1} \) \cr &\times\quad
\( \xi_k^{coal} + \frac{1}{\balpha x} - \sqrt{\frac{1}{(\balpha x)^2} + 1} \)
\bigg] = 0
\mylabel{roots}
\end{align}

For $\balpha>0$,
there are two negative and one positive definite roots given by
Eq.~\ref{roots} for any value of
$x\balpha$, while for $\balpha<0$ it is the other way round.
The sign of $\balpha$ differectiates between clockwise and
anti-clockwise encircling of the EP. 
As the studied dynamics is time-symmetric (see previous Sections),
same results must be obtained for
the two cases. Therefore we will restrain our study to
one case only, in particular assuming $\balpha>0$.

Let us focus on the positive definite root, $\xi_k^{coal}>0$, defined as
\begin{align}
\xi_k^{coal} = \sqrt{\frac{1}{(\balpha x)^2} + 1}  -\frac{1}{\balpha x} ,
\mylabel{zcoal}
\end{align}
which occurs when the two TPs $s_{0i}$, $\bar s_{0i}$ become 
a single point, Fig.~\ref{FIGcoal}.
We will see below that the TPs which are have negative imaginary
parts (including the negative roots here) are not relevant to
the complex contour integration which will be eventually
used to solve the dynamical equations Eq.~\ref{V03},
more precisely the first-order perturbation integral.
%Interestingly, one can see that the point appears in the interval
%\begin{align}
%0\le \xi_k^{coal} < 1
%\end{align}
%for any value of $\balpha x$. 

\subsection{Separator in laser parameter plane}
\noindent
By substituting Eq.~\ref{zcoal} into Eq.~\ref{e3}, we obtain
a {\it separator in the laser parameter space} $[\balpha,x]$,
which divides between the laser parameters associated
with imaginary/complex TPs.
The limits on this curve can be calculated analytically;
First, if $\balpha x =0$ then $z=0$ according to Eq.~\ref{zcoal}, while Eq.~\ref{e3}
specifies that $x=1$ and $\balpha=0$.
Second, if $\balpha x \to \infty$ then $z=1$ according to Eq.~\ref{zcoal}, while Eq.~\ref{e3}
specifies that $\balpha = 2\sqrt{e}$ at $x\to\infty$.
This is illustrated in Fig.~\ref{Fxa}.
\begin{figure}[h!]
	\center{
		\includegraphics[width=3in]{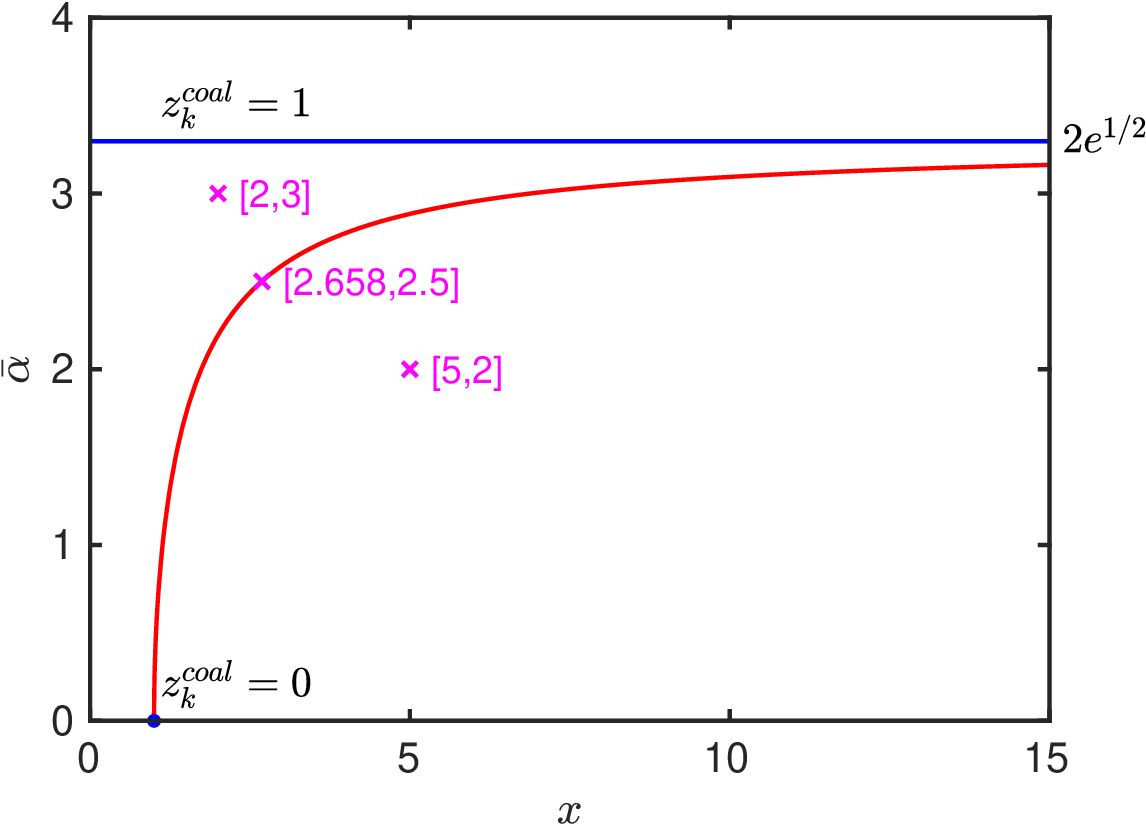}
	}

	\begin{tabular*}{0.45\textwidth}{@{\extracolsep{\fill} }  c  c  }
		\includegraphics[width=1.2in]{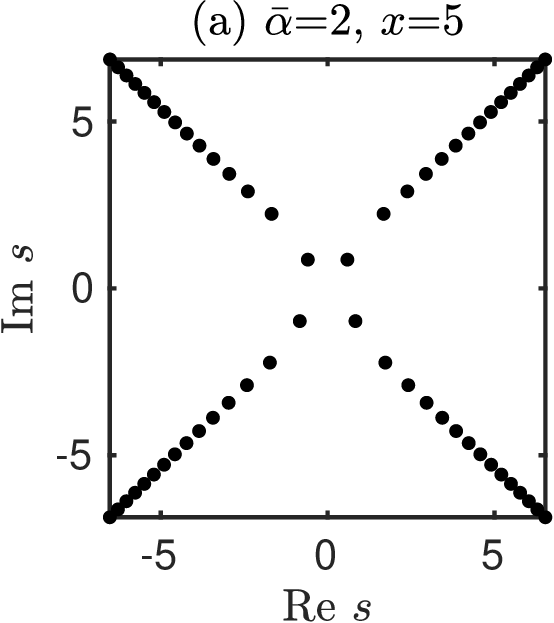} & \includegraphics[width=1.2in]{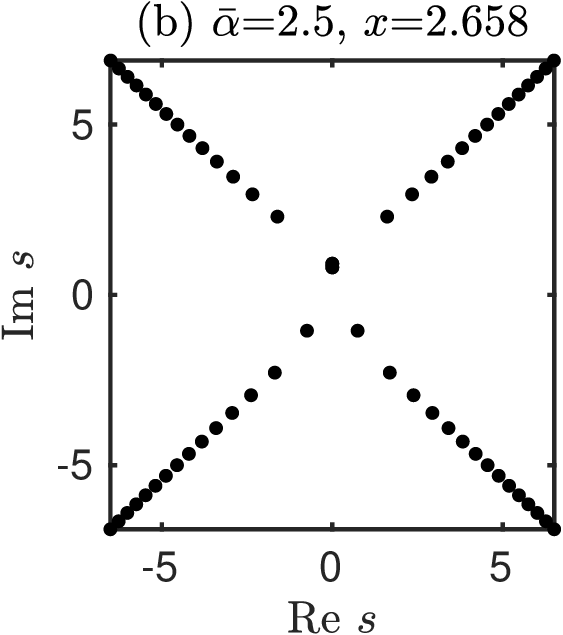} \\
		\includegraphics[width=1.2in]{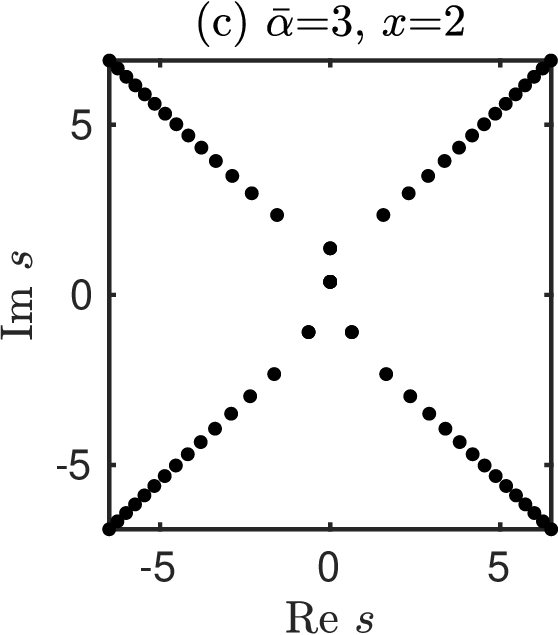} &  
	\end{tabular*}

	\caption{
		The laser parameters $[\balpha,x]$ for which the pair of TPs in the complex time plane
		coalesce on the imaginary axis are represented by the red curve in the main panel.
		The points $[2,3]$ and $[5,2]$ represent examples of laser parameters, where
		the TPs lie/lie not on the imaginary axis, respectively.
		In the limit $x\to\infty$, which is the limit of the rapid adiabatic passage (RAP),
		the two TPs coalesce for $s_k=i$ ($\xi_k^{coal}=1$). By substituting this to Eq.~\ref{e3}
		and applying again the limit $x\to\infty$ we obtain the asymptotic value of $\balpha=2e^{1/2}$.
		This implies a phenomenologically based relation for the occurence of a stable RAP 
		for bound-to-bound transitions (obtained using Eqs.~\ref{EQdefalpb} and \ref{EQth}):
		$\alpha\tau^2/\theta > \sqrt{e/2\pi}$.
		The other limit is represented by $\balpha=0$, where Eq.~\ref{zcoal} implies that the
		coalescence of TPs occurs on the real axis for $s_k=0$ ($\xi_k^{coal}=0$). The corresponding
		laser parameters $[1,0]$ follow from substituting this value to Eq.~\ref{e3}.
		The three lower panels (a-c) display TPs on the complex time plane for the 
		laser parameters corresponding to the particular points indicated in the main panel.
		It is demonstrated how the central pair of TPs
		lay either apart (a) or on (b,c) the imaginary axis; the
		TPs coalesce in the case (b).
		Apart from the cental pair of TPs, we also notice the asymptotic
		series of TPs which seemingly lay on nearly straigh lines.
	}
	\mylabel{Fxa}
\end{figure}

Let us add here a comment on the physical meaning of the two possible
configurations of the TPs that we have just shown taking place in
our physical problem.
First of all, the two configurations of the TPs reflect what
we saw already on the real axis, Fig.~\ref{FIGreals1}, where
we observed either a single minimum (blue curve)
or a double minimum (red curve) on the quasi-energy split. 
Which one of the
situations takes place depends on the laser parameters.
The minima of the quasi-energy split indicate nothing else
then avoided crossings associated with the
increased probability of a non-adiabatic jump.

The problems of non-adiabatic jumps in avoided crossings
have been widely studied.
The single avoided crossing implies that Landau-Zener
formula is applicable in the semiclassical limit~\cite{Zener:1932,Wittig:2005,Vutha:2010},
while the two subsequent avoided crossings implicate
the so called St\"uckelberg oscillations~\cite{Shevchenko:2010,Thorson:1971,Delos:1972,Child:1978,Ota:2018}.

Dykhne, Davis, and Pechukas studied a quadratic coupling
model for the avoided crossings~\cite{Dykhne:1962,Davis:1975}.
They found 
two different possible configurations of the TPs, which
closely corresponds to our findings above.
Upon developping the complex plane method, they associated
the two different configurations of the TPs
with the Landau-Zener and St\"uckelberg
type of non-adiabatic jumps, respectively.
In our physical problem, 
the Landau-Zener regime is manifested as the rapid adiabatic passage (RAP)~\cite{Tannor,Vitanov:1998,Yan:2010},
while the St\"uckelberg regime corresponds with the regime of
Rabi oscillations.
This fact will become clearly apparent in the final results of
our analysis.

\subsection{\myslabel{ASTP}Asymptotic series of transition points}

\noindent
Let us suppose that the term $i/x$ in Eq.~\ref{e1} is negligibly small.
This case occurs if the process is Hermitian or the contour
is far from the EP, i.e. $x\to\infty$, or just if $|s_k|$ is large 
($|s_k|\to\infty$) supposed that $\balpha$ is non-zero.
Then Eq.~\ref{e1} is approximated by
\begin{align}
	e^{-s_k^2} = -\(\frac{\balpha}{2}\)^2 s_k^2 ,
	\mylabel{EQ67}
\end{align}
using
\begin{align}
	\frac{\balpha}{2}\,|s_k| \gg \frac{1}{|x|} .
\end{align}
The left hand side of Eq.~\ref{EQ67} is periodic in the
complex time plane due to the exponential of imaginary
components of $s^2$, thus there is a series of TPs
$s_k$, where the difference between subsequent TPs, $s_{k+1}^2$ and $s_k^2$, is
approximately given by $2i\pi$.
In a more precise approximation (derived in Appendix~\ref{APasymtp}),
\begin{align}
	&s_k^2 = k\cdot 2i\pi + i\pi/2 - \ln(2k\pi)-2\ln\(\frac{\balpha}{2}\) ,\cr
	&\quad k\to\infty.
	\mylabel{EQQ81}
\end{align}
as one can verify by substituting this to Eq.~\ref{EQ67}.
From here we get,
\begin{align}
	&s_k = \sqrt{k\cdot 2i\pi} \[ 1  + \frac{i}{4k\pi}\ln\(\frac{\balpha^2}{2}\,k\pi\) + \frac{1}{8k}\] ,
\cr
	&\quad k\to\infty.
	\mylabel{EQQ82}
\end{align}

The time-symmetry of the problem defined in Eq.~\ref{EQ8d10}
implies that the TPs appear in pairs (let us denote them $s_k$ and $\bar s_k$) that are mutually  
related as 
\begin{align}
	&\bar s_{k} = -s_k^*, 
	&\re\, s_k >0, \quad \re\, \bar s_k <0 .
\mylabel{sym}
\end{align}
(The only situation where there is not a pair of TPs 
occurs for the cases where the TPs lie on the imaginary axis.)
The adjoint series to Eqs.~\ref{EQQ81} and \ref{EQQ82}
are given by,
\begin{align}
&\bar s_{k}^2 = -k\cdot 2i\pi - i\pi/2 - \ln(2k\pi) -2\ln\(\frac{\balpha}{2}\),\cr
&k\to\infty.
\mylabel{EQQ81m}
\end{align}
and
\begin{align}
&\bar s_k = \sqrt{k\cdot 2i\pi} \[ -i  + \frac{1}{4k\pi}\ln\(\frac{\balpha^2}{2}\,k\pi\) - \frac{i}{8k}\] ,
\cr
&\quad k\to\infty.
\mylabel{EQQ82m}
\end{align}

Although the results given in Eqs.~\ref{EQQ81}--\ref{EQQ82m} have been derived for EPs in the asymptotic
limit $k\to\infty$, in practise they represent a good approximation
starting from $k=2$ as we demonstrate in Fig.~\ref{FigEPasym}.
\begin{figure}
	\begin{center}
	\includegraphics[width=3in
	]{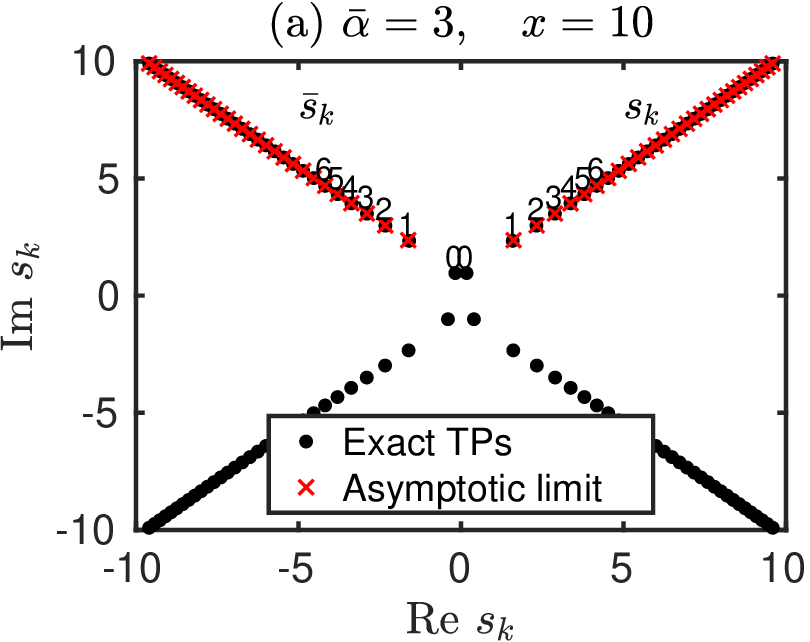}
	\includegraphics[width=3in
	]{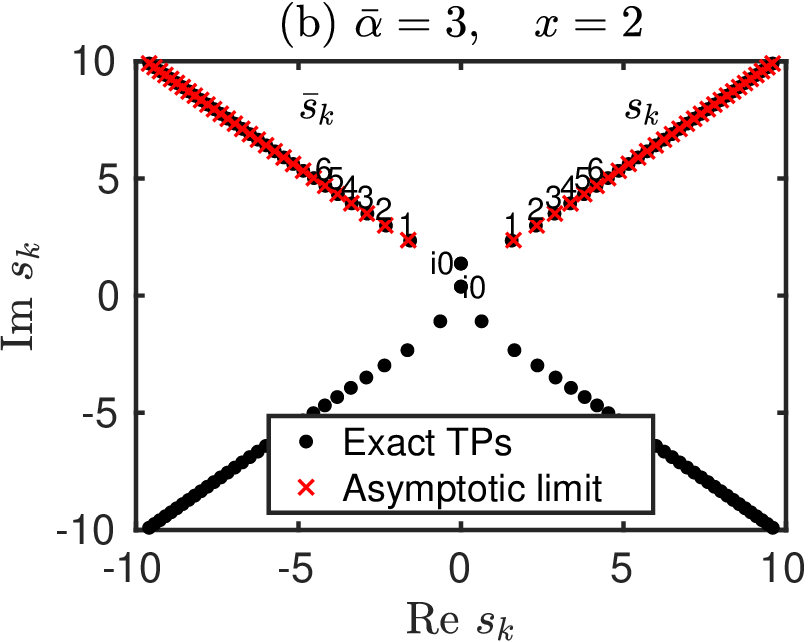}
	\end{center}
\caption{
	Exceptional points $s_k$ and $\bar s_k$ 
	in the complex plane of dimension-less adiabatic time $s$
	are compared with their approximate
	positions obtained for the asymptotic limit $|s_k|\to\infty$ 
	(Eqs.~\ref{EQQ82} and \ref{EQQ82m}).
	This approximation is valid for $k>0$, where the symmetrical rule given in Eq.~\ref{sym}
	yields the two sets of points, $s_k$ and $\bar s_k$.
	In contrast, the points $s_0$ and $\bar s_0$ are related by Eq.~\ref{sym} only for the 
	case of even layout (a), as discussed in Section~\ref{coal}, and the asymptotic approximation
	given Eqs.~\ref{EQQ82} and \ref{EQQ82m}
	is not well defined for $k=0$.
}
\mylabel{FigEPasym}
\end{figure}

\section{Puiseux expansion on the complex time-plane}

\subsection{Product expansion of the quasi-energy split}

\noindent
We have just proved that there are zeros in the complex time-plane
where each one represents a distinct branchpoint.
As we have discussed already for the EPs, the potential
energy split near any branchpoints is given by the
Poiseux expansion. We assume that the quasi-energy split
as a function of time, $\tilde\delta(s)$,
can be also expressed using this expansion such that
\begin{align}
\tilde\delta(s)=\prod\limits_{k} \alpha_k (s-s_k)^{1/2} ,
\mylabel{Puiseux0}
\end{align}
where all TPs are included, namely both $s_k$ and $\bar s_k$
as well as the TPs in the lower
imaginary half-plain.

\subsection{Association of TPs to the EPs for positive and negative laser amplitude}

\noindent
Despite of the fact that the encircling contour in the
frequency and laser amplitude plane is directly associated with one
of the EPs, namely the one with positive laser amplitude
(we refer here to the discussion in Sections~\ref{SEP1} and \ref{SEP2}),
it is an intriguing mathematical fact
that the TPs that occur in the complex time-plane
are associated with both of these EPs.

Let us start by assuming that the quasi-energy split $\tilde{\delta}(s)$
includes two different kinds of zeros 
of either $\tilde\delta_+(s)$ or $\tilde \delta_-(s)$,
respectively, where the functions $\tilde\delta_\pm(s)$ are  
time-dependent equivalents of $\delta_\pm(\omega,\varepsilon_0)$ 
(Eq.~\ref{EQ32}, see also Eq.~\ref{EQdeltil} for the relation between $\tilde \delta$ and $\bar \delta$). 
Namely, $\tilde\delta_\pm(s)$ are given by,
\begin{align}
\tilde \delta^2_\pm(s) 
& = \frac{\mu\varepsilon_0^{max}}{\hbar} \[ \bar \Omega(s) \pm i \bar \Delta(s) \] ,
\mylabel{EQdelpm2}
\end{align}
which in the particular studied case reads,
\begin{align}
\tilde \delta^2_\pm(s) = e^{-s^2/2} \pm i \(\frac{\balp}{2} s + \frac{i}{x}\) .
\end{align}
%Note that $\tilde\delta_\pm(s)$ are complex conjugate as $x\to\infty$,
%which implies a symmetry the TPs in the two half-planes for the Hermitian
%case, whereas the positions of the zeros $s_k$ are non-symmetric
%in the two half-planes in the non-Hermitian case.

For the TPs in the positive imaginary half-plane,
the even indices of $s_k$ and $\bar s_k$, 
$k\in\{0,2,\cdots\}$, represent the zeros of
$\tilde\delta_+$, while the odd ones, $k\in\{1,3,\cdots\}$,
are associated with $\tilde\delta_-$. Let us prove this statement
for the central TPs, $k=0$, first.
Based on the substitution used already above in Eq.~\ref{zdef}
we obtain,
\begin{align}
\tilde \delta^2_\pm(\xi_k) = e^{\xi_k^2/2} \mp  \(\frac{\balp}{2} \xi_k + \frac{1}{x}\) .
\end{align}
Clearly, the exponential and the term in the bracket are both
positive defined, assuming that the central TPs lie on the
imaginary axis. The only way how they can add to zero is
by substraction, which occurs for $\tilde\delta_+$.
Also in the coalescence,  $\tilde\delta_+=0$ while
$\tilde\delta_-\ne 0$, from which we deduce that even
when the central TPs are not purely imaginary, they are
still associated with the zeros of $\tilde\delta_+$, not
$\tilde\delta_-$.

Now, we will explore the higher zeros, $k>0$.
According to our previous discussion in Section~\ref{ASTP},
we can neglect the contribution $1/x$. An equivalent
of Eq.~\ref{EQ67} based on $\tilde \delta_\pm$ rather
then $\tilde\delta$ reads,
\begin{align}
\tilde \delta^2_\pm(s_k) = e^{-s_k^2/2} \pm  i \frac{\balp}{2} s_k  .
\end{align}
Now, we substitute for $s_k$ using Eqs.~\ref{EQQ81} and
\ref{EQQ82} which
apply for the limit $k\to\infty$. By using $s_k^2=2ik\pi + i\pi/2$
in the exponential we see immediately that the exponential is 
given by $\pm e^{-i\pi/4}$ for even/odd values of $k$.
As this is compared with the linear term, one can see
that the even values of $k$ are associated with the zeros
of $\tilde\delta_+$, while the odd values of $k$ with
$\tilde\delta_-$. The same could be shown for $\bar s_k$.

In the negative halfplane, $\im\, s_k < 0$, we could show
that odd/even $k$-s (defined in analogy to the positive halfplane)
correspond to $\tilde \delta_+$, $\tilde \delta_-$, respectively.
We will use the knowledge discussed in this Section
below when defining non-adiabatic coupling.

\subsection{Non-adiabatic coupling as a sum of complex poles}

\noindent
As we have discussed in Section~\ref{TP}, the non-adiabatic coupling 
element $\bar N(s)$ includes poles at the TPs.
The Puiseux expansion of the quasi-energy split on the complex
time-plane defined in Eq.~\ref{Puiseux0} allows us to relate
the non-adiabatic coupling $\bar N(s)$ (Eq.~\ref{EQNAf}) to 
the complex poles more explicitely, namely by rewriting the
non-adiabatic coupling as a sum over the poles.

The non-adiabatic coupling element can be written using $\tilde \delta^2_\pm(s)$ such that
\begin{align}
&
\bar N(s) = 
\frac{1}{4i}\(\frac{d \ln \tilde \delta^2_+}{ds} 
- \frac{d \ln \tilde \delta^2_-}{ds} \) ,
\mylabel{EQNApm}
\end{align}
which one can prove by substitution from Eq.~\ref{EQdelpm2} and then
comparing with Eq.~\ref{EQNAf}:
\begin{align}
&\frac{\frac{d \tilde \delta^2_+}{ds}}{\tilde\delta^2_+} 
- \frac{\frac{d \tilde \delta^2_-}{ds}}{\tilde\delta^2_-} 
= 
\frac{\frac{d\bar \Omega}{ds} + i \frac{d\bar \Delta}{ds}}
{ \bar \Omega + i \bar \Delta } -
\frac{\frac{d\bar \Omega}{ds} - i \frac{d\bar \Delta}{ds}}
{ \bar \Omega - i \bar \Delta }
=\cr
& \quad
2i\frac{
	\frac{d\bar \Delta}{ds} \bar\Omega
	-\frac{d\bar \Omega}{ds} \bar\Delta
}
{\bar \Omega^2 + \bar \Delta^2}=2i \frac{d\frac{\bar\Delta}{\bar\Omega}}{ds}\frac{\bar\Omega^2}
{\bar \Omega^2 + \bar \Delta^2}
= \cr & \quad 2i\, \frac{d\frac{\bar\Delta}{\bar\Omega}}{ds} \frac{1}{1+\frac{\bar\Delta^2}{\bar\Omega^2}} .
\end{align}

The Puiseux expansions for $\tilde \delta_\pm(s)$ functions derive from
Eq.~\ref{Puiseux0}, but now only the TPs with positive/negative imaginary
parts are used in the individual expansions,
\begin{align}
\tilde \delta_\pm^2 (s) = \prod\limits_k \alpha^2_{k\pm} \, (s-s_{k\pm}) .
\end{align}
When these expressions for $\tilde \delta_\pm(s)$ are substituted
to the definition of the non-adiabatic coupling in Eq.~\ref{EQNApm},
we obtain the equation
\begin{align}
&
\bar N(s) = 
\frac{1}{4i} \sum\limits_k \frac{z_k}{s-s_{k}} , \cr
&
z_k = e^{i\, k \pi} \cdot \sign{\im\, s_k}
.
\mylabel{EQNApoles}
\end{align}
Note that this expression is general for two-by-two
Hamiltonians and it is not limited to the particular 
case studied here. 
The general expression Eq.~\ref{EQNApoles}
clearly demonstrates the behavior of $\bar N(s)$
near the poles $s_k$, which can be derived
also using the particular present Hamiltonian
as we do in Appendix~\ref{APNA1}.
It is also obvious that the sum over the first order poles
Eq.~\ref{EQNApoles} implies a decaying asymptotic behavior
of $\bar N(s)$, though it may be hard to prove it
analytically (perhaps using the expressions for $s_k$ such
as Eqs.~\ref{EQQ82} and \ref{EQQ82m})
that the asymptotics 
for the present case is actually given
by the second order exponential, compare Appendix~\ref{NAsection}.
We illustrate the function $\bar N(s)$ for
the present case evaluated numerically in Fig.~\ref{FNA}.
\begin{figure}[h!]
	\includegraphics[width=2in]{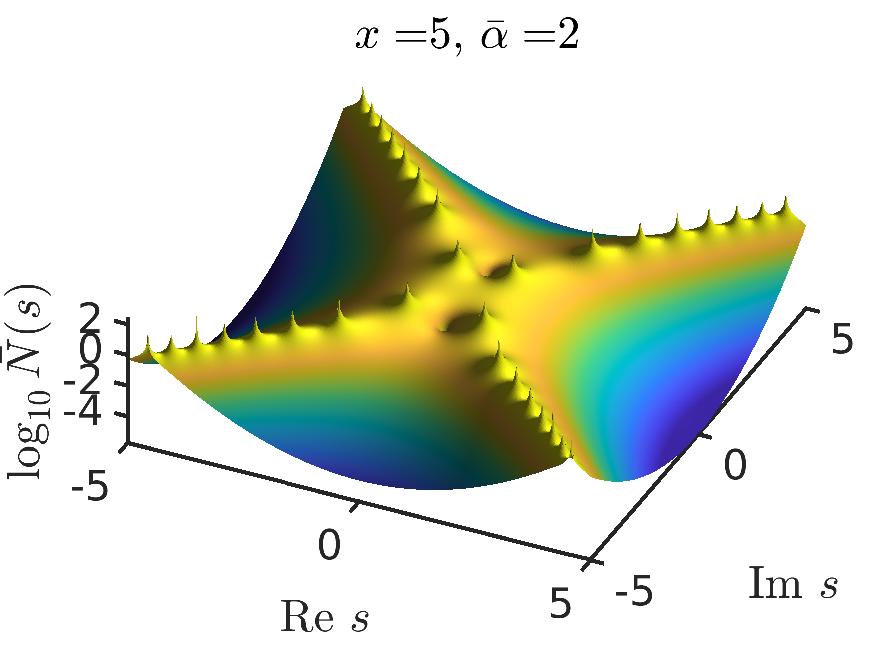}
	\includegraphics[width=2in]{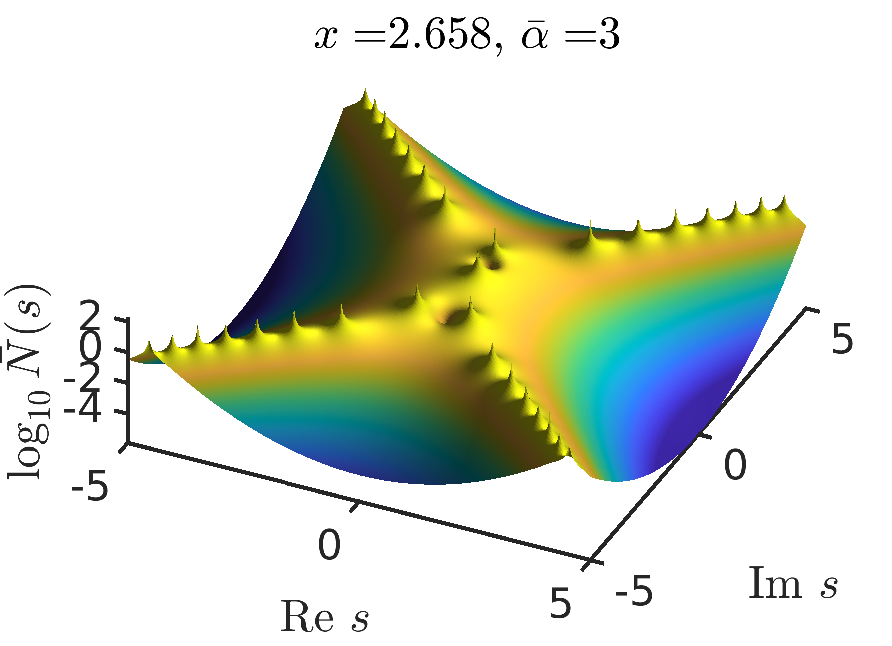}
	\includegraphics[width=2in]{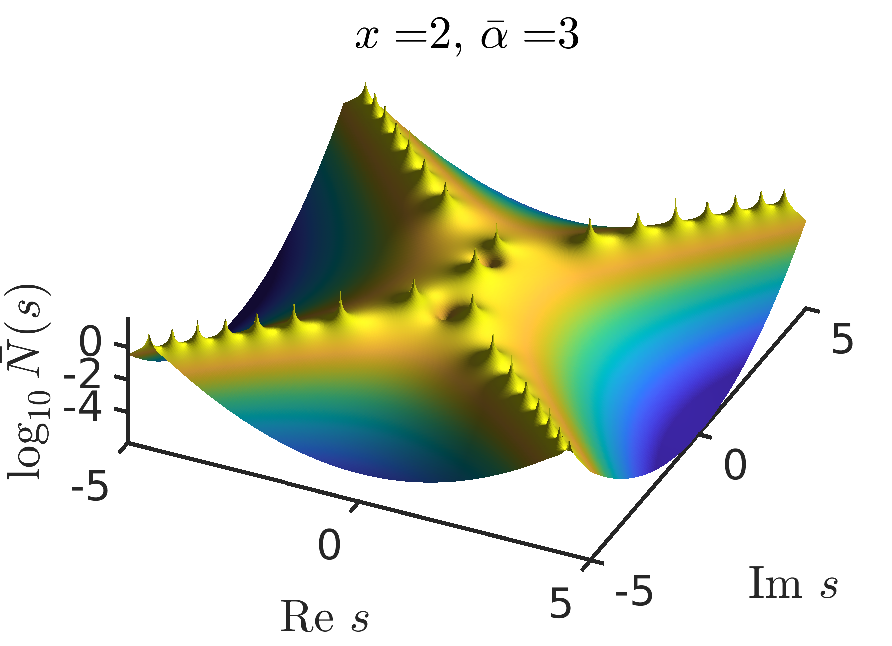}
	\caption{These logarithmic plots obtained by numerical evaluation
		(Appendix~\ref{APNA0}, Eq.~\ref{NA}) illustrate 
		non-adiabatic coupling $\bar N(s)$ in the complex time-plane.
		$\bar N(s)$, has poles at the TPs which is a general
		behavior for two-by-two Hamiltonians, Eq.~\ref{EQNApoles}.
		This figure also illustrates that $\bar N(s)$
		decays in all asymptotes of the complex plane $|s|\to\infty$,
		compare Appendix~\ref{NAsection}.
		The three different choices
		of laser parameters ($x$ and $\balpha$ defined in text)
		correspond with different layouts of
		TPs.}
	\mylabel{FNA}
\end{figure}

\subsection{Local Poiseux expansion coefficients}

\noindent
The Poiseux expansion which includes all TPs can be rewritten to
describe the close neighborhoods of separate TPs. Such expansions are defined as
\begin{align}
\tilde \delta(s) \bigg|_{s\approx s_k} = \beta_k^{(1)} (s-s_k)^{1/2} + \beta_k^{(2)} (s-s_k) + \cdots ,
\mylabel{Puiseux1}
\end{align}
First we prove that for most TPs, the even coefficients $\beta_k^{(2)}$
have zero values.
For this sake we rewrite the above general expansion as,
\begin{align}
\tilde \delta(s) =  f_1(s) \cdot (s-s_k)^{1/2} + f_2(s) .
\mylabel{Puiseuxcompact1}
\end{align}
where $f_1(s)$ and $f_2(s)$ are polynomial expansions 
which include either the odd or even expansion coefficients,
respectively, being defined by the expressions,
\begin{align}
& f_1(s) = \sum\limits_{n=0}^\infty \beta_k^{(2n+1)} (s-s_k)^{n}, \cr
& f_2(s) = \sum\limits_{n=1}^\infty \beta_k^{(2n)} (s-s_k)^{n} .
\mylabel{Puiseuxcompact2}
\end{align}
The square of the quasi-energy split, $\tilde \delta^2(s)$,
using this definition is  given by
\begin{align}
\tilde \delta^2(s) =&  f_1^2(s) \cdot (s-s_k) \cr
&+ 2 f_1(s)\, f_2(s) \cdot (s-s_k)^{1/2} 
+ f_2^2(s).
\end{align}
As this is compared to Eq.~\ref{Puiseux0}, it is clear that
the second term in the equation above must be zero.
This may happen for either $f_1(s)$ of $f_2(s)$ being zero.
As long as we assume as first order TP with non-zero
first-order expansion coefficient $\beta_k^{(1)}$, Eq.~\ref{Puiseux1},
$f_1(s)$ is non-zero,  i.e. $f_2(s)=0$.
This can be true only if all even expansion coefficient 
are zero, 
\begin{align}
\beta_k^{(2n)} = 0. 
\end{align}
An exception to this rule occurs when two TPs coalesc; this intriguing
special situation will be discussed below.

The coefficients $\beta_k^{(2n+1)}$ are related to $\alpha_k$ such that
\begin{align}
\tilde \delta^2(s) &= \alpha_k^2 (s-s_k) \prod\limits_{l\ne k} \alpha_l^2 (s-s_l) \cr
&= (s-s_k) \cdot \(\sum\limits_{n=0} \beta_k^{(2n+1)} (s-s_k)^n\)^2
\end{align}
\begin{align}
\frac{d\tilde \delta^2}{ds} &= \alpha_k^2  \prod\limits_{l\ne k} \alpha_l^2 (s-s_l) \cr
&+ \alpha_k^2 (s-s_k) \(\sum\limits_{l\ne k} \frac{1}{s-s_l} \) \prod\limits_{l\ne k} \alpha_l^2 (s-s_l) 
\cr
\end{align}
\begin{align}
\frac{d\tilde \delta^2}{ds} &=  \(\sum\limits_{n=0} \beta_k^{(2n+1)} (s-s_k)^n\)^2 \cr
&+ 2 \, (s-s_k) \cdot \(\sum\limits_{n=0} \beta_k^{(2n+1)} (s-s_k)^n\) \cr &\cdot
\(\sum\limits_{n=0} \beta_k^{(2n+3)} (n+1) (s-s_k)^{n}\)
\end{align}
Now as we set $s=s_k$ we get for the derivative of the square:
\begin{align}
\frac{d\tilde \delta^2}{ds}\bigg |_{s=s_k} = \alpha_k^2  \prod\limits_{l\ne k} \alpha_l^2 (s_k-s_l) 
=
 \( \beta_k^{(1)} \)^2
 \mylabel{EQ138}
\end{align}
Let us continue in this direction and explore the second derivatives of $\tilde\delta^2$
at the point $s=s_k$. 
\begin{align}
\frac{d^2\tilde \delta^2}{ds^2}\bigg |_{s=s_k} &= 
\alpha_k^2 \(\sum\limits_{l\ne k} \frac{1}{s-s_l} \) \prod\limits_{l\ne k} \alpha_l^2 (s-s_l) 
\cr &=
4\,\beta_k^{(1)}\,\beta_k^{(3)}  .
\end{align}
And we could continue further to obtain definitions all local
Poiseux coefficients $\beta_k^{(n)}$ using the positions of TPs $s_k$ and
their Poiseux coefficients $\alpha_k$. Here we obtained
the expressions,
\begin{align}
& \beta_k^{(1)} = \alpha_k \prod\limits_{l\ne k} \alpha_l (s_k - s_l)^{1/2}, \cr
& \beta_k^{(3)} = \frac{\beta_k^{(1)}}{4} \, \sum\limits_{l\ne k} \frac{1}{s_k-s_l} .
\mylabel{EQ133}
\end{align}
The local expansions effectively describe the quasi-energy split
in the vicinity of the particular TP $s_k$, but in principle
they apply for the whole complex time-plane.

The coefficients for the local expansion, however, become
singular when two TPs become very near, see Eq.~\ref{EQ133}
for $\beta_k^{(3)}$, when $s_l \to s_k$, while
the first coefficient $\beta_k^{(1)}$ goes to zero.
Note that we have shown above that for some specific pulse parameters $\balp$ and $x$,
two purely imaginary roots of $\tilde \delta(s)$  acquire the same 
value $s_{0i} = \bar s_{0i} = s_{0i}^{coal}$, the TPs actually coalesc.
 We can assess that near the coalescence of TPs,
the local Puiseux expansion is slow converging and
possibly efficient and applicable only at infinitesimal
distances from the nearly coalescing TPs.

\subsection{Coalescence of transition points}
\noindent
Let us discuss the Poiseux expansions at the situation
when two TPs, namely $s_{0i}$ and $\bar s_{0i}$, actually coalesc.
This brings about a change in the order of the Poiseux expansion of $\tilde \delta(s)$,
Eq.~\ref{Puiseux0} which takes the new form
\begin{align}
\tilde\delta(s)_{coal}=\alpha_{0i}^{(2)} (s-s_{0i}^{coal})\cdot \prod\limits_{k\ne {0}} \alpha_k (s-s_k)^{1/2} ,
\mylabel{EQcoalC1}
\end{align}
where $\alpha_{0i}^{(2)}$ is the higher-order Puiseux expansion coefficient,
while apart from the coalescence the same expansion takes the form,
\begin{align}
\tilde\delta(s)=\alpha_{0} (s-s_{0})^{1/2} \bar\alpha_{0} (s-\bar s_{0})^{1/2}\cdot \prod\limits_{k\ne {0}} \alpha_k (s-s_k)^{1/2} ,
\end{align}
where the subscript $(k=0)$ stands also for $(k=0i)$, where the difference of the two cases has been defined above.

In local Puiseux expansion, Eq.~\ref{Puiseux1}, the first 
expansion coefficient $\beta_0^{(1)}$ is equal to zero while
the second-order coefficient $\beta_0^{(2)}$ in non-zero.
Using the same argument as for the ususal TPs based on
the compact version of the Puiseux expansion
where the odd and even coefficients participate in
two distinct polynomials $f_1(s)$ and $f_2(s)$, respectively,
see Eqs.~\ref{Puiseuxcompact1}-\ref{Puiseuxcompact2}, we
assess that $f_1(s)=0$, implying that at the
coalescence, all odd coefficients in the local
Puiseux expansion are equal to zero,
\begin{align}
\beta_0^{(2n+1)} = 0.
\end{align}

Again, by comparing the derivatives of $\tilde \delta^2(s)_{coal}$
based on the product and local Puiseux expansions, we
obtain the relation between the higher-order expansion
coefficients $\beta_k^{(2n)}$ and the TPs positions $s_k$ and
the coefficients $\alpha_k$.
The first derivative of $\tilde \delta^2(s)_{coal}$ with respect to $s$
is equal to zero for $s=s_{0i}^{coal}$, 
Eq.~\ref{EQcoalC1}, which is in harmony with our previous findings concerning 
the TPs on the imaginary time axis (Section~\ref{coal}).
The second derivative is given by,
\begin{align}
&\frac{d^2 \tilde \delta^2(s)_{coal} }{ds^2} \bigg |_{s=s_{0i}^{coal}}
= 2  \alpha_{0}^{(2)} \[\prod\limits_{k\ne 0} \alpha_k^2 (s_{0i}^{coal}-s_k)\] \cr
&=2 \(\beta_0^{(2)}\)^2
\mylabel{EQ144}
\end{align}
And the third derivative by
\begin{align}
&\frac{d^3 \tilde \delta^2(s)_{coal} }{ds^3} \bigg |_{s=s_{0i}^{coal}} \cr
& = 6  \alpha_{0}^{(2)} 
\( \sum\limits_{k\ne 0} \frac{1}{s_{0i}^{coal}-s_k}
\)  \[\prod\limits_{k\ne 0} \alpha_k^2 (s_{0i}^{coal}-s_k)\] \cr
&= 12 \, \beta_0^{(2)} \, \beta_k^{(4)} .
\end{align}
From here we obtain the relations for the expansion coefficients
given by
\begin{align}
& \beta_0^{(2)} = \sqrt{\alpha_{0}^{(2)}} \, \prod\limits_{k\ne 0} \alpha_k (s_{0i}^{coal}-s_k)^{1/2}, \cr
& \beta_0^{(4)} = \frac{\beta_k^{(2)}}{2} \, \sum\limits_{k\ne 0} \frac{1}{s_{0i}^{coal}-s_k} .
\end{align}

\subsection{Time-symmetry relations for local Puiseux expansion coefficients}
\noindent
In the case of time-symmetric systems defined above by Eq.~\ref{EQ8d10}
we can define a relation between the Poiseux expansion coefficients
corresponding to the symmetrical pairs of TPs, $s_k$ and $\bar s_k$,
Eq.~\ref{sym}.

Let us start with the relation between the first order expansion
coefficients $\beta_k^{(1)}$ vs. $\bar \beta_k^{(1)}$.
Any point $s$ near $s_k$ can be symmetrically projected to
a point $\bar s$ near $\bar s_k$ as we show in Fig.~\ref{nakressym}.
The value of the quasienergy split at the point $s$ is given by
\begin{align}
\tilde \delta(s) = \beta_k^{(1)} (s-s_k)^{1/2}
+ \beta_k^{(3)} (s-s_k)^{3/2} + \cdots ,
\mylabel{EQ140}
\end{align}
whereas the value at the point $\bar s$ is given by
\begin{align}
\tilde \delta(\bar s) = \bar \beta_k^{(1)} (\bar s-\bar s_k)^{1/2}
+ \bar \beta_k^{(3)} (\bar s-\bar s_k)^{3/2} + \cdots .
\end{align}
Based on Eq.~\ref{EQ8d10} we can substitute in the second equation such that
\begin{align}
[\tilde \delta(s)]^* = \bar \beta_k^{(1)} (\bar s-\bar s_k)^{1/2}
+ \bar \beta_k^{(3)} (\bar s-\bar s_k)^{3/2} + \cdots ,
\end{align}
which can be rewriten such that
\begin{align}
&\tilde \delta(s) = \bar \beta_k^{(1)*} (-)^{1/2} (s-s_k)^{1/2}
\cr & \quad
+  \bar \beta_k^{(3)*} (-)^{3/2} (s-s_k)^{3/2} + \cdots .
\end{align}
By merely comparing this with Eq.~\ref{EQ140} we get
\begin{align}
& \beta_k^{(1)} = \bar \beta_k^{(1)*} (-)^{1/2}, \cr
& \beta_k^{(3)} = \bar \beta_k^{(3)*} (-)^{3/2}, \cdots ,
\end{align}
where the square root $(-)^{1/2}$ has an unknown sign, corresponding
both to $\pm i$.
Based on the fact that the complex number $(s-s_k)^{1/2}$
corresponds to $i(\bar s-\bar s_k)^{1/2}$, 
and not $-i(\bar s-\bar s_k)^{1/2}$,
see Fig.~\ref{nakressym},
it is defined that the sign of the imaginary unit must be positive,
should the square root $(s-s_k)^{1/2}$ be defined as shown in Fig.~\ref{nakressym}.
So that the final time-symmetric relation for the Puiseux
expansion is given by
\begin{align}
\bar \beta_k^{(2n+1)} = -i\,  \beta_k^{(2n+1)*} (-)^{n} .
\mylabel{EQSYMB}
\end{align}

\begin{figure}
	\begin{center}
		\includegraphics[width=3in
		]{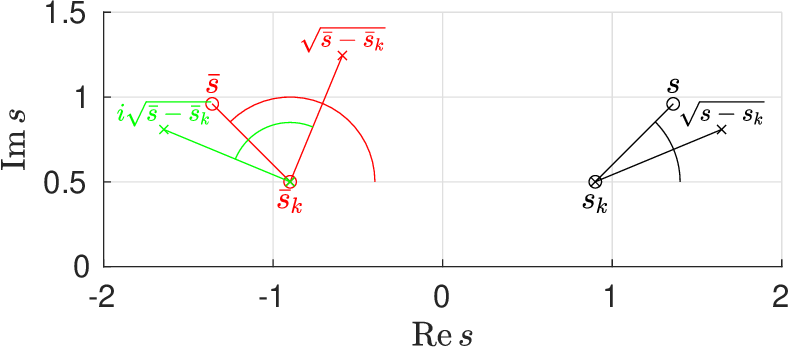}
	\end{center}
	\caption{
	We define the sign of the square root $\sqrt{s-s_k}$
	using the half of the argument of the complex vector $s-s_k$.
	Importantly, the symmetric counterparts $\bar s$ and $\bar s_k$ are
	also shown. 
	We demonstrate that the argument $\arg (\bar s - \bar s_k)$ is
	represented by $\arg ( s -  s_k)$ substracted from $\pi$.
	The complex number defined as the
	square root   $\sqrt{\bar s- \bar s_k}$
	is shown as a vector originating at $\bar s_k$
	to demonstrate that it actually does not point to any
	location symmetrically related to the original vector
	   $\sqrt{s- s_k}$ started at $s_k$.
	However, when $\sqrt{\bar s- \bar s_k}$ is multiplied by
	the imaginary unit $(+i)$, then it does represent the symmetric counterpart
	related to $\sqrt{s- s_k}$.
	}
	\mylabel{nakressym}
\end{figure}

\subsection{\myslabel{argbeta}Complex phase of local Puiseux coefficients}
\noindent
We have proved that the quasi-energy split can
be expressed using Puiseux series defined locally
based on a single chosen reference TP.
The practical way to obtain the local expansion
coefficients $\beta_k^{(n)}$ is using the analytical
expression of the quasi-energy split $\tilde{\delta(s)}$
such as the one given in Eq.~\ref{EQdelred2}.
By differentiating this definition with respect to $s$
and using Eqs.~\ref{EQ138} and \ref{EQ144}
we obtain the main coefficients for the
first and second order TPs:
\begin{align}
	& \beta_{k}^{(1)} = 
\sqrt{\balpha \(\frac{\balpha }{2}s_k + \frac{i}{x}\) - 2s_k\, e^{-s_k^2}}, \cr
	& \beta_{0}^{(2)} = \frac{1}{2} \sqrt{ \balpha^2 - 4e^{-(s_{0i}^{coal})^2}[1-2 (s_{0i}^{coal})^2]} .
	\mylabel{EQ106}
\end{align}

In the course of further derivations, 
we will especially need to know what is the complex
phase of the first order coefficient, $\beta_k^{(1)}$.
Such an analysis, which we show in the Appendix~\ref{PuiseuxBeta1},
requires determining the sign of the root in Eq.~\ref{EQ106}.
This sign depends on how we define the roots $(s-s_k)^{1/2}$,
see Fig.~\ref{nakressym}, and also on the sign of
the quasi-energy split on the real time axis.

The complex phases of the local Puiseux coefficients
obtained via Eq.~\ref{EQ106} and using the sign determined
in the Appendix~\ref{PuiseuxBeta1} in some particular
limiting cases, are illustrated in Fig.~\ref{FigEPbeta}.
Let us summarize here the limiting cases explicitely:
\begin{align}
& \arg \beta_{0i,low}^{(1)} = -\frac{3\pi}{4}, 
& \arg \beta_{0i,upp}^{(1)} = -\frac{\pi}{4}, \cr
& \arg \beta_{0,coal}^{(1)} = -\frac{\pi}{2}, 
& \arg \bar \beta_{0,coal}^{(1)} = 0, \cr
& \arg \beta_{k\to\infty}^{(1)} = -\frac{5\pi}{8}, 
& \arg \bar \beta_{k\to\infty}^{(1)} = \frac{\pi}{8} .\cr
&\mylabel{EQargbeta}
\end{align}
The index $(0,coal)$ refers to the central TPs $s_0$, $\bar s_0$
when they are infinitesimally close to the point of their
coalescence. Fig.~\ref{FigEPbeta}a shows how the complex phase
is modified for different TPs all the way from this very point 
($k=0$) to the limit $k\to\infty$.
Fig.~\ref{FigEPbeta}b illustrates the 
complex phase of the local Puiseux coefficients 
where the central TPs reside on the imaginary time axis.
Comparing the complex phases of $\beta_{0i,low}^{(1)}$ and
$\beta_{0i,upp}^{(1)}$ with the phases of
$\beta_{0,coal}^{(1)}$ and $\bar \beta_{0,coal}^{(1)}$
shows a sudden change  of the complex phases (by $-\pi/4$) of the
local Puiseux coefficients upon the coalescence. 
\begin{figure}
	\begin{center}
		\includegraphics[width=3in
		]{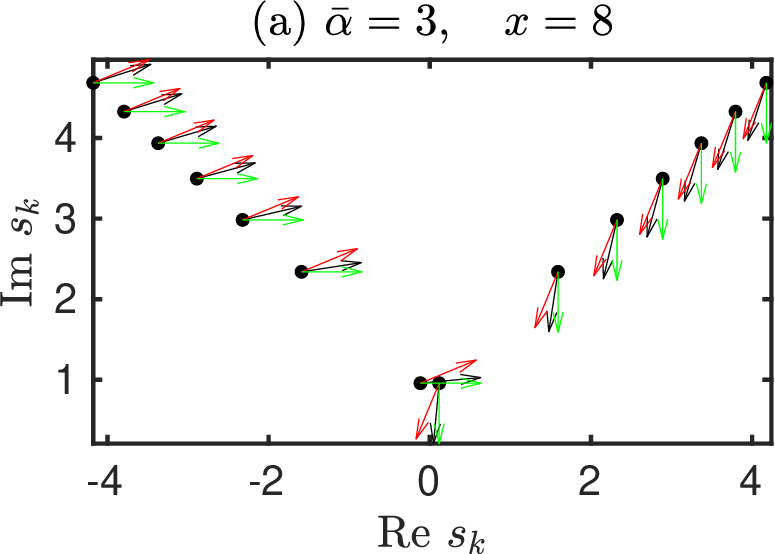}
		\includegraphics[width=3in
		]{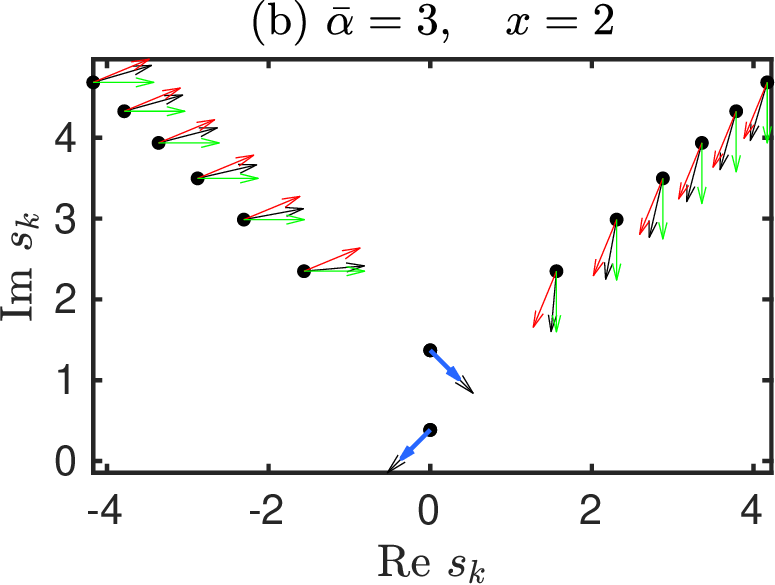}
	\end{center}
	\caption{Complex phases of the local Puiseux coefficients
	$\beta_k^{(1)}$ for two
	different cases of laser pulse parameters $\balpha$ and $x$
	are shown by the black arrows near the
	corresponding TPs $s_k$ and $\bar s_k$, respectively.
	The red arrows show the analytical values for the
	asymptotic limit, $k\to \infty$.
	The green arrows show the analytical values for
	the limit where $s_0$ and $\bar s_0$ approach the
	coalescence.
	The blue arrows show the analytical values for the
	case where the TPs lie on the imaginary axis.
	}
	\mylabel{FigEPbeta}
\end{figure}

\section{Residua}

%\subsection{Auxiliary variables defining quasienergy split and its time-integral}
\noindent
In this Section we calculate the residua at the central TPs numerically,
while giving an analytical estimate for the residua at the distant TPs. 
Before doing so 
we will define a few auxiliary quantities which are useful in such a study
and which will be also used throughout the rest of this paper.
Namely, we will separate the real and imaginary components
of the quasi-energy split,
into effective functions $P(s)$ and $Q(s)$,
respectively,
\begin{align}
&P(s) = \frac{2\hbar}{\Gamma} \re\, \bar \delta(s) \equiv x\, \re \tilde \delta(s),\cr
&Q(s) = \frac{2\hbar}{\Gamma}\im\, \bar \delta(s) \equiv x\, \im \tilde \delta(s).
\mylabel{Palp}
\end{align}
where we used both definitions
of the quasi-energy split in Eq.~\ref{deltaGL1} and 
of the reduced quasi-energy split in Eqs.~\ref{EQdeltil} and \ref{EQdelred2}.
Based on the state exchange in the stroboscopic
encircling (see Section~\ref{Sencirc}, Eqs.~\ref{deltaini} and \ref{deltafin}),
we get the boundary conditions for $Q(s)$ in the simple form
\begin{align}
Q(s\to\mp\infty) = \pm 1 .
\mylabel{Gb}
\end{align}
%which is in accord with the time-symmetry relation Eq.~\ref{EQ8d10} which implies
%\begin{align}
%&P(-s) =  P(s), 
%&Q(-s) = - Q(s) .
%\end{align}

We also define the integral over the 
quasi-energy split such that
\begin{align}
&
\phi(s) = \frac{2\hbar}{\Gamma} \re\[\int\limits_{\tt}^{s} ds'\,\bar\delta(s')\] \equiv
x\, \re\[\int\limits_{\tt}^{s} ds'\,\tilde\delta(s')\] , 
\cr
&
\gamma(s) = \frac{2\hbar}{\Gamma} \im\[\int\limits_{\tt}^{s} ds'\, \bar\delta(s')\] \equiv
x\, \im\[\int\limits_{\tt}^{s} ds'\, \tilde\delta(s')\] ,
\mylabel{gamma}
\end{align}
The time-symmetry relation Eq.~\ref{EQsymenc} is rewritten using 
the functions $\gamma(s)$ and $\phi(s)$ such that
\begin{align}
&\gamma(s) = \gamma(-s^*), 
&\phi(s) = - \phi(-s^*) .
\mylabel{sym21}
\end{align}

%\subsection{\label{CH8}Values of integral quasi-energy split in TPs}

%\noindent
%As we shall see in the course of this paper, the positions of the central EPs, which
%we will denote by the index zero ($s_{i0}$ if they are on the imaginary axis and $s_0$ otherwise),
%define the basic behavior of the survival probability, particularly for the limit
%of a large pulse area $\theta$. For this reason we show here briefly an
%algorithm to numerically obtain these points.

\subsection{Residua of the central transition points}

\head{Position of central transition points}
\noindent
We solve Eq.~\ref{e3} for given parameters $[\balpha,x]$ numerically. First, we calculate 
an initial guess for $\xi_0$ 
(where we note that $s_0 = i \xi_0$, Eq.~\ref{zdef})
using the equation
\begin{align}
\xi_0 \approx \pm \frac{1}{2} \(\balpha \pm \sqrt{\balpha^2 +8\(\frac{1}{x}-1\)}\) ,
\end{align}
which is a solution to
\begin{align}
& x \(1 + \frac{\xi_0^2}{2}\) \approx \pm\(\frac{\balpha x}{2} \xi_0 + 1\) ,
\end{align}
which has been obtained from Eq.~\ref{e3} by applying the square root on both sides
of the equation and then approximating the exponential as a parabole.
This approximation is correct given the fact that $\xi_0^2/2$ is rather small, see
Figs.~\ref{FigEP0} (note that $\xi_0 \equiv -\im s_0$). 
This guess is substituted to a standard minimization procedure to solve Eq.~\ref{e3}
numerically exactly.
Yet, in some cases it is better to use a starting guess using some known
point for similar laser parameters rather then using this quadratic approximation.
Numerically calculated positions of the central TPs are displayed in Fig.~\ref{FigEP0}.
\begin{figure}
	\begin{center}
		\includegraphics[width=3in
		]{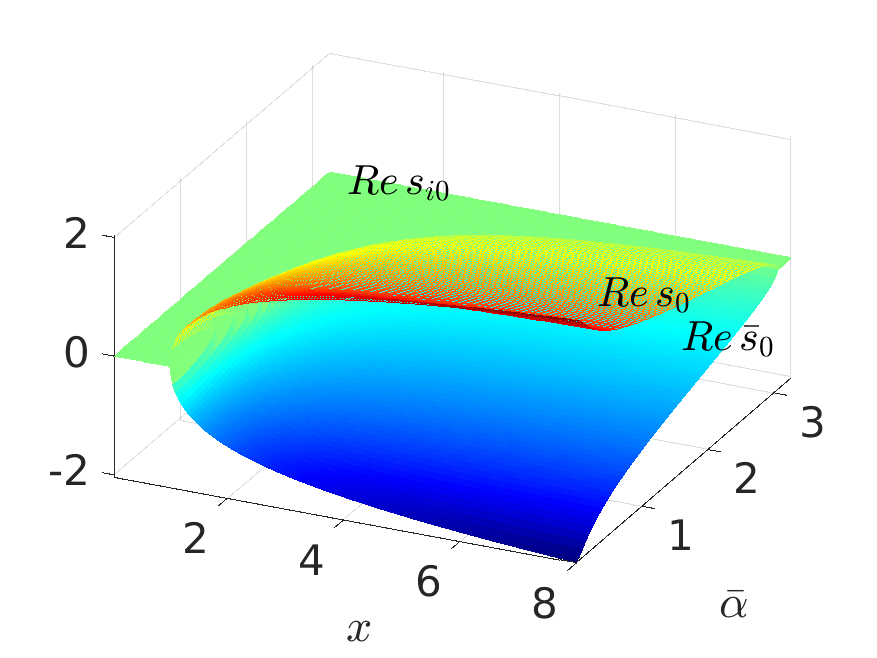}
		\includegraphics[width=3in
		]{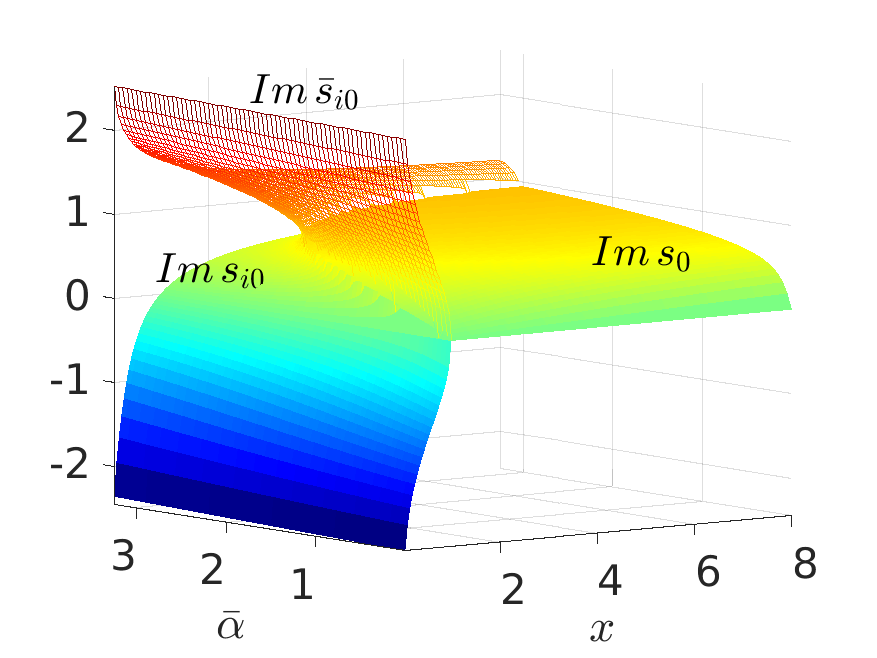}
	\end{center}
	\caption{
		Positions of the central exception points 
		$s_0$, $\bar s_0$ or
		$s_{i0}$, $\bar s_{i0}$ for different laser parameters $x$ and $\balpha$.
	}
	\mylabel{FigEP0}
\end{figure}

\head{Residua of central transition points}
\noindent
The functions $\gamma(s_0)$ and $\phi(s_0)$ for the central TPs
$s_0$, $\bar s_0$ or
$s_{i0}$, $\bar s_{i0}$ 
are calculated using a numerical integration using the numerically obtained values
$s_0$, $\bar s_0$ or
$s_{i0}$, $\bar s_{i0}$.
Fig.~\ref{Figgamphi} confirms numerically
the symmetry relations Eqs.~\ref{sym21}
for the case of $s_0$ and $\bar s_0$
In the case of $s_{i0}$ and $\bar s_{i0}$ these relations are not applicable
because the two TPs do not satisfy Eq.~\ref{sym}. However, the numerical
calculation indicates that  
\begin{align}
\gamma(s_{i0})=\gamma(\bar s_{i0}) 
\mylabel{EQgami0}
\end{align}
holds.
This relation will be explained below where we define the
equivalue lines as the lines
 which include points in the complex plane
 with the same value of $\gamma(s)$.
Namely, it will be shown
that the TPs $s_{i0}$ and $\bar s_{i0}$ are connected
by such a line.
\begin{figure}
	\begin{center}
		\includegraphics[width=3in
		]{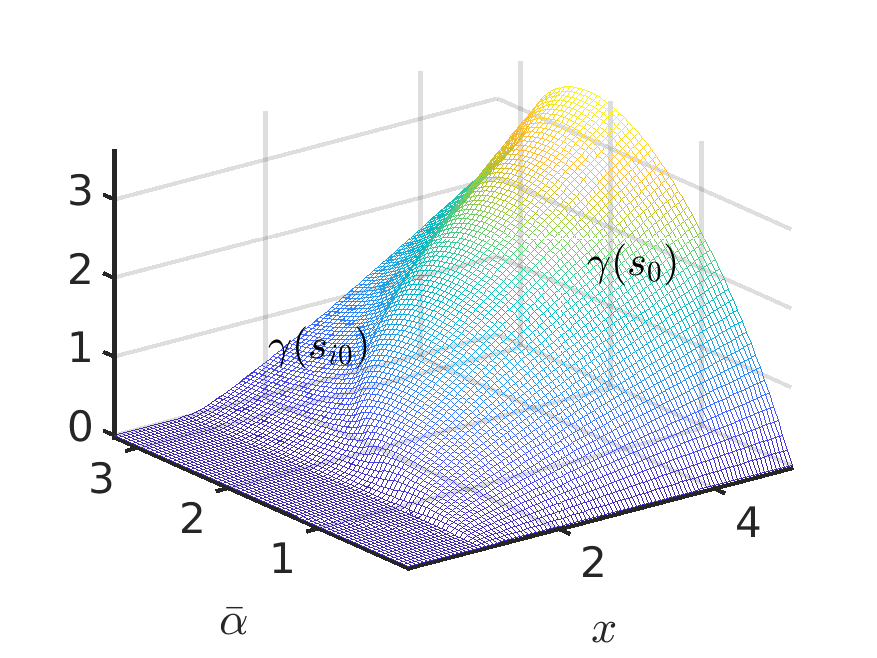}
		\includegraphics[width=3in
		]{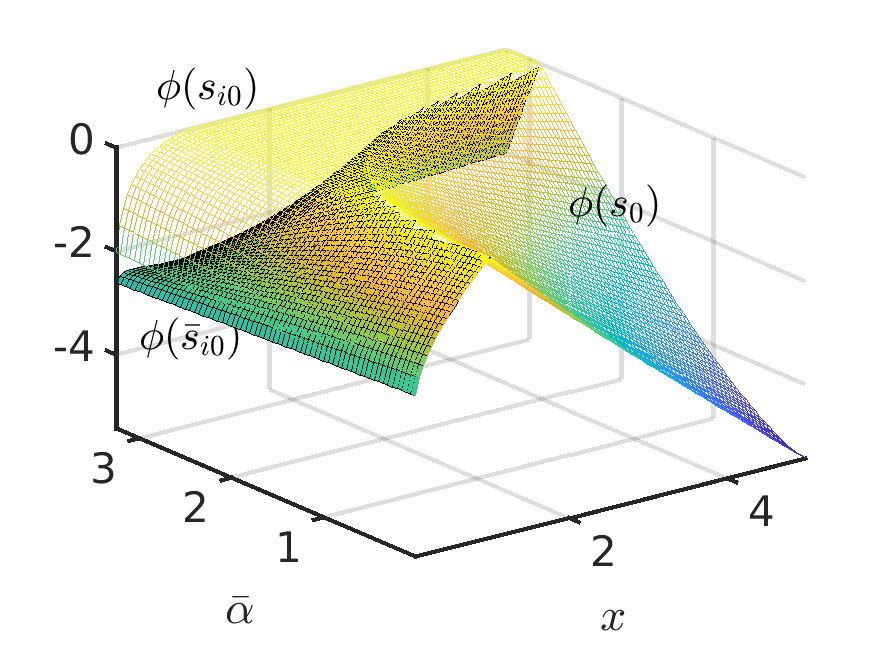}
	\end{center}
	\caption{
		Functions $\gamma(s_0)$ and $\phi(s_0)$ at the central exception points 
		$s_0$, $\bar s_0$ or
		$s_{i0}$, $\bar s_{i0}$ for different laser parameters $x$ and $\balpha$.
	}
	\mylabel{Figgamphi}
\end{figure}

\subsection{Residua of distant transition points ($k\to\infty$)}

\noindent
In the case of other then the central TPs, $s_k$ ($\bar s_k$) lie far from
the center whereas we can assume
\begin{align}
|s_k| \gg 0,
\quad
\arg s_k \approx \frac{\pi}{4}.
\mylabel{asymsk}
\end{align}
Note that the asymptotic value of the argument given
in Eq.~\ref{asymsk}
can be derived immediately using Eq.~\ref{EQQ81}.

We will calculate the residual components $\gamma(s_k)$ and $\phi(s_k)$
using the general equations,
\begin{align}
& \gamma(s_k) = \int\limits_{0}^{s_k} \[ (\im ds) P(s) + (\re ds) Q(s) \], \cr
& \phi(s_k) = \int\limits_{0}^{s_k} \[ (\re ds) P(s) - (\im ds) Q(s) \], 
\mylabel{EQ88}
\end{align}
assuming that the complex integration contour avoids any non-analytical point and
it does not cross any branchcut.
This condition is applicable for the contour which consists of two straight lines --
the first part along the real axis, $0\le s \le \re s_k$, and then the
second part along the imaginary axis, $\re s_k \le s \le s_k$:
\begin{align}
& \gamma(s_k) = \int\limits_{0}^{\re s_k} ds Q(s) + \int\limits_{\re s_k }^{s_k} (\im ds) P(s) , \cr
& \phi(s_k) = \int\limits_{0}^{\re s_k} ds P(s)  - \int\limits_{\re s_k }^{s_k} (\im ds) Q(s) . 
\mylabel{EQ89}
\end{align}

$Q(s)$ and $P(s)$ are generally nontrivial functions which we would get
by substituting
from Eq.~\ref{EQdelred2} to
 Eqs.~\ref{Palp}.
However, we may define the asymptotic forms $Q_\infty(s)$ and $P_\infty(s)$ 
which are applicable
far from the center of axis along our present integration
contour, namely for $s \gg 0$ and $\arg s \le \frac{\pi}{4}$
(see Eq.~\ref{asymsk}).
The asymptotic forms are given by, 
\begin{align}
& P_\infty(s) = \frac{\balpha x}{2} \cdot |\re s|, 
& Q_\infty(s) = \frac{\balpha x}{2} \cdot \im s \cdot \frac{\re s}{|\re s|}.
\mylabel{EQ82}
\end{align}
Let us partition Eq.~\ref{EQ89} using the asymptotic forms such that
\begin{align}
& \gamma(s_k) = \int\limits_{0}^{\re s_k} ds [Q(s) - Q_\infty(s)] \cr
& + \int\limits_{\re s_k}^{s_k} (\im ds) [P(s) - P_\infty(s)] \cr
& + \int\limits_{0}^{\re s_k} ds Q_\infty(s) 
+ \int\limits_{\re s_k }^{s_k} (\im ds) P_\infty(s) , 
\end{align}
and
\begin{align}
& \phi(s_k) = \int\limits_{0}^{\re s_k} ds [P(s) - P_\infty(s)] \cr
& - \int\limits_{\re s_k }^{s_k} (\im ds) [Q(s) - Q_\infty(s)] \cr
& + \int\limits_{0}^{\re s_k} ds P_\infty(s)
  - \int\limits_{\re s_k }^{s_k} (\im ds) Q_\infty(s) . 
\end{align}
After evaluation of the contribution of the asymptotic behavior
we obtain,
\begin{align}
& \gamma(s_k) = \frac{\balpha x}{4} \cdot  \frac{|\re s_k|}{\re s_k} \cdot \im (s_k)^2  \cr & +
\int\limits_{0}^{\re s_k} ds [Q(s) - Q_\infty(s)] \cr
& + \int\limits_{\re s_k}^{s_k} (\im ds) [P(s) - P_\infty(s)] , 
\mylabel{EQ90a}
\end{align}
and
\begin{align}
& \phi(s_k) = 
\frac{\balpha x}{4} \cdot \frac{|\re s_k|}{\re s_k} \cdot \re (s_k)^2 \cr &+
\int\limits_{0}^{\re s_k} ds [P(s) - P_\infty(s)] \cr
& - \int\limits_{\re s_k }^{s_k} (\im ds) [Q(s) - Q_\infty(s)] .
\mylabel{EQ90b}
\end{align}

Let us discuss the nontrivial part of 
the integration including $Q(s)-Q_\infty(s)$ and
$P(s)-P_\infty(s)$ as the integrands. 
As for the integration along the real axis of $s$
we refer the reader to Fig.~\ref{FIGreals1},
noting that the
functions $Q(s)$ and $P(s)$ are related to real and imaginary
components of $\tilde \delta(s)$ via Eqs.~\ref{Palp}.
By substracting the asymptotic forms, $Q(s)-Q_\infty(s)$ and
$P(s)-P_\infty(s)$, we have constructed integrands which are zeros in the
asymptotes $\re s \to \pm \infty$.
Assuming that the upper bound $\re s_k$ represents the asymptotic limit,
$\re s_k \to \infty$, these integrals can be approximated by the
assymptotic expressions which we define as,
\begin{align}
& \gamma_\infty(x,\balpha) = \int\limits_{0}^{\infty} ds [Q(s;x,\balpha) - Q_\infty(s)], \cr
& \phi_\infty(x,\balpha) = \int\limits_{0}^{\infty} ds [P(s;x,\balpha) - P_\infty(s)] ,
\end{align}
and which are independent on the position of the transition point $s_k$.
The integrals $\gamma_\infty(x,\balpha)$ and $\phi_\infty(x,\balpha)$ only
depend on the pulse parameters $x$ and $\balpha$ through the dependence
of $Q(s)$ and $P(s)$ on these parameters.

Now we shall explore the integrals in the imaginary axis
which ends up at the TP $s_k$. Due to the fact that the complex argument
of $s$ is limited along the integration contour 
such that
\begin{align}
0 \le \arg s < \frac{\pi}{4} ,
\end{align}
compare Eq.~\ref{asymsk}, the absolute value of the
exponential term which is included in 
the defintion of $\tilde \delta(s)$ (Eq.~\ref{EQdelred2})
is always smaller or equal to one.
Therefore $\tilde \delta(s)$ can be approximated using the first
order of the Taylor expansion in the
asymptotic limit (Eqs.~\ref{asymsk}) such that
\begin{align}
\tilde \delta(s) \approx \(z\frac{\balpha}{2} \cdot s + \frac{i}{x}\)
+ \frac{e^{-s^2}}{2}\(z\frac{\balpha}{2} \cdot s + \frac{i}{x}\)^{-1},
\end{align}
where
\begin{align}
z=\frac{\re s}{|\re s|} .
\end{align}
The exponential term can be written as
\begin{align}
e^{-s^2} = e^{-|s|^2 \ \cos (2\arg s) }\ 
e^{-i|s|^2 \ \sin(2\arg s)} ,
\end{align}
which shows clearly that the absolute value of the exponential
is negligible as $|s|\to\infty$.
Therefore the integrals along the imaginary axis in
Eqs.~\ref{EQ90a} and \ref{EQ90b} can be omitted.
Finally, we obtain,
\begin{align}
\gamma(s_k) = \frac{\balpha x}{4} \cdot  \frac{|\re s_k|}{\re s_k} \cdot \im (s_k)^2 +
\gamma_\infty(x,\balpha) , 
\mylabel{EQ91a}
\end{align}
and
\begin{align}
\phi(s_k) = 
\frac{\balpha x}{4} \cdot \frac{|\re s_k|}{\re s_k} \cdot \re (s_k)^2 +
\phi_\infty(x,\balpha) . 
\mylabel{EQ91b}
\end{align}

Now it is possible to substitute to Eqs.~\ref{EQ91a} and
\ref{EQ91b} using the result in Eq.~\ref{EQQ81} such that
\begin{align}
& \gamma(s_k) =  \frac{\balpha x}{4} \pi\(2k + \frac{1}{2}\)+
\gamma_\infty(x,\balpha), \cr
& \phi(s_k) = - \frac{\balpha x}{4} \ln\(k\pi\frac{\balpha^2}{2}\)+
\phi_\infty(x,\balpha),\cr &
\quad\quad k\to\infty.
\mylabel{EQ8d7}
\end{align}
using the fact that $\re s_k > 0$.

The TPs in the negative real half-axis are denoted
as $\bar s_k$ as defined in Eq.~\ref{sym}; an asymptotic approximation $k\to\infty$
for the points $\bar s_k^2$ is given in Eq.~\ref{EQQ81m}.
By substituting from this equation to Eqs.~\ref{EQ91a} 
and \ref{EQ91b} and using $\re \bar s_k<0$, we obtain
\begin{align}
& \gamma(\bar s_k) =  \frac{\balpha x}{4} \pi\(2k + \frac{1}{2}\)+
\gamma_\infty(x,\balpha), \cr
& \phi(\bar s_k) = \frac{\balpha x}{4} \ln\(k\pi\frac{\balpha^2}{2}\)-
\phi_\infty(x,\balpha),\cr &
\quad\quad k\to\infty.
&\mylabel{EQ8d8}
\end{align}
Note that the results satisfy the symmetry relations
\begin{align}
&\gamma(s_k) = \gamma(\bar s_k), \cr
&\phi(s_k) = - \phi(\bar s_k), & k\to\infty,
\mylabel{sym2a}
\end{align}
which is in harmony with Eq.~\ref{sym21}.

\section{\label{EQVL}Equivalue lines of the residua}

\subsection{Residuum theorem in the presence of branch cuts}
\noindent
The calculation of the integral Eq.~\ref{v1} using residuum theorem
assumes integrating along a complex contour which avoids the
poles, namely the TPs. 
However, as the TPs represent not only the poles, but also the
branchpoints of the quasi-energy split, the
infinitesimal circles 
\begin{align}
s_k + r \cdot e^{i\varphi}, \quad r\to 0, \ 0\le \varphi < 2\pi,
\end{align}
drawn by such an integration contour
near each one of the TPs, suffer from the discontinuity,
\begin{align}
\tilde\delta(s_k + r\cdot e^{i\varphi}) \ne \tilde\delta(s_k + r\cdot e^{i\varphi+2\pi}),
\end{align}
which follows from the Puiseux expansion (Eq.~\ref{Puiseux1}).
Note that this point has been discussed in the context of
encircling contour in Section~\ref{Sencirc} whereas a
similar argument applies here for the integration contour.

The dicontinuity of $\tilde \delta(s)$
is reflected in the discontinuity of $\gamma(s)$ and $\phi(s)$
(Eq.~\ref{gamma}) which appear in the exponent of
the integrand (Eq.~\ref{v1}).
This problem can be solved by defining the complex contour
which includes also integration along the branchcuts,
represented by curves starting at each TP.
Such an integration contour will be discussed in more detail below,
see Fig.~\ref{FigCont} for the typical integration contours.

The introduction of the branchcuts into the integration
contour implies that the complex plane
is cut by a line which starts at the TP and proceeds all the way to
an asymptote. We have demonstrated, see Eq.~\ref{EQ82}, that $Q(s)$ and $P(s)$
functions are linearly dependent on $s$, which may easily lead to
a divergent behavior when integrated up to infinity ($|s|\to\infty$).
This could lead to a rather clumsy if not impossible application
of the suggested complex contour integration.

It is possible, however, to define curves, associated with a
particular TP ($s=s_k$), with the specific behavior that $\gamma(s)$
is constant along such curves, while all the change of the integrated
function $\tilde \delta(s)$ is reflected into the variations of
$\phi(s)$. Because $\phi(s)$ is responsible only for the phase
of the integrated function (Eq.~\ref{v1}), the complex contour integration
along such curves avoids the problem of divergency.
Such curves are therefore most suitable to define the intended integration path.

These curves have been proposed before for integration 
including branchpoints and
are refered to either as the {\it equivalue lines} or the {\it Stoke's lines},
Refs.~\cite{Dykhne:1962,Davis:1975}.
In this Section we explore the asymptotic analytical behavior of
the equivalue lines for the particular studied case
of the symmetric EP encircling based on the Gaussian contour.

\subsection{Definition of equivalue lines}

\noindent
The so called equivalue lines are defined
as the curves which

\noindent
(i) start at the positions of TPs in the complex plane of, $s=s_k$, and

\noindent
(ii) satisfy the condition 
\begin{align}
	\gamma(s) - \gamma(s_{k}) = 0
\end{align}
at any point $s$ on the curve. This means that the exponential term
of the integrand (Eq.~\ref{v1}) given by
\begin{align*}
	\exp{\[ -\frac{\tau\Gamma}{2\hbar}\[\gamma(s) - i \phi(s_k)\]\]} .
\end{align*}
has a constant absolute value along the equivalue lines, which is defined
by 
\begin{align}
	\exp{\[ -\frac{\tau\Gamma}{2\hbar}\gamma(s_k)\]} .
\end{align}
The equivalue lines are defined by the condition:
\begin{equation}
\im \oint\limits_{s_k}^{s} ds' \tilde\delta(s') = 0 .
\mylabel{EQ_BC}
\end{equation}
%The equivalue lines will be used in a definition of the complex contour,
%because they allow to avoid exponential blowing up of the integrated
%function along the contour.
%
%
%The curves defined in Eq.~\ref{EQ_BC} 
%have specific asymptotic limits, which coincide either with the
%imaginary axis of $s$ or with the line $\im s = - 2/\balpha x$,
%parallel to the real axis of $s$ (as we will show below).

\subsection{\myslabel{EQLPHIN}Equivalue lines behavior near the TPs}

\noindent
The condition given by Eq.~\ref{EQ_BC} may be satisfied by different
curves emanating from the given TP ($s_k$) in the complex time plane
depending on the Puiseux order of the TP.
Dykhne, Davis, and Pechukas show three different equivalue lines
launching from the TPs in their studies~\cite{Dykhne:1962,Davis:1975}.

Let us define an infinitesimal contour closely
encircling the TP and find for how many points on this circle
Eq.~\ref{EQ_BC} is satisfied. 
We substitute for $\tilde\delta(s)$ in Eq.~\ref{EQ_BC} by using the first term
in the Puiseux series, Eqs.~\ref{Puiseux1}:
\begin{align}
\im \oint\limits_{s_k}^s ds' \beta_k^{(m)} (s-s_k)^{\frac{m}{2}} = 0,
\end{align}
where $m = 2$ for the case of coalescent TP and $m=1$ otherwise.
We solve this integral and get
\begin{align}
\im \[\frac{2\beta_k^{(m)}}{m+2} (s-s_k)^{\frac{m}{2}+1}\] = 0.
\end{align}
The imaginary component is equal to zero if the argument of the expression
satisfies:
\begin{align}
&\arg{\(\frac{2\beta_k^{(m)}}{m+2}\)} + \(\frac{m}{2}+1\) \arg{(s-s_k)} = n\pi, 
\cr
&\quad n\in Z .
\end{align}
$\arg(s-s_k)$ represents an angle which defines the slope
of the equivalue line as it emanates from the TP. The
angle is actually defined by the point which lies on the infinitesimal
circular contour and satisfies the
condition given by Eq.~\ref{EQ_BC}. Using the
above equation we obtain the analytical expression
for this angle, namely
\begin{align}
&\arg{(s-s_k)} = 2\pi\,\frac{n}{m + 2} - \frac{2}{m+2}\,\arg{\(\frac{2\beta_k^{(m)}}{m+2}\)} , \cr
& \quad 1\le n \le (m+2) .
\mylabel{BC1}
\end{align}
Obviously, there are $(m+2)$ possible equivalue lines for every TP.
Namely, there are {\it three} equivalue lines
of non-coalescent TPs ($m=1$), and {\it four} equivalue lines
for the special case of the coalescent TP ($m=2$).

Let us denote the angles of the emanating equivalue lines 
defined in Eq.~\ref{BC1} as $\varphi_{kn}$,
\begin{align}
\varphi_{kn} \equiv \arg (s-s_k) ,
\end{align}
where the integer number $n$ represents a unique identification
for each equivalue line. 
The angles $\varphi_{kn}$ of 
the most common first Puiseux order TPs  are defined as
\begin{align}
\varphi_{kn} = \frac{2 \pi}{3} n - \frac{2}{3}\arg \beta_k^{(1)}\ ,
\mylabel{EQ157}
\end{align}
where $n=\{0,1,2\}$. Clearly, the angles
are equally distributed along the circle divided by $2\pi/3$,
while the first angle associated with $n=0$ is
simply given by $-2 \arg \beta_k^{(1)} /3$.

The complex phases of the first order local Puiseux coefficients
have been discussed and calculated in Section~\ref{argbeta},
see particularly Fig.~\ref{FigEPbeta} and Eq.~\ref{EQargbeta}.
Using the numerical results given in Fig.~\ref{FigEPbeta} and
Eq.~\ref{EQ157} we calculate the angles $\varphi_{kn}$
as shown in Fig.~\ref{F3}.
Note that the number cycle given by $n\in\{0,1,2\}$ 
fails to reflect the symmetry of the problem. Namely,
if we assume the pairs of $s_k$ and $\bar s_k$, the equivalue lines
numbered by the same values of $n$ do not correspond to the 
pairs of angles $\varphi_{kn}$ and $\pi - \varphi_{kn}$, as one
would expect based on Fig.~\ref{nakressym}.
Yet the number cycle proves correct --
let us define
\begin{align}
\bar{\varphi}_{kn} = \arg(s-\bar s_k) = \pi - \varphi_{kn} .
\end{align}
Using the new definition of $\varphi_{kn}$ in Eq.~\ref{EQ157}
we get
\begin{align}
\bar{\varphi}_{kn} = \pi - \frac{2 \pi}{3} n + \frac{2}{3}\arg \beta_k^{(1)} .
\end{align}
Let us use the argument of the local Puiseux coefficient associated
with $\bar s_k$ rather then $s_k$, Eq.~\ref{EQSYMB}; we obtain,
 \begin{align}
 \bar{\varphi}_{kn} = \pi - \frac{2 \pi}{3} n - \frac{\pi}{3} - \frac{2}{3}\arg \bar \beta_k^{(1)} .
 \end{align}
From here we see that the symmetric equivalue lines must be started
at different parts of the cycle. Supposed that we define the beginning
of the cycle for the symmetric counterpart using the symbol $\bar n$ such
that
\begin{align}
\bar{\varphi}_{kn} = \frac{2 \pi}{3} \bar n - \frac{2}{3}\arg \bar \beta_k^{(1)} .
\end{align}
we get
\begin{align}
\bar n = 1 - n ,
\mylabel{EQnsym}
\end{align}
which is in agreement with the asymmetry of numbering apparent from Fig.~\ref{F3}.

%First, we examine the TPs 
%$s_{0i,low}$,
%$s_{0i,upp}$ ,
%$s_{0,coal}$,
%$\bar s_{0,coal}$,
%$s_{k\to\infty}$, and
%$\bar s_{k\to\infty}$
%which represent
% limiting cases for which  the complex phases of 
% the first order Puiseux coefficients $\beta_k^{(1)}$
% are known analytically, see Section~\ref{argbeta} and Eqs.~\ref{EQargbeta}.
% The results are listed in Table~\ref{TABphi}.
%\begin{table}
%\begin{tabular*}{0.4\textwidth}{ @{\extracolsep{\fill} }c | c c c }
%	\backslashbox{$s_k$}{$n$} & 1 \hfill & 2 \hfill& 0 \\ \hline
%	$s_{0i,low}$ & $-\frac{5\pi}{6}$ & $-\frac{\pi}{6}$ & $\frac{\pi}{2}$ \\
%	$s_{0i,upp}$ & $\frac{5\pi}{6}$ & $-\frac{\pi}{2}$ & $\frac{\pi}{6}$ \\
%	$s_{0,coal}$ & $\pi$ & $-\frac{\pi}{3}$ & $\frac{\pi}{3}$ \\
%	$\bar s_{0,coal}$ & $\frac{2\pi}{3}$ & $-\frac{2\pi}{3}$ & $0$ \\
%	$s_{k\to\infty}$ & $-\frac{11\pi}{12}$ & $-\frac{\pi}{4}$ & $\frac{5\pi}{12}$ \\
%	$\bar s_{k\to\infty}$ & $\frac{7\pi}{12}$ & $-\frac{3\pi}{4}$ & $-\frac{\pi}{12}$
%\end{tabular*}
%\caption{Angels $\varphi_{nk}$ of the equivalue lines as they emanate from the
%indicated TPs. These results obtained analytically correspond well
%with the numerically calculated equivalue lines shown in Fig.~\ref{F3}.
%}
%\mylabel{TABphi}
%\end{table}

%Importantly, we obtained the relation between the values
%of $(n)$ and the angles $\varphi_{nk}$ of the lines which will be
%necessary to solve the complex contour integration below.

%\subsection{\label{EQL0}Numerical calculation of the equivalue lines}

%\noindent
The equivalue lines far from the TPs, which are also shown
in Fig.~\ref{F3}, represent curves 
satisfying Eq.~\ref{EQ_BC}. They are calculated numerically
as the initial value problem using a propagation scheme.
We define a real propagation parameter $\lambda$, where the points on the equivalue line
$s_{BC}$ satisfy:
\begin{align}
\frac{d\, s_{BC}(\lambda)}{d\lambda} = \pm\exp\(-i\arg\tilde\delta[s_{BC}(\lambda)]\) ,
\mylabel{BC2}
\end{align}
where $d\lambda = |ds_{BC}|$.
The initial point $s_{BC}(d\lambda)$ is obtained using Eq.~\ref{BC1}:
\begin{align}
	&
s_{BC}(d\lambda) = s_k + \cr
	&\quad d\lambda \cdot
\exp\[i\(2\pi \,\frac{n}{m + 2} - \frac{2}{m+2}\,\arg{\(\frac{2\beta_k^{(m)}}{m+2}\)}\)\] .\cr
	&\mylabel{EQ105}
\end{align}
\begin{figure}[h!]
	\begin{center}
(a)
\includegraphics[width=2.8in
]{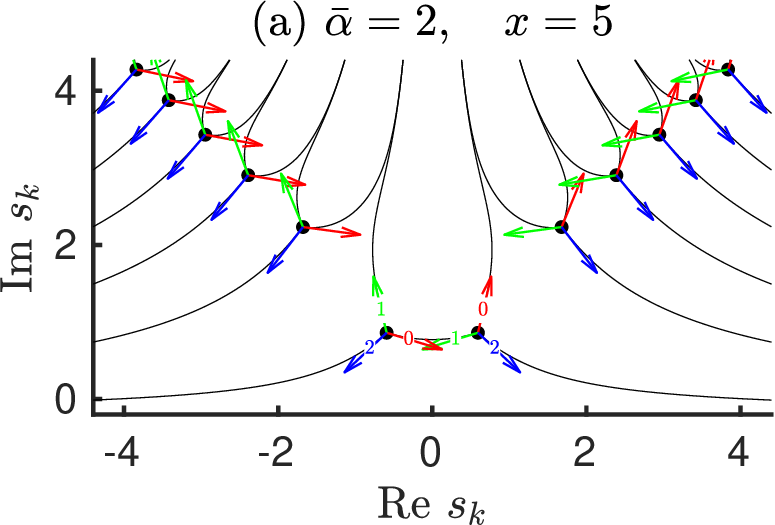}
\vskip 0.5 in
(b)
\includegraphics[width=2.8in
]{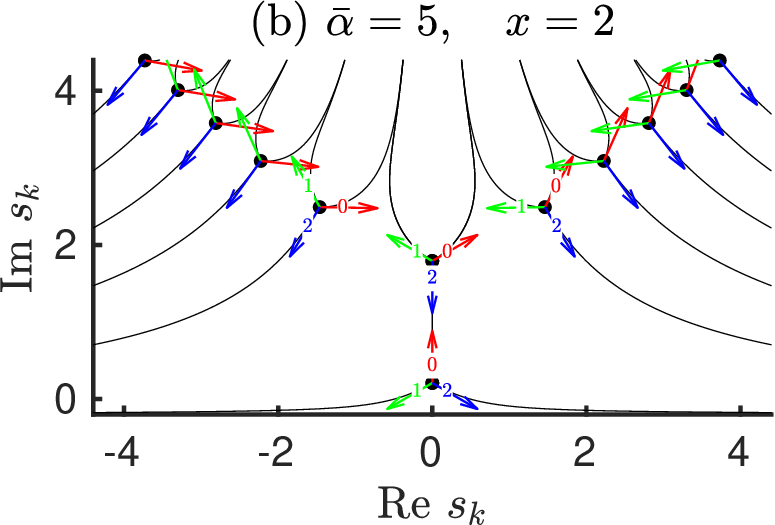}
\end{center}
\caption{
	Equivalue lines (black curves) represent lines where the
	integral over quasienergy split has the same imaginary
	value as in the corresponding TP.
	(Such lines will be very important below for the definition of integration
	contours in the complex time plane as the integrand is
	non-divergent along these particular lines.)
	The number of equivalue lines associated with a
	given TP is defined by the TP Puiseux order. Mostly
	encountered first order TPs  are
	associated with three equivalue lines.
	The three equivalue lines are uniquely identified by the
	index $n$, which is associated with the complex
	phase of the local Puiseux coefficient and the angle of
	the emanating equivalue line (denoted by the colored
	arrows). 
	The values of $n$ are denoted by different colors ($n=0$ -- red,
	$n=1$ -- green, $n=2$ -- blue).
}
\mylabel{F3}
\end{figure}

\subsection{\myslabel{EQLAS}Asymptotic behavior of the equivalues lines}

\noindent
As one can see from the numerical calculation
 shown in Fig.~\ref{F3}, the 
equivalue lines approach the real axis of $s$ (in the case of the ``green'' set) or the imaginary axis
of $s$ (in the case of the ``red'' and ``blue'' sets)
in the asymptotes. 
Below we show the exact analytical asymptotic relations.

Let us start with the equivalue lines approaching the
imaginary time axis.
In this limit, $\im\, s\to\infty$, $\tilde \delta(s)$ is approximated by the exponential term,
\begin{align}
\tilde \delta(s) \approx \pm  e^{-s^2/2} ,
\mylabel{EQ9d11}
\end{align}
as follows from Eq.~\ref{EQdelred2}.
The asymptotic form of 
$\tilde \delta(s)$ enters the definition of the equivalue line 
on the right hand side of Eq.~\ref{BC2} through its complex phase.
By performing appropriate analytical steps which are
given in Appendix~\ref{APEQL1},
we get the asymptotic expression 
\begin{align}
\re s_{BC} = \frac{(n\pm 1/2)\pi}{\im s_{BC}}  .
\mylabel{EQ718}
\end{align}
The asymptotic expression given by Eq.~\ref{EQ718} is illustrated
in Fig.~\ref{FigBCasym}.

Now, a very similar procedure is used for the equivalue lines
which are parallel to the real axis. 
Here we use the assumption $\re\, s\to\infty$ and
$\tilde \delta(s)$ is approximated by,
\begin{align}
\tilde \delta(s) \approx \pm \(\frac{\balpha}{2} \cdot s + \frac{i}{x} \) .
\mylabel{EQG8}
\end{align}
based on its definition Eq.~\ref{EQdelred2}.
By doing the appropriate analytical derivation as described
in Appendix~\ref{APEQL2},
we obtain the asymptotic relation given by,
\begin{align}
{\im s} = \frac{c_k}{\re s} -\frac{2}{\balpha x} ,\quad \re s>0 ,
\mylabel{EQ114}
\end{align}
where the constant $c_k$ remains undetermined.
Eq.~\ref{EQ114} shows that the equivalue lines approach
asymptotically to the line parallel to the real time axis 
lying at $\im\,s = -2/(\balpha x)$ which is in accord
with the numerical result (Fig.~\ref{F3}).

The unknown constants $c_k$ would determine the points $s_{k0}$
where the equivalue lines cross with the real axis such that
\begin{align}
s_{0k} =(\pm) \frac{\balpha\, x\, c_k}{2} .
\mylabel{EQsrecr}
\end{align}
The constants $c_k$ can be determined for large indices $k$,
where one can assume that the asymptotic form of
equivalue lines Eq.~\ref{EQ114} is valid including the
TP $s_k$ itself. Using the knowledge of the positions of $s_k$
for the asymptotic regime $k\gg 1$ as given in Eq.~\ref{EQQ82},
we get an analytical relation for the constants $c_k$ given by
\begin{align}
& c_k \approx (k\pi + \frac{\pi}{4}) 
+ (k\pi + \frac{\pi}{4})^{1/2} 
\( 1 - \frac{\ln(k\balpha\pi)}{4\pi k} \).
\mylabel{EQ940}
\end{align}
The details of the derivation are given in Appendix~\ref{APEQL3}.
The qualitative and even quantitative
precision of such an approximation is illustrated in Fig.~\ref{FigBCasym}.
\begin{figure}[h!]
	\begin{center}
		\includegraphics[width=3in]{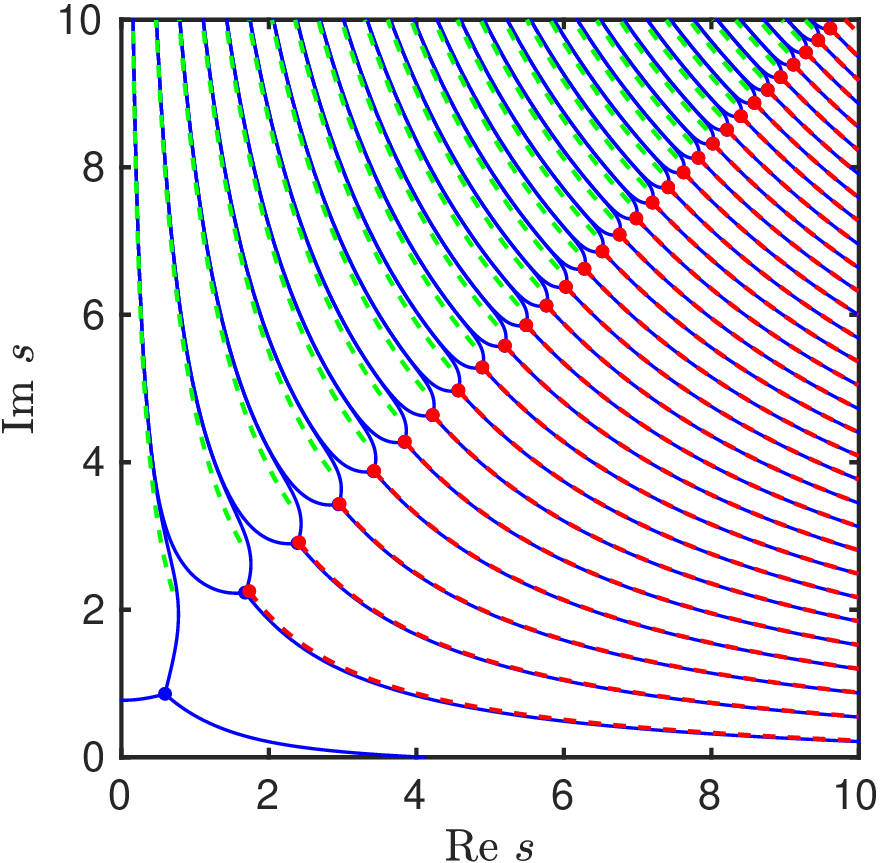}
	\end{center}
	\caption{Asymptotic approximations to the equivalue lines given by
		Eqs.~\ref{EQ718} (the minus sign before $1/2$
		is used, $n=0,1,...$) and \ref{EQ114} together with Eq.~\ref{EQ940} ($k=1,2,...$).}
	\mylabel{FigBCasym}
\end{figure}

We point out that the crossing points of the equivalue lines
with the real axis $s_{0k}$ depend
linearly on $k$ in the asymptote, while their distance
is determined by the increment $\balpha\,x/2$, which
gets infinite in the Hermitian case where the limit $x\to\infty$
is applicable.
Additionally, the equivalue lines asymptotically 
approach to the real time axis in the Hermitian limit
as follows from Eq.~\ref{EQ114} as long as $-2/(\balpha\,x)\to 0$.

\section{\label{contour}New integration contours for non-Hermitian problem}

\subsection{Integration contours in Hermitian vs. non-Hermitian cases\myslabel{contoura}}

\noindent
This is the key Section of the present paper. 
We derive here the {\it complex time method
for dissipative systems}, which include
all cases where the EP (which lies in the frequency--field-amplitude
plane at the non-zero field-amplitude) is dynamically encircled.

Our present approach is based on the first-order 
perturbation approximation of the perturbation series
discussed above.
In this approximation, the sum in Eq.~\ref{EQa10}
is replaced only by the first term such that
\begin{align}
v\equiv \sum\limits_{j=1,3,\cdots} \ v^{(j)}(t\to \infty)
\approx 
v^{(1)}(t\to \infty) .
\end{align}
To simplify the notation, we denote this expression using
the symbol $v$.
Thus, the following parts of this study will be based on
an analysis of the one-dimensional integral given by
\begin{align}
	v = -\int\limits_{-\infty}^\infty ds e^{-\frac{\tau\Gamma}{2\hbar}[\gamma(s)-i\phi(s)]} \bar N(s) .
	\mylabel{v1}
\end{align}
It has been shown by Davis and Pechukas that the first order 
approximation leads to an error given by a factor of
$3/\pi$, where this result applies in the semiclassical limit, here
defined by the large pulse area limit, $\theta\to\infty$~\cite{Davis:1975}. 
Their derivation, highly non-trivial as it is,
is applicable only for the simpliest case including a single TP.
Yet, more then one TP contributions will be included in the analysis of the non-Hermitian case below.
Based on these facts, here we limit ourselves to using an ad hoc
approach, assuming the first-order approximation having the 
approximate error of the same factor, 
which is in agreement with our numerical verifications (Appendix~\ref{AP:APTnum}).

Davis and Pechukas defined a complex contour for
Hermitian systems which takes into account only contributions of the TPs that
are the closest to the real axis.
Such complex contour has been advocated for use also 
in dissipative systems in Refs.~\cite{Schilling:2006,Dridi:2010,Dridi:2012}.
The discussed complex contours are depicted in Figs.~\ref{FigContDDP}) for
the odd and even layouts of the TPs.
The contours lead along the equivalue lines that start at the central TPs and end
asymptotically at $\re\, t\to \pm\infty$.
In this asymptotic limit, the imaginary part is given by
$\im\, t \to -2/\bar\alpha x$, Eq.~\ref{EQ114}, which is non-zero for dissipative systems
therefore the contour ends up {\it below} the real axis.
In order to make this contour useful,
the interval on the real axis must shrinked
to a smaller interval designated by the points of intersections
with the equivalue lines defined by $s_{00}$, see Fig.~\ref{FigContDDP}.
Only supposed that contribution of $|s|>s_{00}$ is negligible, this type of complex contour
integration may be justified as an approximation for dissipative systems.
%In other words, while the DPP formula is exactly justified for Hermitian Hamiltonians,
%the same contour is not correct for dissipative systems (such as ours).

Here we present a different approach to this problem.
We define new complex contours based on brachcuts that are placed
along other equivalue lines, namely those that end up in the asymptotic limit $\im t \to \infty$
as shown in Figs.~\ref{FigCont}.
This leads to a different perception of the quantum dynamics, from a
two or single TP-controlled one to the one controlled by the whole
complex time plane, because now $all$ TPs contribute
to the non-adiabatic amplitudes.

This brings about two implications.
First, Hermitian problems represent a special case where the contributions
of the TPs inside of the complex plane are mutually canceled 
due to a destructive interference.
Second, only when the pulse area is
asymptotically large, the contributions of the two central TPs prevail,
and for this matter the use of the Dykhne, Davis, and Pechukas solution is justified
in dissipative cases in this limit.
\begin{figure}[h!]
	\begin{center}
		\includegraphics[width=3in]{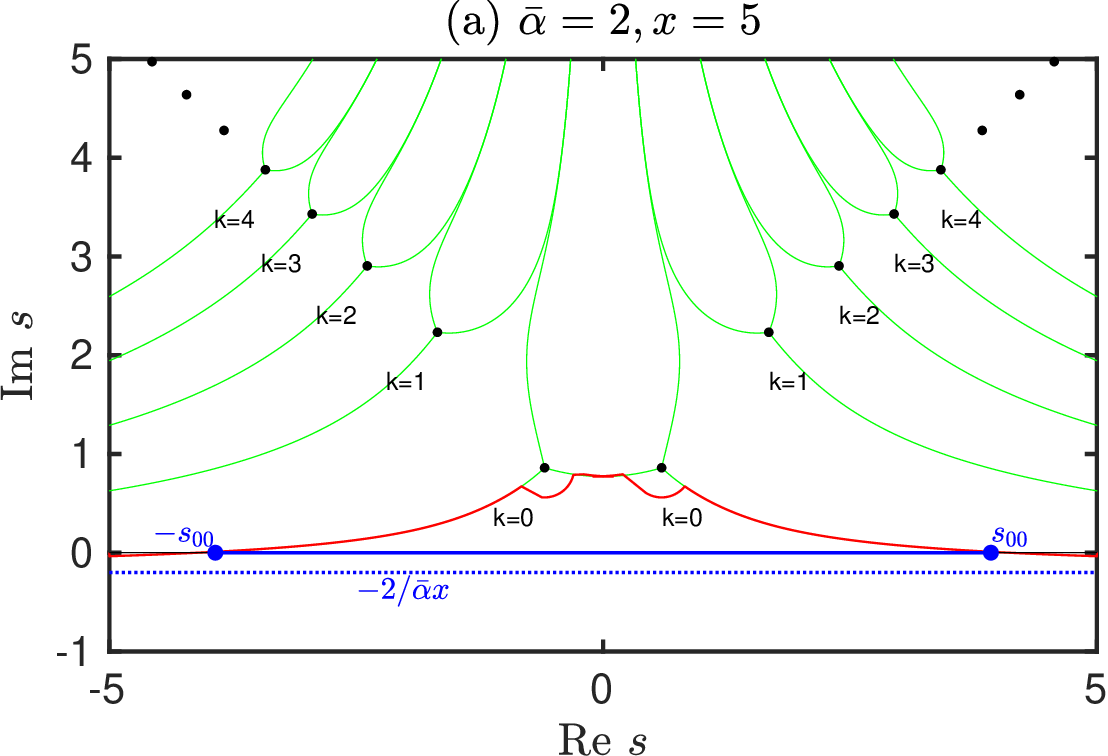}
		\includegraphics[width=3in]{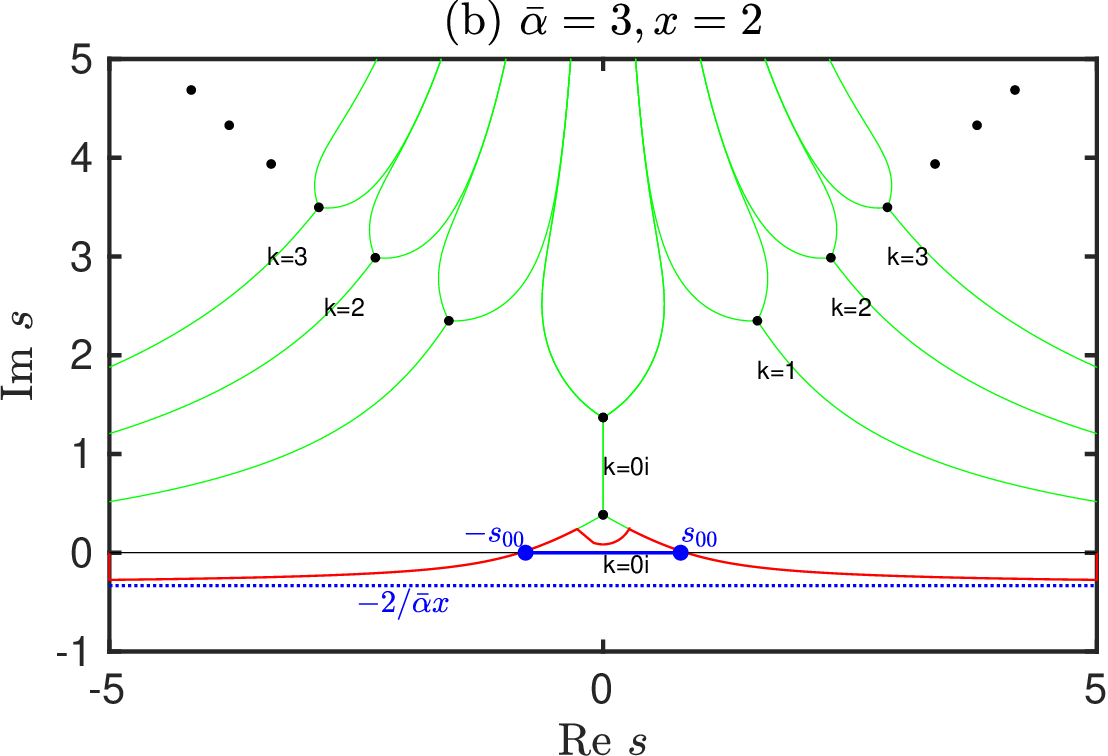}
	\end{center}
	\caption{
	Complex contours originally proposed by Davis and Pechukas as they are applied for the non-Hermitian case.
	It is shown that only the TP(s) that are nearest to the real axis are touched by the contour,
	therefore other TPs that lie further from the real axis are irrelevant to the result obtained
	using this integration.
	Further it is shown that the complex contour crosses the real axis at the points
	$s=\pm s_{00}$ as, in the asymptotic limits $s\to\pm \infty$,
	this contour clings to the line defined by $s=-2/\bar \alpha x$ which is colinear
	with the real axis.
	This implies that the integration along this contour is not equal to the
	integral along the real axis. Yet, this integration may represent a good approximation
	if the integration along the real axis can be reduced to the finite interval
	designated by $-s_{00}\le s \le s_{00}$.
	}
\mylabel{FigContDDP}
\end{figure}

\subsection{\myslabel{CONT}New integration contours for the two basic layouts of TPs}

\noindent
In Fig.~\ref{FigCont} we show a complex contour proposition for two possible different
layouts of the central two TPs discussed in Section~\ref{coal}.
To make a distinction between the two possible layouts
we will refer to them as the {\it even layout} -- where
the central TPs $s_k$ and $\bar s_k$ lie apart from the imaginary
axis (Fig.~\ref{FigCont}a), and the {\it odd layout} -- where the central TPs $s_{0i,low}$
and $s_{0i,upp}$ lie on the imaginary axis (Fig.~\ref{FigCont}b).
Two different integration contours are defined for these
layouts.

The proposed contours involve many TPs. The last included TPs $\bar s_k$
and $s_k$, $k\to\infty$, are represented by $k=4$
in  Figs.~\ref{FigCont} to give a simple illustration.
The integrals along such contours consists of the following
different types of contributions:
\begin{enumerate}
	\item integration along circles fully encircling the TPs $s_k$ ({\it contributions of residua}). These involve  all TPs except the last
	included ones ($k=4$) in
	the case the even layout, Fig.~\ref{FigCont}a.
	In the case of the odd layout, another exception is represented by
	the TP $s_{0i,upp}$, Fig.~\ref{FigCont}b.
	Partial residual contributions are
	calculated for these exceptions.

	\item integration along the two sides of the same 
	branchcuts ({\it branchcut contributions}).
	The branchcut contributions involve both incomming and outgoing 
	integration contours.
	Such contributions
	are relevant for all TPs except for the last included TPs
	($k=4$).
	 The branchcuts are defined by particular equivalue lines 
	which are identified by both the $k$ index
	of the TP and the $n$ index of the equivalue line, see Fig.~\ref{F3}. Typically, $n=0$ for the branchcuts in the positive real half-plane and $n=3$ in the negative real half-plane, compare
	Figs.~\ref{FigContDDP} and \ref{F3}.
	In the case of the odd layout, the branchcut for the TPs lying on the
	imaginary time axis, $s_{0i,low}$ and $s_{0i,upp}$, is
	defined as the equivalue line connecting between them.
	This line is characterized by $n=0$ for the lower TP, while
	$n=2$ for the upper TP.

	\item integration along the closing contours connecting the real and imaginary axes. Evaluation of these contributions is very similar to the two sided branchcut integration. The main difference is represented by the different value of $n$ of the lines connecting to the real axis which is given by $n=2$.
	
	\item integration along the equivalue lines $n=1$ and $n=0$ emanating from $s_{0i,upp}$ which are not branchcuts (in the case of the odd layout). 
	Again, the integration is performed in a very similar way to the integration along the two-sided branchcuts. 
	
	\item integration along a connecting contour between two branchcuts in the asymptote $\im s \to \infty$. Note that the branchcuts end up asymptotically in the imaginary axis, $s\to i\infty$,
	as we show in Section~\ref{EQLAS}, therefore 
	the complex integration contour avoiding these branchcuts passes through the asymptote,
	see Fig.~\ref{FigCont}.
	 
\end{enumerate}

\begin{figure}[h!]
	\begin{center}
		\includegraphics[width=3in]{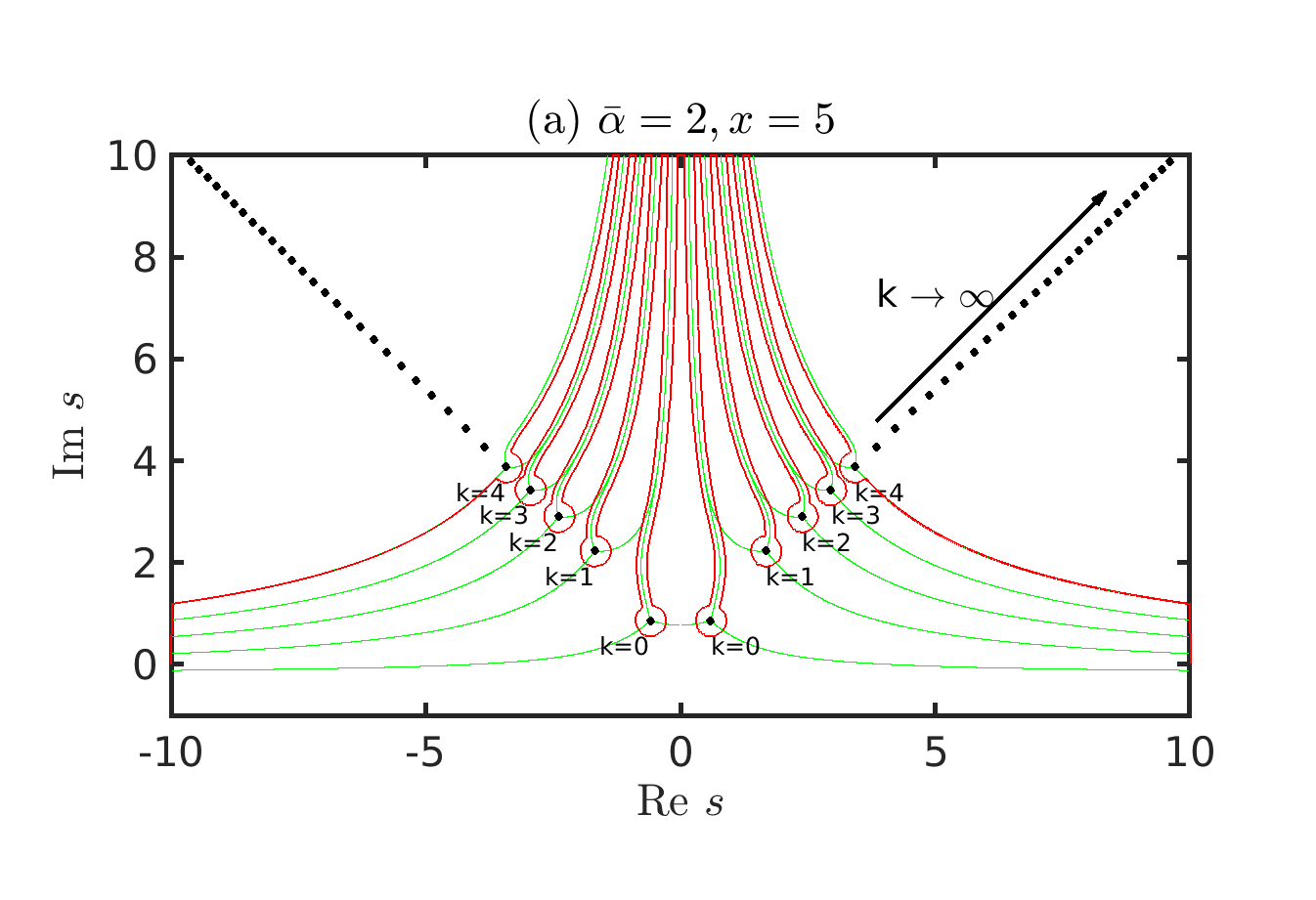}
		\includegraphics[width=3in]{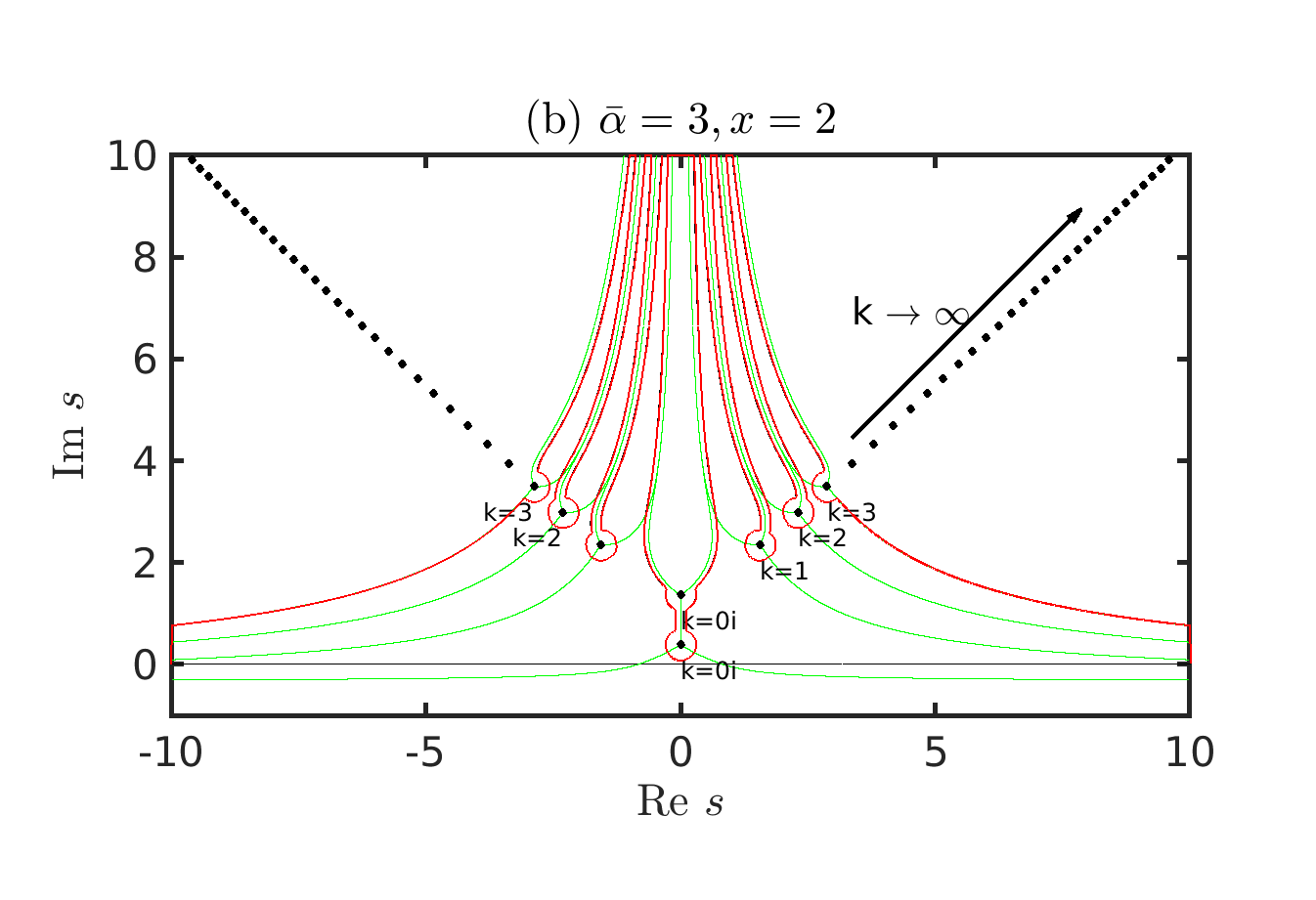}
	\end{center}
	\caption{New proposed complex integration contour for dissipative systems for
	(a) even layout, (b) odd layout. 
	The contours start on the real axis at $-s_{k0}$ and end
	at $+s_{k0}$, where $k=4$. These contours encompass a finite integration interval on the
	real axis $-s_{k0}\le s\le s_{k0}$, 
	where this interval can be increased by taking the last contributing $k$ to infinity,
	$k \to \infty$.
	The region encompassed by the contour 
	excludes all discontinuities on the integrated
	function, which are represented by the TPs (black circles)
	and the branchcuts which are intentionally chosen along the equivalue lines (green color).
	}
\mylabel{FigCont}
\end{figure}

The contributions of the connecting contours in the asymptotic
limit ($\im s \to \infty$) is new with the new type of 
integration contours.
We prove that the corresponding contributions to the integral
is zero in the Appendix~\ref{connect}.

\section{Analytical expressions for branchcut and residual contributions}

\subsection{\myslabel{BC}Branchcut contributions}

\noindent
The BC contributions $b_k$ are associated with integrals over the lines, which emanate
from the TPs:
\begin{align}
	b_k^{(j)} = \int\limits_{s_k}^{{\rm along\  path}\  k(j)} e^{-\frac{\tau\Gamma}{2\hbar}[\gamma(s)-i\phi(s)]} \, \bar N(s)\, ds.
	\mylabel{EQ155}
\end{align}
Analytical expressions for
these contributions can be obtained for the large pulse area limit.

\subsubsection{\label{BC-SRL}Large pulse area limit}
\noindent
Let us express the integrated function near the TP ($s=s_k$) using the first order expansions for
the quasi-energy split $\tilde \delta(s)$ as given in Eq.~\ref{Puiseux1}
where $\beta_{k}^{(1)}$ is given by Eq.~\ref{EQ106};
$\bar N(s)$ is defined in Eq.~\ref{EQNApoles}.
%\subsubsection{\label{BC-INT}Integration along a straigh line emanating from EP}
%\noindent
In accord with this {\it short range limit}, 
we define a straight line emanating from the TP as
\begin{align}
	s = s_k + r\cdot e^{i \varphi_{kn}} ,
	\mylabel{EQ11d18}
\end{align}
where $\varphi_{kn}$ represents the angle of the integration contour
defined by one of the three equivalue lines as
given by Eq.~\ref{EQ157}
\begin{align}
	\varphi_{kn} = \frac{2\pi n}{3} - \frac{2}{3}\arg \beta_k^{(1)},
	\revisited{EQ157}
\end{align}
where $n\in\{0,1,2\}$ determines particularly which one of the three equivalue lines is used.

Let us express the integrated function in the short-range limit
using the contour parameter $r$. The non-adiabatic coupling element near the
pole is given by Eq.~\ref{EQNApoles}. By substituting from Eq.~\ref{EQ11d18} we get,
\begin{align}
	\bar N(s) = \frac{z_k\,e^{-i\varphi_{kn}}}{4ir}.
	\mylabel{EQ13d4}
\end{align}

The exponent is given by (see definition of $\gamma(s)$ and $\phi(s)$ in Eq.~\ref{gamma})
\begin{align}
	&\gamma(s) -i\phi(s) = \gamma(s_k) -i\phi(s_k) -i x \int\limits_{s_k}^s \tilde\delta(s)\,ds .
\end{align}
The energy split near the pole is given by the Puiseux series, Eq.~\ref{Puiseux1},
\begin{align}
\tilde \delta(s) = \beta_{k}^{(1)}\cdot(s-s_k)^{1/2}\, .
\revisited{Puiseux1}
\end{align}
By a substitution from Eq.~\ref{EQ11d18} we get,
\begin{align}
	\tilde \delta(s) = \beta_{k}^{(1)}\cdot r^{1/2} e^{i\varphi_{kn}/2},
	\mylabel{EQdeltaR}
\end{align}
where $\beta_k^{(1)}$ is defined in Eq.~\ref{EQ106}
From here
\begin{align}
	& \gamma(s)-i\phi(s) = \cr
	&\gamma(s_k) -i\phi(s_k) - i x \beta_k^{(1)}\cdot e^{3i\varphi_{kn}/2}\int\limits_{0}^r (r')^{1/2}\,dr' =\cr
	& \gamma(s_k) -i\phi(s_k) - \frac{2ix}{3}\beta_k^{(1)} \cdot e^{3i\varphi_{kn}/2} r^{3/2} .
	\mylabel{EQ161}
\end{align}
Now if we substitute for $\varphi_{kn}$ from Eq.~\ref{EQ157} we get
\begin{align}
	&\gamma(s) -i\phi(s) = \gamma(s_k) -i\phi(s_k) - e^{in\pi} \frac{2ix}{3}|\beta_{k}^{(1)}| r^{3/2} ,
\end{align}
which can be further simplified such that
\begin{align}
	&\gamma(s) -i\phi(s) = \gamma(s_k) -i\phi(s_k) - (-)^n \frac{2ix}{3}|\beta_{k}^{(1)}| r^{3/2} .
	\mylabel{EQ162}
\end{align}
Note that according to our Eq.~\ref{EQ162},
where we explored the short range limit,
the function $\gamma(s)$ which is defined by the real
part of the right hand side, remains constant along the integration contour,
$\gamma(s)=\gamma(s_k)$. This obtained result is correct as it is in accord with the
definition of the equivalue lines.

Let us evaluate integral Eq.~\ref{EQ155} along the straight line, which is correct very near the TP: First we
define
\begin{align}
	b_k^{(n)}(s') = \int\limits_{s_k}^{s'({\rm in\ direction\ }\varphi_{kn})} e^{-\frac{\tau\Gamma}{2\hbar}[\gamma(s)-i\phi(s)]} \, \bar N(s)\, ds,
\end{align}
and then
\begin{align}
	&\lim\limits_{s'\to s_k} b_k^{(n)}(s') = \cr
&\frac{z_k }{4i}\,
	e^{-\frac{\tau\Gamma}{2\hbar}[\gamma(s_k)-i\phi(s_k)]}
	\int\limits_{0}^{|s'|} 
	\frac{e^{(-)^n\frac{i|\beta_{k}^{(1)}|x\tau\Gamma}{3\hbar}r^{3/2}}}{r}  dr, \cr
\end{align}
note that the phase $\exp(-i\varphi_{kn})$ 
in the definition of $\bar N(s)$ (Eq.~\ref{EQ13d4})
is canceled due to the change 
of integration variables ($ds = dr \cdot \exp(i\varphi_{kn})$).
This integral is simplified by using the substitution $\rho=r^{3/2}$, which leads to
\begin{align}
	&\lim\limits_{s'\to s_k} b_k^{(n)}(s') = \cr
	&\frac{z_k }{6i}
	e^{-\frac{\tau\Gamma}{2\hbar}[\gamma(s_k)-i\phi(s_k)]}
	\int\limits_{0}^{|s'|^{3/2}} 
	\frac{e^{(-)^n\frac{i|\beta_{k}^{(1)}|x\tau\Gamma}{3\hbar}\rho}}{\rho}  d\rho.
\end{align}
This is simplified as
\begin{align}
	&
	\lim\limits_{s'\to s_k} b_k^{(n)}(s') = \frac{z_k }{6i}
	e^{-\frac{\tau\Gamma}{2\hbar}[\gamma(s_k)-i\phi(s_k)]}
	\cr&\times
	\int\limits_{0}^{|s'|^{3/2}} 
	\frac{
	\cos{\(\frac{|\beta_{k}^{(1)}|x\tau\Gamma}{3\hbar}\rho\)}
	+ (-)^n\,i
	\sin{\(\frac{|\beta_{k}^{(1)}|x\tau\Gamma}{3\hbar}\rho\)}
	}{\rho}  d\rho \cr
	&
	= \frac{z_k }{6}
	e^{-\frac{\tau\Gamma}{2\hbar}[\gamma(s_k)-i\phi(s_k)]}
	\bigg[-i\,
	\int\limits_{0}^{|s'|^{3/2}} 
	\frac{
	\cos{\(\frac{|\beta_{k}^{(1)}|x\tau\Gamma}{3\hbar}\rho\)}
	}\rho\,d\rho
	\cr & 
	+ (-)^n\,
	\Si{\(\frac{|\beta_{k}^{(1)}|x\tau\Gamma}{3\hbar}\,|s'|^{3/2}\)}
		\bigg] .
	\mylabel{EQ165}
\end{align}

Even though Eq.~\ref{EQ165} represents merely a
short range integration, it gives us a sufficient information 
assuming that the factor,
\begin{align}
	\frac{|\beta_{k}^{(1)}|x\tau\Gamma}{3\hbar} \to \infty,
	\mylabel{EQ11d30}
\end{align}
is very large. In such a case, the sine integral function aquires its limiting
values at $+\infty$ for $s'$ still very close to the TP $s' \to s_k$.
At the same time, the integration over the cosine function is diverging.
Eq.~\ref{EQ11d30} which can be also written as
\begin{align}
	\theta \sqrt{\frac{2}{\pi}}\, \frac{|\beta_{k}^{(1)}|}{3} \to \infty,
\mylabel{EQthinf}
\end{align}
defines the {\it large pulse area limit}, see
Appendix~\ref{THINF} for some concrete equations
which apply for the system under the present study.

\subsubsection{Integration along two sides of the same branchcut}
\noindent
The branchcut contribution associated with the TP is 
defined by the incomming and outgoing contours, which
happen to be on the different sides of the branchcut.
Eq.~\ref{EQ165} defines the contribution of the path {\it out} of the TP.
Then the contribution of the path {\it towards} the TP has the opposite sign 
and also $n$ must be taken as $n\to n+3$, which changes the sign second time
but now only for the contribution of the sine function in Eq.~\ref{EQ165},
while the diverging contributions cancel out.
The contribution from the {\it two sides} of the branchcut is thus given by
\begin{align}
	&b_{kn} = b_k^{(n)} - b_k^{(n-3)} =
	\cr
	& \frac{z_k\, (-)^n}{3}
	e^{-\frac{\tau\Gamma}{2\hbar}[\gamma(s_k)-i\phi(s_k)]}
	\,
	\Si{\(\frac{|\beta_{k}^{(1)}|x\tau\Gamma}{3\hbar}\,|s'|^{3/2}\)}
		 .
		 \cr
		 \mylabel{EQBC0}
\end{align}
Now, as the divergent part has been taken out, 
the large pulse area limit defined by Eq.~\ref{EQ11d30} or
equivalently Eq.~\ref{EQthinf} can be readily applied
by substituting for the sine integral its limiting value
given by $\pi/2$ such that
\begin{align}
	&b_{kn} = 
	\frac{z_k\, (-)^n\, \pi}{6} \, 
	e^{-\frac{\tau\Gamma}{2\hbar}[\gamma(s_k)-i\phi(s_k)]} \ .
	\mylabel{EQBC}
\end{align}
Eq.~\ref{EQBC} represents the branchcut contribution assuming
that the branchcut is defined by one of the three equivalue lines. 
Eq.~\ref{EQBC} shows that the branchcut contribution is
proportional to the residuum. The sign given by $n$ is altered
based on the angle of the equivalue line associated with the
branchcut, see Eq.~\ref{EQ157}.

\subsubsection{Closing contours connection the complex contour to the real axis}
\noindent
The complex contour includes equivalue lines starting at each side of the real axis,
passing around the last included TP, and continuing along the inner side of the branchcut
to a proximity of imaginary axis of time. The contribution of 
this part of the contour, excluding the semi-circular part near the TP, 
will be evaluated now. 

The contribution of the integration path {\it out} of the end TP is given by 
$b_k^{(n)}$, while the integration path {\it to} the TP is given by the same equation with the opposite sign. Unlike in the other TPs, now
the path winds around the TP not by a full circle but only by
$2/3$ of the circle, namely $n\to n-2$, such that the sum
of the {\it in} and {\it out } contributions is defined as
\begin{align}
b_{k-end} = b_k^{(n)} - b_k^{(n-2)} .
\end{align}
Using the definition Eq.~\ref{EQ165} we find out that the
two contributions cancel out, therefore the {\it integration
paths associated with the end TPs  do not contribute},
\begin{align}
b_{k-end} = 0 .
\end{align}
Again, by using Eq.~\ref{EQ165} we effectively assumed only the
close neighborhood of the TP, which is correct in the
large pulse area limit.

\subsection{Contributions of residua}

\noindent
A contribution associated with encircling a pole is given by 
\begin{align}
	r_k = -\oint\limits_{\varphi_{kn} \to \varphi_{kn'}} e^{-\frac{\tau\Gamma}{2\hbar}[\gamma(s)-i\phi(s)]}
	\bar N(s)\, ds ,
	\mylabel{EQ166b}
\end{align}
where $n$ and $n'$ determine the starting and final integration points, as defined by the
equivalue lines. The $(-)$ sign is included as it is assumed that the path of $s$ has the
clockwise direction, whereas the true integration path has the anti-clockwise direction.
The angles $\varphi_{kn}$ are given by Eq.~\ref{EQ157}. 
To evaluate this integral, one can use all equations applying for the short-range limit
as introduced in Section~\ref{BC-SRL}.

\subsubsection{Integrand along closely encircling contour}
\noindent
Following the logic of Section~\ref{BC}, we define a position $s$ near the pole such that,
\begin{align}
	s = s_k + r \cdot e^{i(\varphi_{kn}+\varphi)}, \quad\quad r>0,
	\mylabel{EQ13d33}
\end{align}
where $\balpha_{1k}$ is given in Eq.~\ref{EQ106}.
We assume that
the partial integration contour in Eq.~\ref{EQ166b} encircles the pole at the distance
of $r=R$, where $R\to 0$.
The non-adiabatic coupling element is given by Eq.~\ref{EQ13d4} such that
\begin{align}
	\bar N(s) = \frac{z_k e^{-i(\varphi_{kn}+\varphi)}}{4iR},
	\mylabel{EQ168}
\end{align}
and the exponent in Eq.~\ref{EQ166b} is given by
\begin{align}
	&\gamma(s) - i\phi(s) = \gamma(s_k) - i\phi(s_k) \cr
	&- ix \int\limits_{s_k}^{s_k+R\cdot e^{i\varphi_{kn}}} \bar \delta(s)\, ds 
	\cr
	&
	- ix \oint\limits_{s_k+R\cdot e^{i\varphi_{kn}}}^{s_k+R\cdot e^{i(\varphi_{kn}+\varphi)}} 
	\bar\delta(s)\, ds
	.
	\mylabel{EQ169b}
\end{align}
The energy split near the pole is given by the Puiseux series (Eq.~\ref{Puiseux1}),
where we substituted from Eq.~\ref{EQ13d33} using the distance from the pole $r=R$,
\begin{align}
	\tilde \delta(s) = \beta_k^{(1)}\, R^{1/2}\, e^{i\varphi_{kn}/2}\, e^{i\varphi/2} ,
\end{align}
which is substituted to Eq.~\ref{EQ169b}. The first integral is evaluated following 
the derivation in Eqs.~\ref{EQ161}-\ref{EQ162}:
\begin{align}
	&
	\gamma(s) - i\phi(s) = \gamma(s_k) - i\phi(s_k) 
	-e^{in\pi}\frac{2ix}{3} |\beta_k^{(1)}|\, R^{3/2}
	\cr & \quad
	-ix \oint\limits_{s_k+R\cdot e^{i\varphi_{kn}}}^{s_k+R\cdot e^{i(\varphi_{kn}+\varphi)}} 
	\tilde\delta(s)\, ds
	.
\end{align}
The second integral is simplified such that
\begin{align}
	&\gamma(s) - i\phi(s) = \gamma(s_k) - i\phi(s_k) 
	-e^{in\pi}\frac{2ix}{3} |\beta_k^{(1)}|\, R^{3/2}
	\cr &
	-ix\, R^{1/2}\beta_k^{(1)}
	\int\limits_{0}^{\varphi} 
	e^{i(\varphi'+\varphi_{kn})/2}\, d\varphi'\, (iR\cdot e^{i(\varphi'+\varphi_{kn})})
	,
\end{align}
where the expression after $d\varphi'$ represents the volume element. This simplifies to
\begin{align}
	&\gamma(s) - i\phi(s) = \gamma(s_k) - i\phi(s_k) 
	-e^{in\pi}\frac{2ix}{3} |\beta_k^{(1)}|\, R^{3/2}
	\cr &\quad
	-i^2 x R^{3/2} e^{3i\varphi_{kn}/2} \beta_k^{(1)}
	\int\limits_{0}^{\varphi} 
	e^{3i\varphi'/2}\, d\varphi'
	.
\end{align}
By using Eq.~\ref{EQ157} one gets
\begin{align}
	&\gamma(s) - i\phi(s) = \gamma(s_k) - i\phi(s_k) 
	-e^{in\pi}\frac{2ix}{3} |\beta_k^{(1)}|\, R^{3/2}
	\cr
	&\quad
	-e^{in\pi} i^2 x  |\beta_k^{(1)}|\, R^{3/2}
	\int\limits_{0}^{\varphi} 
	e^{3i\varphi'/2}\, d\varphi'
	.
\end{align}
By solving the integral one gets,
\begin{align}
	&
	\gamma(s) - i\phi(s) = \gamma(s_k) - i\phi(s_k) 
	-e^{in\pi}\frac{2i x}{3} |\beta_k^{(1)}|\, R^{3/2}
	\cr
	&\quad
	-e^{in\pi}\frac{2ix}{3}  |\beta_k^{(1)}|\, R^{3/2}
	(e^{3i\varphi/2} - 1)
	,
\end{align}
which simplifies by term cancellation to:
\begin{align}
	&
	\gamma(s) - i\phi(s) = \gamma(s_k) - i\phi(s_k) 
	\cr &\quad
	-e^{in\pi}\frac{2ix}{3}  |\beta_k^{(1)}|\, R^{3/2}
	e^{3i\varphi/2} 
	.
	\mylabel{EQ176}
\end{align}

\subsubsection{Integration}
\noindent
Integral Eq.~\ref{EQ166b} is simplified in the limit $R\to 0$ using Eqs.~\ref{EQ168}
and \ref{EQ176} such that
\begin{align}
	&
	r_k = -\frac{z_k}{4iR}
	e^{-\frac{\tau\Gamma}{2\hbar}[\gamma(s_k)-i\phi(s_k)]}
	\ 
	(iR\cdot e^{i(\varphi+\varphi_{kn})})
	\cr
	&\times
	\int\limits_{\varphi_{kn}}^{\varphi_{kn'}} 
	e^{e^{in\pi}\frac{ix\tau\Gamma}{3\hbar}|\beta_k^{(1)}| R^{3/2} e^{3i\varphi/2}}
	e^{-i(\varphi+\varphi_{kn})}\,
	d\varphi\, ,
\end{align}
where the expression after $d\varphi$ is the volume element.
The first exponential in the integral goes to one as $R\to 0$, therefore
it can be removed. By simplifying we get 
\begin{align}
	r_k = -\frac{z_k}{4}
	e^{-\frac{\tau\Gamma}{2\hbar}[\gamma(s_k)-i\phi(s_k)]}
	\int\limits_{\varphi_{kn}}^{\varphi_{kn'}} 
	d\varphi .
\end{align}
If we substitute for the limits using Eq.~\ref{EQ157} we get
\begin{align}
	r_k = -z_k\,\frac{\pi(n'-n)}{6}
	e^{-\frac{\tau\Gamma}{2\hbar}[\gamma(s_k)-i\phi(s_k)]} .
\end{align}
The expression is proportional to the integer difference $(n'-n)$,
which determines how many thirds $\Delta_n$ of a circle are used by the complex
contour, such that $\Delta_n = n'-n$ and clearly, $\Delta_n \in \{1, 2, 3\}$.
\begin{align}
	r_k = -z_k\,\frac{\pi\Delta_n}{6}
	e^{-\frac{\tau\Gamma}{2\hbar}[\gamma(s_k)-i\phi(s_k)]} .
	\mylabel{EQ12d15}
\end{align}

\section{Overall contributions of the central vs. asymptotic transition points}

\subsection{\myslabel{INTSYM}Contributions of time-symmetric pairs of TPs}

\noindent
As we solve here the time-symmetric problem, also the complex
integration contour is time symmetric with respect to the
sign change of the real part of time $s$.
The contributions of both residua and branchcuts may be
collected according to the pairs of the time-symmetric TPs.

Let us examine the symmetric relation between the branchcut
contributions given by $b_k^{(n)}$ associated with $s_k$
\begin{align}
&b_{kn} = 
\frac{z_k\, (-)^n\, \pi}{6} \, 
e^{-\frac{\tau\Gamma}{2\hbar}[\gamma(s_k)-i\phi(s_k)]} \ ,
\revisited{EQBC}
\end{align}
and $\bar b_{kn}$ associated with $\bar s_k$,
where $\bar s_k$ has been defined in Eq.~\ref{sym}.
$\bar b_{kn}$ is obtained by the substitution $s_k \to \bar s_k$ into Eq.~\ref{EQBC}:
\begin{align}
&
\bar b_{kn} = 
\frac{\bar z_k (-)^{\bar n} \pi}{6}
e^{-\frac{\tau\Gamma}{2\hbar}[\gamma(\bar s_k)-i\phi(\bar s_k)]}
\ .
\end{align}
$\gamma(\bar s_k)$ and $\phi(\bar s_k)$ are directly related
to $\gamma(s_k)$ and $\phi(s_k)$
based on the symmetric relations given in Eq.~\ref{sym21} such that
\begin{align}
&\gamma(s_k) = \gamma(\bar s_k), 
&\phi(s_k) = - \phi(\bar s_k) .
\mylabel{sym22}
\end{align}
The sign $\bar z_k$ is defined in Eq.~\ref{EQNApoles},
show that it is same for the two symmetric TPs.
This is also confirmed by substituting $\bar s_k$ into Eq.~\ref{aEQ91}
defining $z_k$ in the Appendix~\ref{APNA},  
\begin{align}
\bar z_k = \( -\frac{1}{x} + \frac{\balpha}{2}\, i\bar s_k\)\, e^{\bar s_k^2/2} = z_k^*=z_k .
\mylabel{sym3}
\end{align}
As for $n$ and $\bar n$, which define the branchcut
equivalue lines, they have been defined above 
(Eq.~\ref{EQnsym})
such that
\begin{align}
& n=0,
& \bar n = 1,
\end{align}
however it is also permitted that either $n$ or $\bar n$ is 
augmented by adding one full cycle such that $n\to n+ 3$ or $\bar n \to \bar n + 3$,
which of course would change the sign of one or both contributions $b_{kn}$
and $\bar b_{kn}$, respectively.
Due to symmetry reasons we assume that $n=0$ for $\re\, s_k>0$, whereas
\begin{align}
	\bar n = 1+3 = 4 .
\end{align}
By putting this together we get,
\begin{align}
&b_{kn} = \bar b_{kn}^* =
\frac{z_k\, \pi}{6} \, 
e^{-\frac{\tau\Gamma}{2\hbar}[\gamma(s_k)-i\phi(s_k)]} \ , \cr
& {\rm for}\ \re s_k > 0 .
\end{align}
The two contributions sum up to:
\begin{align}
b_{kn}^{pair} = b_{kn} + \bar b_{kn}, 
\end{align}
such that
\begin{align}
&b_{kn}^{pair} = \frac{z_k\, \pi}{3} \, 
e^{-\frac{\tau\Gamma}{2\hbar}\,\gamma(s_k)} \, 
\cos{\[\frac{\tau\Gamma}{2\hbar}\,\phi(s_k)\]} \ , 
\cr
& {\rm for}\ \re s_k > 0 .
	\mylabel{symBC}
\end{align}

Let us define the symmetrized residual contributions from the pair of
poles at $s_k$ and $\bar s_k$ (Eq.~\ref{sym}) such that
\begin{align}
	r_k^{pair} = r_k + \bar r_k,
\end{align}
where $\bar r_k$ corresponds to the pole at $\bar s_k$ and is formally
defined as $r_k$ (see Eq.~\ref{EQ12d15}):
\begin{align}
	\bar r_k = -\bar z_k\,\frac{\pi \bar\Delta_{n}}{6}
	e^{-\frac{\tau\Gamma}{2\hbar}[\gamma(\bar s_k)-i\phi(\bar s_k)]} .
\end{align}
Clearly, $\bar \Delta_n$ is equal to $\Delta_n$ due to the symmetry. 
In the examined pairs the full circle is encompased by the
integration contour such that,
\begin{align}
\Delta_n = \bar \Delta_n = 3.
\end{align}
The other
relations for the symmetric counterparts such as $\gamma(\bar s_k)$, $\phi(\bar s_k)$, and
$\bar z_k$, are given in Eqs.~\ref{sym22} and \ref{sym3}. By sustitution from these
equations we get
\begin{align}
	\bar r_k = r_k^*
	\mylabel{symRS}
\end{align}
From here we get:
\begin{align}
	r_k^{pair} = 2\,\re r_k .
	\mylabel{symRS2}
\end{align}
Using Eq.~\ref{EQ12d15} which represents the formula for 
the residual contributions $r_k$ we obtain the
contribution of the symmetric pair of resisua such that
\begin{align}
r_k^{pair} = -z_k\,\pi \,
e^{-\frac{\tau\Gamma}{2\hbar}\,\gamma(s_k)} \ 
\cos \[ \frac{\tau\Gamma}{2\hbar}\,\phi(s_k) \]
.
\mylabel{symRS3}
\end{align}

We may now evaluate the sum of the residual and branchcut
contributions for each symmetric pair. 
Clearly, the presence of the branchcut reduces the residuum 
contribution by $1/3$.
Let us denote the overall contributions of the symmetric
pairs of the TPs as $v_k^{pair}$. It is defined as
\begin{align}
&v_k^{pair} = - \frac{z_k\, 2\pi}{3} \, 
e^{-\frac{\tau\Gamma}{2\hbar}\,\gamma(s_k)} \, 
\cos{\[\frac{\tau\Gamma}{2\hbar}\,\phi(s_k)\]} \ , 
\cr
& {\rm for}\ \re s_k > 0 .
\mylabel{symv}
\end{align}

\subsection{\myslabel{INTIM}Contributions of the TPs on the imaginary time axis}

\noindent
While for the even layout, the integration includes predominantly
the symmetric pair contributions, in the case of the odd layout,
the contributions of the TPs on the imaginary time axis
are the most important ones. Note that the magnitude of the TPs
contributions will be discussed in detail below.

Let us start with the lower TP, $s_{0i,low}$.
The branchcut is given by the equivalue line
in the upward direction which is characterized by $n=0$,
see Fig.~\ref{F3}b.
This value of $n$ is substituted into the definition of
the branchcut contribution, Eq.~\ref{EQBC}.
Likewise, we may substitute for the known value of $\phi(s_{0i,low})$,
which is equal to zero, see Fig.~\ref{Figgamphi}b.
Finally we substitute for the sign $z_{0i,low}$,
which is positive according to Eq.~\ref{EQNApoles}.
By doing this we obtain the definition of the
branchcut contribution of the lower imaginary TP given by,
\begin{align}
&b_{0i,low} = 
\frac{ \pi}{6} \, 
e^{-\frac{\tau\Gamma}{2\hbar}\,\gamma(s_{0i,low})} \ .
\mylabel{BC0ilow}
\end{align}
The residual contribution of the lower TP, $s_{0i,low}$,
is given by the full circle, therefore $\Delta_n$ in 
the definition Eq.~\ref{EQ12d15} is given by $\Delta_n = 3$.
By doing the same substitutions used above we obtain
\begin{align}
r_{0i,low} = -\frac{\pi}{2}
e^{-\frac{\tau\Gamma}{2\hbar}\,\gamma(s_{0i,low})} .
\mylabel{RS0ilow}
\end{align}
The overall TP contribution of $s_{0i,low}$ given by
the sum of the residuum and branchcut contribution is
defined as
\begin{align}
v_{0i,low} = -\frac{\pi}{3}
e^{-\frac{\tau\Gamma}{2\hbar}\,\gamma(s_{0i,low})} .
\mylabel{EQ0ilow}
\end{align}

Now, we will evaluate the contribution of the upper
TP $s_{0i,upp}$ to the integration.
The branchcut for $s_{0i,upp}$ is given by the
equivalue line in the downward direction ($n=2$ according
to Fig.~\ref{F3}).
Because of how this branchcut is placed, the value
of $\phi(s_{0i,upp})$ is not uniquely defined,
but it differs for the two sides of the branchcut:
Let us denote these values as
$\phi(s_{0i,upp-})$ and $\phi(s_{0i,upp+})$
according to the respective real-time half plane.
For this reason we also cannot use Eq.~\ref{EQBC}
to evaluate the two side branchcut contribution unlike
in the other cases.
Instead, we will evaluate the sum of the contributions of the
equvalue lines pertaining to each of the half-planes
separately.

Let us start with the left real half plane ($\re s < 0$).
First we calculate the sum of the contributions of the equivalue lines
defined by $n=1$ (for the incomming contour) and $n=2$ 
(for the outgoing contour). The outgoing contribution is
defined directly by Eq.~\ref{EQ165}, while the incomming contour
contributes with the opposite sign, therefore
the sum is given by
\begin{align}
b_{0i,upp-} = b_{0i,upp-}^{(2)} - b_{0i,upp-}^{(1)} \, .
\end{align}
This result corresponds with Eq.~\ref{EQBC0}, which after a
simplification (including the application of the large pulse area limit)
yields a contribution equivalent to the
two side branchcut integral, Eq.~\ref{EQBC}.
Here $z_{0i,upp}=1$ and $n=2$ such that
\begin{align}
b_{0i,upp-} = \frac{\pi}{6} \, 
e^{-\frac{\tau\Gamma}{2\hbar}[\gamma(s_k)-i\phi_{0i,upp-}(s_k)]} \ .
\end{align}
The upper TP is partially encircled by the integration contour
in the left half plane of time,
where $\Delta_n = 1$ in the definition of the partial
residual contribution in Eq.~\ref{EQ12d15}.
The partial residual contribution of the upper pole in the
left half plane of time is given by
\begin{align}
&r_{0i,upp-} = 
- \frac{ \pi}{6} \, 
e^{-\frac{\tau\Gamma}{2\hbar}[\gamma(s_{0i,upp})-i\phi(s_{0i,upp-})]} \ .
\mylabel{RS0iupp}
\end{align}
By putting together the branchcut contribution, the partial residuum,
and the contribution of the incomming contour
following the equivalue line $n=1$, we
find out that the overall contribution of the upper
TP in the left half plane of time is zero,
\begin{align}
v_{0i,upp-} = 0 .
\end{align}

The same is applied for the right half plane of time,
where the incomming contour is associated with $n=2$ (the 
right side of the branchcut), while the outgoing contour
is associated with $n=0$, Fig.~\ref{F3}. 
Note that there is no branchcut crossing when the TP
is semiencircled between the two equivalue lines.
It is thus correct to use $n=3$ instead of $n=0$
for the contribution of the outgoing contour.
The sum of the contributions of the incomming and
outgoing contours is
given by
\begin{align}
b_{0i,upp+} = b_{0i,upp+}^{(3)} - b_{0i,upp+}^{(2)} \, ,
\end{align}
which is simplified to
\begin{align}
b_{0i,upp+} = \frac{\pi}{6} \, 
e^{-\frac{\tau\Gamma}{2\hbar}[\gamma(s_k)-i\phi_{0i,upp+}(s_k)]} \ .
\end{align}
The residuum contributes partially as
\begin{align}
&r_{0i,upp+} = 
- \frac{ \pi}{6} \, 
e^{-\frac{\tau\Gamma}{2\hbar}[\gamma(s_{0i,upp})-i\phi(s_{0i,upp+})]} \ .
\end{align}
Again by summing up all contribution to the integration
in the right half plane of time we obtain that the
upper TP does not contribute to the overall integral,
\begin{align}
v_{0i,upp+} = 0 .
\end{align}

\subsection{Contribution of the TPs in the asymptotic limit}

\noindent
Let us consider the contribution of the pair of TPs
$s_k$, $\bar s_k$ where $k>0$. To some degree of
approximation we may consider those TPs as part of the
asymptotic limit, where the values for the
integrals $\phi(s_k)$ and $\gamma(s_k)$ are given
in Eqs.~\ref{EQ8d7}.
The pair contribution $v_k^{pair}$ defined
in Eq.~\ref{symv} is evaluated based on the asymptotic limit
such that
\begin{align}
&v_k^{pair} \approx - \frac{(-)^k\, 2\pi}{3} \, 
e^{-\frac{\tau\Gamma\gamma_\infty}{2\hbar}}
e^{-\frac{\tau\Gamma}{2\hbar}\,
\frac{\balpha x}{2} \pi\(k + \frac{1}{4}\)
} \, \cr
&
\times\cos{\left\{\frac{\tau\Gamma}{2\hbar}\,
	\[
	\frac{\balpha x}{4} \ln\(k\pi\frac{\balpha^2}{2}\)
	-\phi_\infty
	\]
	\right\}} \ , \cr &
\quad \quad k>0 ,
\mylabel{TPasym}
\end{align}
where we substituted $z_k=(-)^k$ using the definition in Eq.~\ref{EQNApoles}.
The overall contribution of the asymptotic TPs is given by
the sum
\begin{align}
v_\infty = \sum\limits_{k=1}^\infty v_k^{pair}\ ,
\end{align}

In the Hermitian limit, where $x\to\infty$, the contributions
of the asymptotic TPs
are given by zero, which implies that the dynamics is
given only by the positions of the central TPs.
This is in agreement with the theory of Dykhne, Davis, and
Pechukas, who used an integration contour including
only cotributions of the central
TPs in the Hermitian case, see Fig.~\ref{FigContDDP}.

However, as the system is non-Hermitian, the contribution
of the asymptotic TPs is non-zero as we can see.
Eq.~\ref{TPasym} even suggests that the contribution of
the asymptotic TPs is related to the value $2/\balpha x$,
which corresponds to the asymptotic behavior of the
equivalue lines, see Fig.~\ref{FigCont}.
Eq.~\ref{TPasym} suggests that the role of non-Hermitian
contributions thought the asymptotic TPs may be dominant
for small chirps {\it and} small values of $\tau\Gamma$,
\begin{align}
\tau \Gamma \balpha x \to 0 .
\mylabel{DDPcond}
\end{align}
We leave this suggestion an open question.

\subsection{Summary of the complex contour integration}

\noindent
The complex amplitude $v$ defined within the first-order perturbation
approach in Eq.~\ref{v1} has been resolved as a sum of contributions
of the individual TPs.
We have divided the final sum into separate contributions of the
central pair of TPs (either $s_0$, $\bar s_0$ or
$s_{0i,low}$, $s_{0i,upp}$, depending on the layout) on one side, 
and the asymptotic TP pairs ($s_k$, $\bar s_k$, $k>0$) on the other side.

The contribution of the central TPs is defined by Eq.~\ref{symv}
(taking $k=0$ where $z_k=1$) and Eq.~\ref{EQ0ilow} for the
two different layouts. The result can be summarized as,
\begin{align}
&v_0 = - \frac{2\pi}{3} \, 
e^{-\frac{\tau\Gamma}{2\hbar}\,\gamma(s_0)} \times \Phi_0(s_0) \,
\mylabel{EQCENTR0}
\end{align}
where $\Phi_0(s_0)$ represents an interference term, which
is active in the case of even layout, where it depends on
$\phi(s_0)$ such that
\begin{align}
\Phi_{0,even}(s_0) = \cos{\[\frac{\tau\Gamma}{2\hbar}\,\phi(s_0)\]} \ .
\mylabel{EQCENTR1}
\end{align}
Note that $\phi(s_0)$ is indirectly but clearly related to the distance
of the two TPs $s_0$ and $\bar s_0$.
The interference is not present in the case of the odd layout,
where the interference term is simply given by
\begin{align}
\Phi_{0,odd}(s_{0i,low}) = \frac{1}{2} \ .
\mylabel{EQCENTR2}
\end{align}
In the case of the odd layout, 
$s_0$ is formally replaced by $s_{0i,low}$ in Eq.~\ref{EQCENTR0}.
It has been proven that
 the upper central TP $s_{0i,upp}$ does not contribute
to the integral therefore takes no part in the expressions for the amplitude $v$
of the physical process.
This is in harmony with avoiding this TP
when using the integration
contour proposed by Dykhne, Davis, and Davis 
in the Hermitian limit.

We have shown that the contribution of the asymptotic
TPs is non-zero in the non-Hermitian case, therefore the
integration contour by Dykhne, Davis, and Davis is not exact,
see Eq.~\ref{TPasym}.
Yet the contributions of the asymptotic TPs may
be considered negligible in the large pulse area limit
if the product of the chirp and effective intensity
$\balpha x$ is nonzero.

\section{Survival amplitude}

\subsection{Analytical formulas for survival probability}

\noindent
The survival probility $p_1$ 
has been defined in Section~\ref{Ssurv}.
It is obtained from the complex
amplitude $v$ using Eq.~\ref{psurv2} which includes
a normalization factor $f$ defined in Eq.~\ref{EQfdefsym}
and in the text below.
By using the expression for the pulse area 
in the case of Gaussian pulses (Eq.~\ref{EQth})
and the definition of $\gamma(s)$ (Eq.~\ref{gamma}),
we obtain a compact expression for $f$ given by,
\begin{align}
f=\lim\limits_{s\to\infty}
\exp\[ \frac{\theta\[s+\gamma(s)\]}{x}\frac{1}{\sqrt{2\pi}\hbar} \]  .
\mylabel{EQ5d24}
\end{align}

As we have stated in the beginning of Section~\ref{contoura},
we approximate $v$ within the first order of the adiabatic
perturbation expansion, $v$ is thus defined in Eq.~\ref{v1}.

Additionally, we will assume that 
(i) the pulse area is large (Eq.~\ref{EQthinf}),
and (ii)
the non-Hermitian system is
mostly controlled by the central pair of the TPs, i.e.
Eq.~\ref{DDPcond} is {\it not} true.
This allows us to approximate $v$ as $v_0$ given by Eqs.~\ref{EQCENTR0}-\ref{EQCENTR2}.

\subsubsection{Even layout}
\noindent
The survival probability for the even layout is obtained when $v_0$ is defined
by Eqs.~\ref{EQCENTR1} such that
\begin{align}
	&p_{1,even} =  f^2 |v_{0,even}|^2 \cr &=\frac{4 \pi^2}{9} f^2\, e^{-\frac{\tau\Gamma}{\hbar}\gamma(s_0)}
	\,
	\cos^2\[\frac{\tau\Gamma}{2\hbar}\phi(s_0)\] .
\end{align}
Using the 
definition of the pulse area Eq.~\ref{EQthinf} allows us to
remove the system dependent parameter $\Gamma$:
\begin{align}
	&p_{1,even} = \frac{4 \pi^2}{9} f^2\,\exp{\[-\hH\frac{\theta\,\gamma(s_0)}{x}\]} \cr 
	&\times\quad
	\cos^2\[\hh\frac{\theta\,\phi(s_0)}{x}\]
	.
	\mylabel{EQ14d17}
\end{align}
Eq.~\ref{EQ14d17} shows that the behavior of $p_1$  is oscillatory due to the
cosine term. 
The important conclusion following from here is that
the even layout of TPs in the complex plane of adiabatic time
defines an {\it oscillatory subspace in the plane
of the reduced laser parameters $x,\balpha$}.
\begin{figure}
	\begin{center}
	\includegraphics[width=3in
	]{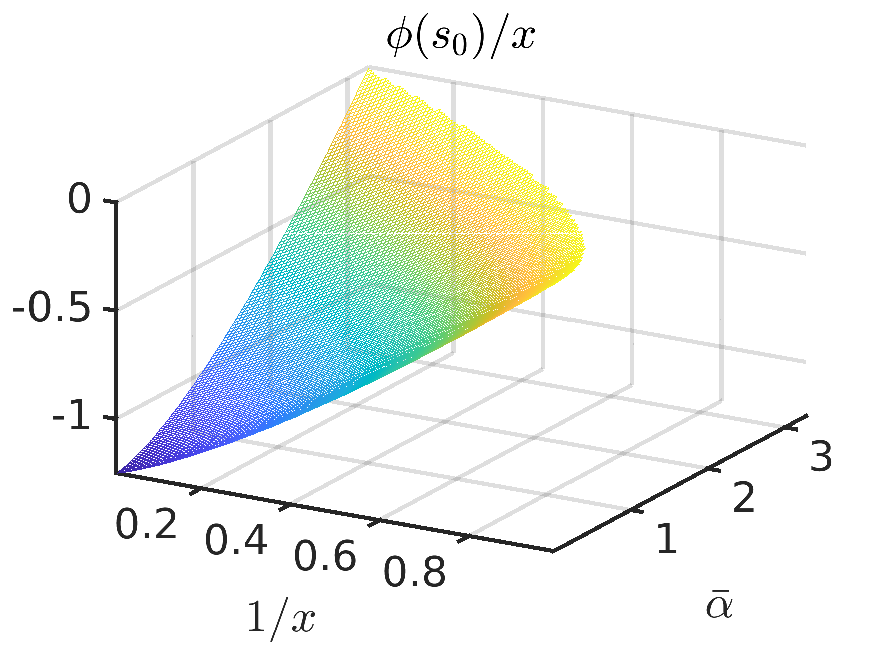}
	\includegraphics[width=3in
	]{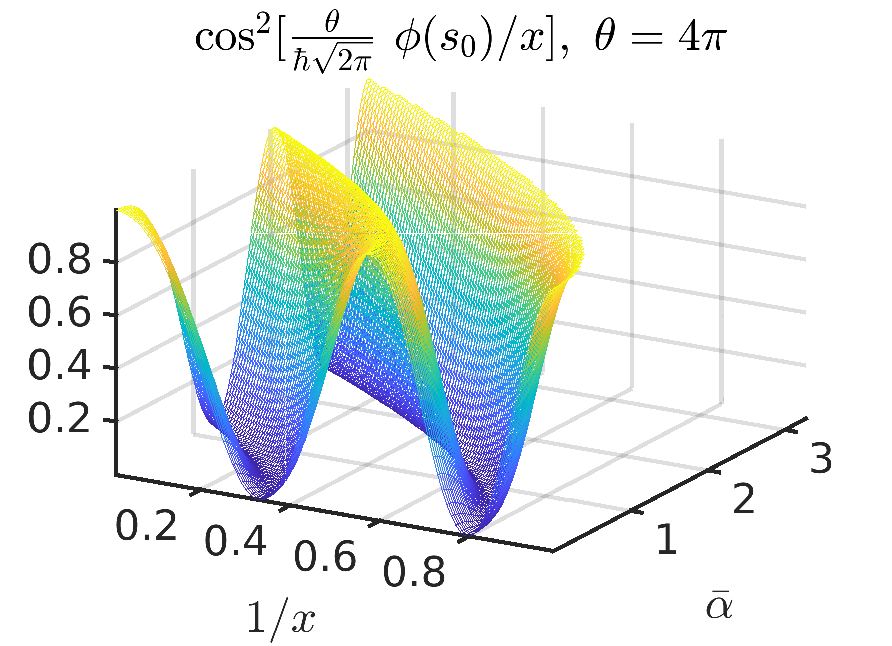}
	\end{center}
\caption{
	This figure illustrates oscillations of the survival probability, Eq.~\ref{EQ14d17}.
	The argument of the oscillatory function is represented by $\phi(s_0)/x$. It is found
	numerically that this function is nearly linear in the plane of $1/x,\balpha$. 
	The pulse area $\theta=4\pi$.
}
\mylabel{Figphix0}
\end{figure}

The exponential modulation of the survival probability
$p_1$ is given by the values of the exponent in Eq.~\ref{EQ14d17},
and additionally, we have to normalize the survival probability by the term $f$, Eq.~\ref{EQ5d24}.
The overall exponential modulation including both factors is given by
\begin{align}
	\bar f=\lim\limits_{s\to\infty}
	\exp\[ 
	\frac{1}{\hbar\sqrt{2\pi}} 
	\frac{\theta\[s+\gamma(s)-\gamma(s_0)\]}{x}
	\]  .
\end{align}
Based on this we define the integral function $\bar\gamma(s_0)$ such that
\begin{align}
	\bar\gamma(s_0) = \lim\limits_{s\to\infty} [s + \gamma(s) - \gamma(s_0)] ,
	\mylabel{EQ14d21}
\end{align}
which controls the exponential decay of the survival probability
in the oscillatory regime.
Such a function is displayed in Fig.~\ref{Figgamx0}.

The final expression for the survival probability in the case of even layout is given by
\begin{align}
	& p_{1,even} = \frac{4 \pi^2}{9} \,\exp{\[-\hH\frac{\theta\,\bar\gamma(s_0)}{x}\]} \cr
	& \times\ \cos^2\[\hh\frac{\theta\,\phi(s_0)}{x}\]
	.
	\mylabel{EQ14d21b}
\end{align}
\begin{figure}
	\begin{center}
	\includegraphics[width=3in
	]{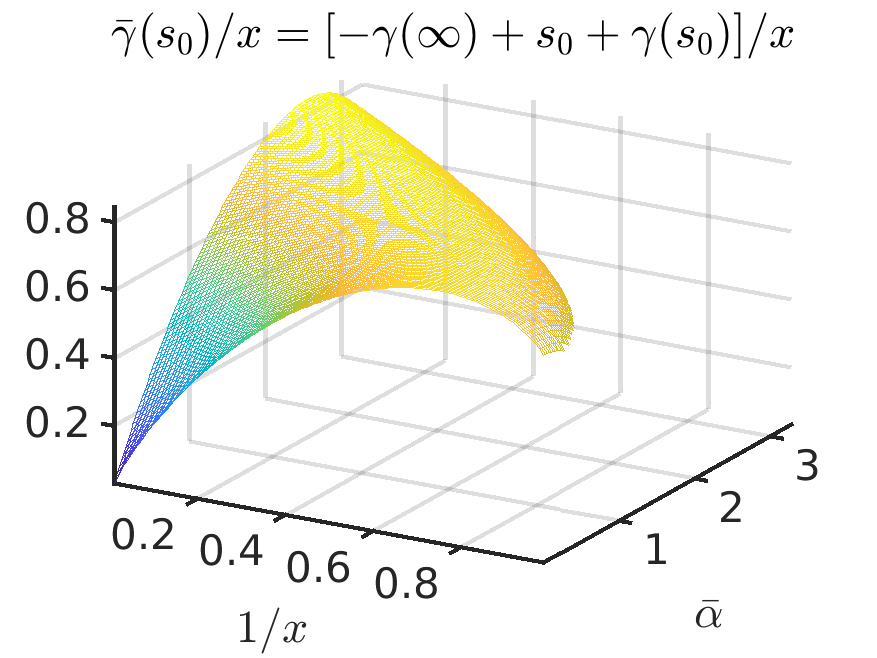}
	\includegraphics[width=3in
	]{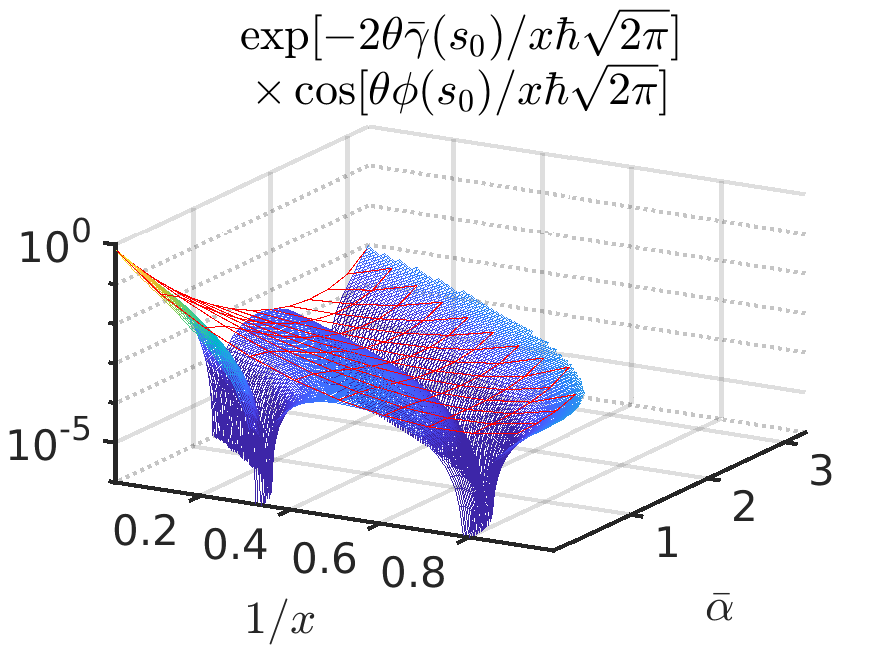}
	\end{center}
	
\caption{
	This figure shows exponential modulation of the survival probability, which is 
	defined by an imaginary part of integrals over the energy split in complex plane,
	$\bar \gamma(s_0)$, see text for details. The first panel shows the exponent,
	while the second panel shows the exponential
	modulation of the basic oscillatory term, which is associated with residua contributions,
	see Fig.~\ref{Figphix0}. The pulse area $\theta=4\pi$.
}
\mylabel{Figgamx0}
\end{figure}

\subsubsection{Odd layout}
\noindent
The survival probability $p_1$ for the odd layout is obtained by analogy
by using $v_0$ defined in Eqs.~\ref{EQCENTR0} 
and \ref{EQCENTR2}
such that
\begin{align}
	p_{1,odd} = f^2 |v_{0,odd}|^2 = f^2 \frac{\pi^2}{9}
	e^{-\frac{\tau \Gamma}{\hbar} \[ \gamma(s_{0i}) \]}.
\end{align}
Using the definition of the pulse area in Eq.~\ref{EQthinf} 
and using the same definition for the normalization
factor $f$ as in the previous case, we get
\begin{align}
	p_{1,odd} = \frac{\pi^2}{9}
	\exp\[-\frac{2\,\theta}{\hbar\sqrt{2\pi}} \frac{\bar\gamma(s_{0i}) }{x}\] .
	\mylabel{EQ14d44}
\end{align}
We can see that in the case of odd layout, the
oscillatory behavior is not present. The
exponential decay is controlled by the
effective integral  $\bar\gamma(s_{0i})$ which
is formally defined by the same equation as before,
Eq.~\ref{EQ14d21}.
The important conclusion following from here is that
the odd layout of TPs in the complex plane of adiabatic time
defines a {\it monotonic subspace in the plane
	of the reduced laser parameters $x,\balpha$}.

\subsubsection{Comparison with a numerical calculation}
\noindent
We demonstrate results obtained using the analytical formulas
Eqs.~\ref{EQ14d21b} and \ref{EQ14d44}
 in comparison with
a numerically implemented first order perturbation method 
based on the integration along the real time axis
as given in Eq.~\ref{v1}.
The comparison is presented
in Fig.~\ref{FigureOsc1}.
The agreement is remarkable even for the relatively small pulse areas $\theta\in\{4\pi,6\pi\}$,
showing some improvement as the pulse area gets larger. 
This favorable result shows that the complex time plane
method can be recomended as an analytical tool for
different problems concerning the EP encircling dynamics.
\begin{figure}
	\begin{center}
	\includegraphics[width=3in
	]{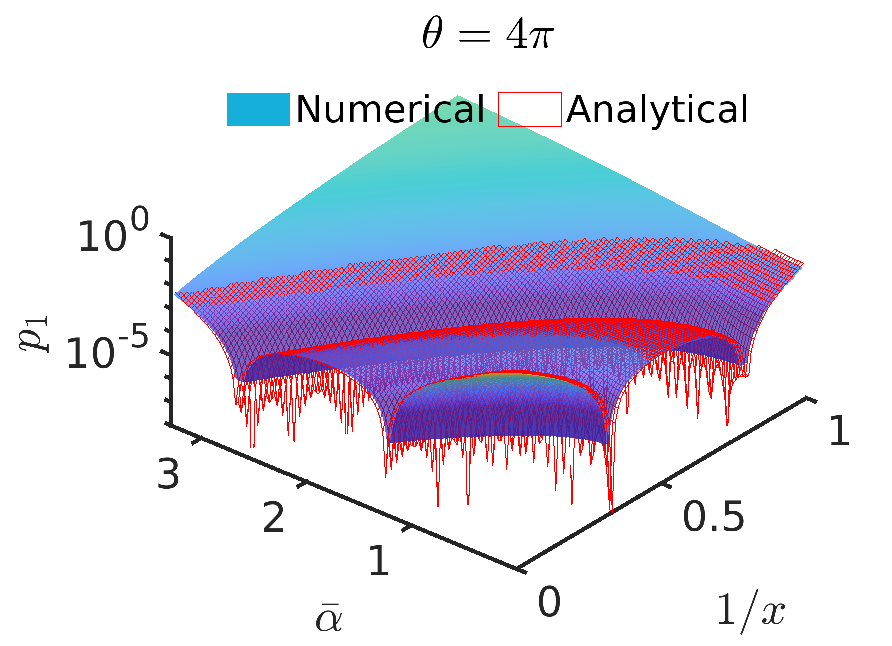}
	\includegraphics[width=3in
	]{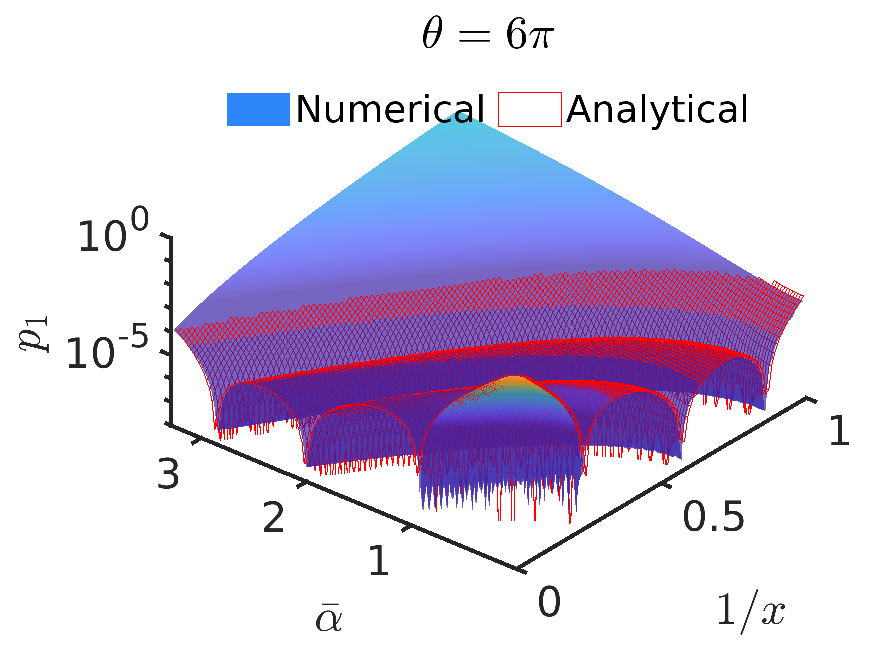}
	\end{center}
\caption{
	The results on these figures showing the survival probability have been obtained
	using the analytical formula Eq.~\ref{EQ14d17}, which is applicable for the
	{\it oscillatory subspace} of the laser parameters $[x,\balpha]$, and the numerical
	first order perturbation method, Eq.~\ref{v1}. The analytical formula derived for the limiting pulse
	length area $\theta\to\infty$ is tested for two finite values of $\theta\in\{4\pi,6\pi\}$.
}
\mylabel{FigureOsc1}
\end{figure}
\begin{figure}
	\begin{center}
	\includegraphics[width=3in
	]{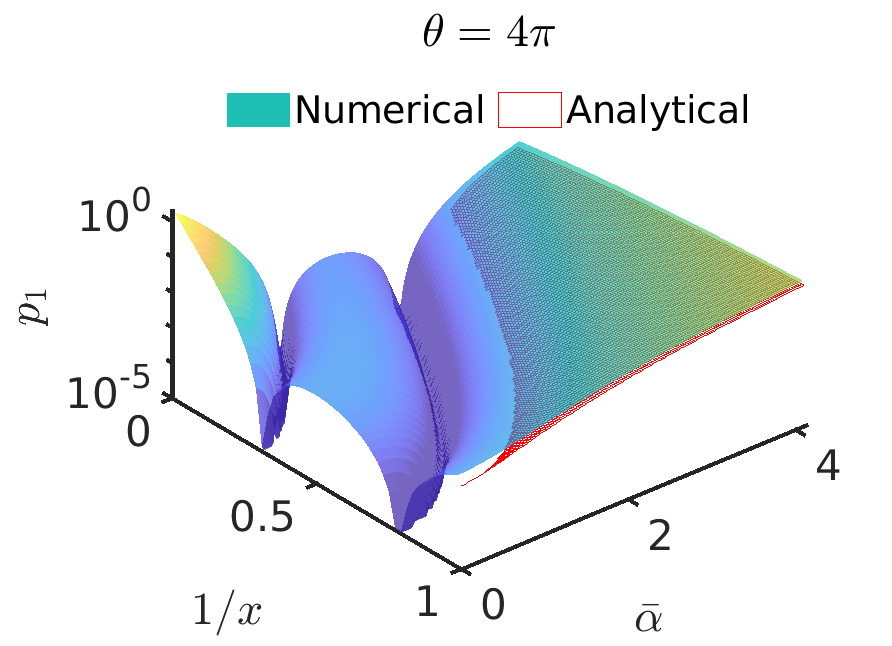}
	\includegraphics[width=3in
	]{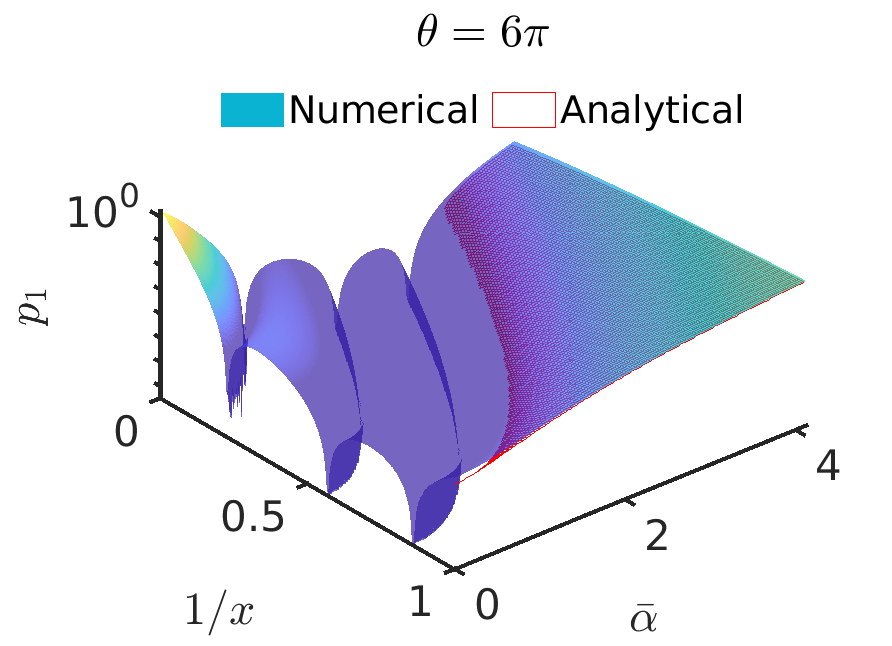}
	\end{center}
\caption{
	The results on these figures showing the survival probability have been obtained
	using the analytical formula Eq.~\ref{EQ14d44}, which is applicable for the
	{\it non-oscillatory subspace} of the laser parameters $[x,\balpha]$, and the numerical
	first order perturbation method, Eq.~\ref{v1}. The analytical formula derived for the limiting pulse
	length area $\theta\to\infty$ is tested for two finite values of $\theta\in\{4\pi,6\pi\}$.
}
\mylabel{FigureExp1}
\end{figure}

\subsubsection{Survival probability at the separator crossing}
\noindent
Let us add a very short comment on the crossing over the separator.
Eqs.~\ref{EQ14d21b} and \ref{EQ14d44} show a discontinuity
that should appear on the survival probability.
To study this,
we performed a comparison of the analytical results with
the numerical solution of first-order adiabatic perturbation
theory which are summarized in Appendix~\ref{Topology}.
The abrupt change of the prefactor seems to be a
specific property of the large pulse area limit.
More importantly, the amplitude of the survival probability
decays exponentially at the crossing of the separator, which
overloads the changes of the prefactor at least in the overall picture.

\subsection{Analytical fit for the functions defining the survival probability}
\noindent
The functions $\bar\gamma[s_0(\varepsilon_0,\balpha)]/x$ and 
$\phi[s_0(\varepsilon_0,\balpha)]/x$ (displayed
in Figs.~\ref{Figphix0} and \ref{Figgamx0})
 control the
very behavior of the survival probability. These functions
must be obtained numerically
through evaluation of the residua at the cental TPs,
which is not very convenient.
It is therefore beneficial, in the view of 
possible future applications, 
to have analytical expressions 
 for the functions $\bar\gamma[s_0(\varepsilon_0,\balpha)]/x$ and 
$\phi[s_0(\varepsilon_0,\balpha)]/x$ which would be
based on a suitable fit to the numerically obtained values.

\subsubsection{Natural coordinates on the laser pulse parameter plane}
\noindent
For the sake of convenience we define
the plane displayed in Figs.~\ref{Figphix0} and \ref{Figgamx0} using new defined
coordinates $X$ and $Y$ such that
\begin{align}
& X=\frac{1}{\bar\varepsilon_0},
& Y=\frac{\balpha}{2\sqrt{e}} .
\end{align}
Note that the oscillations associated
with the even layout of the TPs
take place within the interval
\begin{align}
& 0\le X \le 1,
& 0\le Y \le 1 .
\end{align}

As a matter of fact,
the functions
$\bar\gamma(1/\varepsilon_0,\balpha/2\sqrt{e})/x$ and 
$\phi(1/\varepsilon_0,\balpha/2\sqrt{e})/x$ are almost
radial, namely we can hardly see any angular dependence
in Figs.~\ref{Figphix0} or \ref{Figgamx0}.
We define $\varphi$ as the angle of the plane
and $\eta$ as a decreasing function
with $\varphi$:
\begin{align}
&\varphi=\arctan \frac{X}{Y}\ , \cr
&\eta = \frac{1}{2\sqrt{e}}\(
\sqrt{\(\frac{X}{Y}\)^2 + 4e\,}\,
-\,\frac{X}{Y}
\) \, ,
\end{align}
and a convenient ``radius'' $R$ such that
\begin{align}
R=e^{-\eta^2/2} \cdot \( Y\, \sqrt{e}\ \eta + X\) \, .
\end{align}
This choice of $R$ is in fact based on
the definition of the separator $s$ 
displayed in Fig.~\ref{FIGcoal},
which is the solution of Eq.~\ref{e3}
when using Eq.~\ref{zcoal}.
Namely,  we get the separator of the
plane for $R=1$; for $R<1$ we get the
oscillatory regime (even layout of TPs), for $R>1$ we
get the monotonic regime (odd layout of TPs).

\subsubsection{Fit of $\phi/x$ defining Rabi oscillations}
\noindent
The behavior of $\phi/x$ is roughly linear in $R$
and independent of $\varphi$,
\begin{align}
\frac{\phi}{x} \approx - \sqrt{\frac{\pi}{2}} (1-R) .
\end{align}
To provide a precision fit, we use a quartic polynomial
for the $R$ dependence, adding a mild angular dependence
$f(\varphi,R)$,
such that
\begin{align}
\frac{\phi}{x} = \(- \sqrt{\frac{\pi}{2}}  + \sum\limits_{k=1}^4
a_k R^k \) + f(\varphi,R) ,
\end{align}
where 
\begin{align}
&a_1 = 0.6428, \quad
a_2 = 1.5637, \cr &
a_3 = -1.4929,\quad
a_4 = 0.5398 .
\end{align}
Note that the sum over the polynomial coefficients is equal zero,
\begin{align}
\sum\limits_{k=1}^4
a_k = 0.
\end{align}
The mild angular dependence $f$ is fitted to
the analytical expression
\begin{align}
&f=g(\varphi)\cdot \sin\[\pi(R+\delta_R)\] , \cr
&g=b_0\ \[\(1-\frac{2\varphi}{\pi}\)^3 + \(\frac{b_1}{b_0}\)^3\]^{1/3} ,\cr
&\delta_R = 8 c_0 \, \frac{\varphi}{\pi}\ R\,(R-1) ,
\end{align}
where
\begin{align}
& b_0 = 0.1093, \quad
b_1 = 0.03569, \cr
& c_0 = 0.0912 .
\end{align}
Note that
\begin{align}
f(\varphi,R=0) = f(\varphi,R=1) = 0.
\end{align}
The maximum error of this fit is given by $10^{-4}$.

\subsubsection{Fit of $\bar\gamma/x$ defining exponential damping of the survival probability}

\noindent
Also the character of $\bar \gamma/x$ is determined by 
the $R$-dependence, whereas the angular dependence
is minor, see Fig.~\ref{Figgamx0}.

Via a numerical inspection of the dependence of $\bar \gamma/x$
on $R$ we discovered a very specific analytical behavior at the
criticality, $R=1$.
The found analytical form of $\bar \gamma/x$ is given by,
\begin{align}
& \frac{\bar \gamma}{x} \  = \  a_0\  + \cr
& \quad f_\pm \cdot \ln |R-1| \cdot \tan[ 2 \arctan (R-1) ] , \cr
&\mylabel{EQFitC}
\end{align}
where $f_+$ and $f_-$ represent two different functions.
$f_+$ is applicable in the interval $1<R<2$ and is 
approximately linear.
$f_-$ is applicable for $0<R<1$ and is approximately parabolic.

To achieve a satisfactory precision, however,
$f_-$ is fitted to a fifth order polynomial in $(R-1)$
and second order polynomial is used to
define the angular
dependence of the polynomial coefficients:
\begin{align}
f_-(R,\varphi) = \sum\limits_{k=0}^5 \sum\limits_{l=0}^2 c^{-}_{k,l} \ \varphi^l \cdot (R-1)^k .
\end{align}
The coefficients $c^{-}_{k,l}$ are listed in Table~\ref{T1}.
$f_+$ is fitted to a second order polynomial to get the
radial dependence, and a third order polynomial to
get the angular dependence:
\begin{align}
f_+(R,\varphi) = \sum\limits_{k=0}^2 \sum\limits_{l=0}^3 c^{+}_{k,l} \ \varphi^l \cdot (R-1)^k .
\end{align}
The coefficients $c^{+}_{k,l}$ are listed in Table~\ref{T2}.
$a_0$ is fitted to a second order polynomial to get its
angular dependence:
\begin{align}
a_0 = 0.0207 \ \varphi^2 - 0.0585\  \varphi + 0.685 .
\end{align}

Finally, the asymptotic behavior of $\bar{\gamma}/x$ 
for $R>2$ does not fit well to the form of Eq.~\ref{EQFitC}.
Instead, we use the racional form,
\begin{align}
& \frac{\bar \gamma}{x} \  = \frac{1}{\sum\limits_{k=0}^2 \sum\limits_{l=0}^2 c^{\infty}_{k,l} \ \varphi^l \cdot (R-1)^k} ,
\mylabel{EQFitInf}
\end{align}
where the coefficients $c^{\infty}_{k,l}$ are listed in Table~\ref{T3}.

The maximum error of this analytical fit for $\bar\gamma/x$
on the
covered interval ($R<5$)
is given by $0.0035$ (near $R=0$), whereas the standard 
deviation calculated for the errors  is given by $0.0007$.

\begin{table}
	\begin{tabular*}{0.45\textwidth}{@{\extracolsep{\fill} }  c | c  c  c}
		$c^{-}_{k,l}$	&  $l=2$     &   $l=1$      &   $l=0$   \\
		\hline
		$k=5$ & -0.399   &    0.996   &    0.315  \\
		$k=4$ & -0.893   &     2.14   &    0.476 \\
		$k=3$ & -0.779   &     1.79   &    0.358 \\
		$k=2$ & -0.343   &    0.763   &   -0.641 \\
		$k=1$ & -0.0306  &    0.0341  &     0.133 \\
		$k=0$ & 0.00384  &   -0.0171  &     0.288 \\
	\end{tabular*}
	\caption{Expansion coefficients for $f_-$, 
		which codefines the analytical fit for
		$\bar \gamma/x$ near the critical behavior,
		$0<R<1$, see
		Eq.~\ref{EQFitInf} and equations below.}
	\mylabel{T1}
\end{table}

\begin{table}
	\begin{tabular*}{0.45\textwidth}{@{\extracolsep{\fill} }  c | c  c  c c}
		$c^{+}_{k,l}$	&  $l=3$     &   $l=2$      & $l=1$  & $l=0$ \\  
		\hline
		$k=2$ &  0.02704  &  -0.05689  &  0.02424  & -0.007911 \\
		$k=1$ & -0.03046  &   0.06553  &   -0.0293 &     0.1553 \\
		$k=0$ & 0.004964 & -0.0007442  &   -0.0229 &     0.2883 \\
	\end{tabular*}
	\caption{Expansion coefficients for $f_+$, 
		which codefines the analytical fit for
		$\bar \gamma/x$ near the critical behavior,
		$1<R<2$, see
		Eq.~\ref{EQFitC} and equations below.}
	\mylabel{T2}
\end{table}

\begin{table}
	\begin{tabular*}{0.45\textwidth}{@{\extracolsep{\fill} }  c | c  c  c }
		$c^{\infty}_{k,l}$    &   $l=2$      & $l=1$  & $l=0$ \\  
		\hline		
		$k=2$ & -0.495  &     0.382  &   -0.0106 \\
		$k=1$ & 0.559   &    0.107   &     2.16 \\
		$k=0$ & -0.319  &     0.137  &      1.82 \\
	\end{tabular*}
	\caption{Expansion coefficients for the analytical fit for
		$\bar \gamma/x$ at the asymptotic
		distances $R>2$, see
		Eq.~\ref{EQFitC} and equations below.}
	\mylabel{T3}
\end{table}

\section{Conclusions}

\noindent
We have adjusted the complex time plane method (discovered
originally by Dykhne, David, and Pechukas) 
to study non-Hermitian
quantum dynamics where an exceptional point is encircled.
To do this we proposed  a new integration contour
for non-Hermitian problems.

The key idea of the complex time plane method is based on
the fact that the quasi-energy split as a function
of adiabatic time can be decomposed as a product
first-order Puiseux expansion which is based on
branchpoints in the complex
time plane.
These branchpoints have the same properties as 
exceptional points up to the fact that they occur
in the complex time plane rather then a plane defined
by physical parameters.
(The term TPs goes back to the research of non-adiabatic
jumps at avoided crossings of potential curves in
the scattering theory.)

The complex amplitude of the quantum states in the
end of the encircling loop is expressed via a sum
of contributions associated with all individual TPs.
We derived analytical expressions for these contributions
which are valid
in the semiclassical limit here characterized
by the large pulse area.
Each TP contribution consists of two
different
additive terms given by Eqs.~\ref{EQ12d15} and \ref{EQBC},
which reflect separate residual and
branchcut contributions of such singularities, respectively.

The currect study is limited to solution within the
first order perturbation theory and to the problem
of the survival amplitude if only one state
is initially populated under the additional condition that this
state is the less dissipative one.
An improvement and generalization of the complex time method 
beyond this approximation and to
other cases remains an open problem at this point.

We illustrate the application of the complex time
method in its present form
for the special case where the encircling
dynamics is time symmetric, namely the time-asymmetric
mode switching does not take place. This case is new and
therefore carefully introduced.
The time-symmetric EP encircling is associated with
a new phenomenon which is represented by a
behavior switch between Rabi oscillations and rapid adiabatic passage. 
We discuss that this phenomenon is associated
 with a coalescence of
two TPs of the first Puiseux order upon a 
formation of one TP of the second Puiseux order.

\section*{Acknowledgement}
\noindent
The author wants to thank 
Prof Nimrod Moiseyev and Dr Milan Sindelka for fruitful
scienfic discussions that helped accomplish the present work.
The work was financially supported by the 
Grant Agency of the Czech Republic (Grant 20-21179S).

\appendix

\section{\label{AP1}Derivation of rotating wave approximation (RWA)}

\subsection{Classical dipole approximation}

\noindent
We start from the classical dipole approximation.
The Hamiltonian for the driven atom is given by,
\begin{align}
\hat H(t) = \hat H_0 + \hat V_{int}(t),
\end{align}
where $\hat H_0$ is the Hamiltonian for the field free atom and
$\hat V_{int}(t)$ is the interaction term. We use the length gauge where
\begin{align}
\hat V_{int}(t) = \(\sum\limits_{k=1}^{N_{el}} \hat x_k\) \varepsilon_0(t) \, \cos\[\omega(t)t\] .
\mylabel{A2}
\end{align}

\subsection{Floquet Hamiltonian as a foundation for an adiabatic ansatz}

\noindent
Given the fact that the pulse hosts many optical cycles, 
\begin{align}
\tau \gg \frac {2\pi}{\omega_r} ,
\mylabel{S1}
\end{align}
it is possible to separate the fast and slow time coordinates in the
field interaction term $V_{int}(t)$, Eq.~\ref{S2} and \ref{S3}, where
the fast variation is represented by
 the field oscillations given by the cosine term, while the
slow variation is represented by the change of the pulse strength
$\varepsilon_0(t)$ and frequency $\omega(t)$.
Based on this reasoning we define the instantaneous states $\Phi[\vec r,t';\omega(t),\varepsilon_0(t)]$
of the driven atom in the continuous wave (CW) field corresponding to the given instantaneous
frequency and strength, $[\omega(t),\varepsilon_0(t)]$, where $t'$ is here
the fast coordinate, while $t$ is here assumed only as a parameter of the slow coordinate.
Such states are defined as the solutions of the Floquet Hamiltonian,
\begin{align}
	&\left\{
\hat H_0 + \hat V_{int-cw}[t';\omega(t),\varepsilon_0(t)] - i\hbar \frac{\partial}{\partial t'}\right\} \cr
&\times\quad\Phi[\vec r,t';\omega(t),\varepsilon_0(t)] \cr
&\quad= \epsilon_k[\omega(t),\varepsilon_0(t)] \Phi[\vec r,t';\omega(t),\varepsilon_0(t)] ,
\end{align}
where $\epsilon_k$ are the Floquet energies (quasienergies), which depend on the
instantaneous parameters $[\omega(t),\varepsilon_0(t)]$. The interaction term in the
length gauge is given correspondingly to Eq.~\ref{A2} as
\begin{align}
\hat V_{int-cw}[t';\omega(t),\varepsilon_0(t)]
 = \(\sum\limits_{k=1}^{N_{el}} \hat x_k\) \varepsilon_0(t) \, \cos\[\omega(t)t'\] .
\end{align}

The Floquet Hamiltonian is expanded in the basis set of the
field free atomic states $\psi_k$,
\begin{equation}
\hat H_0 \psi_k = \EE_k \psi_k,
\end{equation}
 and Fourier components $\exp(in\omega t')$:
{

\begin{align}
\[ 
\begin{matrix}
&	&	& \dots	&	&	& \\
&\E{-2}	& \M   	& \O	& \O	& \O	& \\
&\M	& \E{-}	& \M	& \O	& \O	& \\
\dots &\O	& \M	&\vec E	& \M	& \O	 & \dots \\
&\O	& \O	& \M	& \E{+} & \M	& \\
&\O	& \O	& \O	& \M	& \E{+2}	& \\
&	&	& \dots	&	&	& \\
\end{matrix}
\],
\end{align}
}
where $\vec \EE$ is the diagonal matrix of the field free atomic energies,
\begin{align}
\vec \EE =
\[
\begin{matrix}
\EE_1 	&	0	& \\
0	&	\EE_2	& \\
	&		&	\dots \\
\end{matrix}
\] ,
\end{align}
and $\vec \mu$ is the corresponding matrix of the transition dipole operator
in the basis set of the field free atomic states, 
\begin{equation}
\{\vec \mu\}_{i,j} = (\psi_i^{(l)}|\sum\limits_{k=1}^{N_{el}} \hat x_k\left|\psi_j\),
\end{equation}
where $\psi_i^{(l)}$ represents the left vector as solution of
the transposed Hamiltonian operator, $\hat H_0^t$.

\subsection{Rotating wave approximation}

\noindent
The rotating wave approximation (RWA) represents an approximation to
the pair of Floquet states, which correspond to two atomic states
coupled by a foton tuned near their resonance frequency. It is valid
when the coupling elements $\varepsilon_0\vec\mu/2$ are relatively small,
therefore the coupling to other atomic states and higher Fourier
components (multifoton processes) is negligible. The Floquet operator
for these two strongly coupled Floquet vectors simplifies as,
\begin{align}
\[ 
\begin{matrix}
\EE_2  & \frac{\varepsilon_0}{2}\mu_{2,1} \\
\frac{\varepsilon_0}{2}\mu_{1,2} & \EE_1 + \hbar \omega
\end{matrix}
\] + n\hbar\omega 
\[ 
\begin{matrix}
1 & 0 \\
0 & 1 
\end{matrix}
\] ,
\end{align}
where $n$ represents the shift of the Brillouine zone. The Floquet vectors
in different Brillouine zones are equivalent up to the phase shift $e^{in\omega t'}$, 
therefore only one pair of the Floquet vectors needs to be calculated.
In this vector pair we may also exclude the phase factor $e^{-i E_2 t'/\hbar}$,
which leads to the usual form of the RWA Hamiltonian given by,
\begin{align}
\hbar \[ 
\begin{matrix}
0 & \frac{1}{2}\Omega \\
 \frac{1}{2}\Omega & \Delta
\end{matrix}
\],
\mylabel{AH0}
\end{align}
where 
\begin{align}
\Delta=\omega - (\EE_2-\EE_1)/\hbar
\mylabel{Adetun0}
\end{align}
 is the frequency detuning, which may be generally complex if
$\EE_1$ or $\EE_2$ are complex,
and 
\begin{align}
	\Omega=\mu_{1,2}\varepsilon_0/\hbar
\mylabel{rab0}
\end{align}
 is the Rabi frequency, where $\mu_{1,2}$ is the
transition dipole element between states $|1\>$ and $|2\>$. 
Note that this form of Hamiltonian is widely used in quantum optics.
This Floquet Hamiltonian can be precisely written as
\begin{align}
& \hat H_F =
\hat H_0 + \hat V_{int-cw}[t';\omega(t),\varepsilon_0(t)] - i\hbar \frac{\partial}{\partial t'} \cr
& \approx |e^{i\omega t'} e^{iE_2 t'/\hbar} \psi_1)(e^{-i\omega t'} e^{-iE_2 t'/\hbar} \psi_1^{(l)}| \hbar\Delta(t)
+ \cr
& \quad +
\frac{\hbar\Omega(t)}{2}\bigg[|e^{i\omega t'} e^{iE_2 t'/\hbar} \psi_1)(e^{-iE_2 t'/\hbar} \psi_2^{(l)}|
+ \cr
	&\quad
|\psi_2e^{iE_2 t'/\hbar} )(e^{i\omega t'} e^{-iE_2 t'/\hbar} \psi_1^{(l)}| \bigg] + \cdots,
\mylabel{EQ1A}
\end{align}
where the dots represent the other Brillouine zones and the dressed atomic states,
which nearly do not interact with each other nor the coupled pair of states.

Let us apply the RWA to our problem. The detuning can be rewritten using
the definition of energies of the relevant field free states $|1\>$ and $|2\>$
defined in Section~\ref{system} 
\begin{align}
	&\EE_1 = E_1,
	&\EE_2 = E_2 - i\Gamma/2 ,
\end{align}
such that
\begin{align}
	\Delta=\omega - [(E_2-i\Gamma/2)-E_1]/\hbar 
	=\omega - \omega_r + \frac{i\Gamma}{2\hbar} ,
%\mylabel{detun}
\end{align}
where we used the resonance frequency $\omega_r$ defined in Eq.~\ref{omegar}.
We will also simplify the notation of the transition dipole element between
the two field-free states -- Appendix~\ref{mu12} -- such that
\begin{align}
\mu=\mu_{1,2}=\mu_{2,1}.
\mylabel{mu}
\end{align}
By using this notation we rewrite the definition of the Rabi frequency such that
\begin{align}
	\Omega=\mu\,\varepsilon_0/\hbar .
%\mylabel{rab}
\end{align}

\section{\label{mu12}Transition dipole element}

\noindent
The transition dipole moment
$\mu_{1,2}$ is directly related to the oscillator strength $f_{1\to 2}$ of the transition
$|1\> \to |2\>$ through the relation
\begin{align}
|\mu_{1,2}|^2 = f_{1\to 2} \cdot \frac{3\hbar d}{2m_e\omega_r},
\end{align}
where $d$ represents a factor taking care for the rotational degeneracies of
the atomic levels, $d=\max(L_1,L_2)/(2L_2+1)$, where $L_1,L_2$ are the main
rotational numbers of the initial and excited states, respectively.

The transition dipole moment which couples non-Hermitian resonance states
to bound states may acquire complex values in general. However, here 
we restrict ourselves to transitions which are characterized by approximately
{\it real defined} transition dipole elements:
\begin{align}
	\mu_{1,2} \in \Re \ .
	\mylabel{muRe}
\end{align}
Note that a relation between the complex phase of the transition dipole element and
asymmetric absorption lines exists as discussed in Ref.~\cite{Pick:2019}.

\section{\label{APADIAB}Instantaneous adiabatic states}
\noindent
The instantaneous adiabatic states are defined based on the solution of the
time-dependent Schr\"odinger equation,
\begin{align}
	&i\hbar \frac{\partial}{\partial t'} \psi[\vec r,t';\omega(t),\varepsilon_0(t)]
=
\cr
	&\quad
\left \{ \hat H_0 + \hat V_{int-cw}[t';\omega(t),\varepsilon_0(t)]  \right\} 
	\cr
	&\times\quad
	\psi[\vec r,t';\omega(t),\varepsilon_0(t)] ,
\end{align}
where the time evolution is performed considering only the fast field
oscillations as a part of the interaction Hamiltonian (the fast time is denoted as $t'$),
while the slowly varying instantaneous frequency $\omega(t)$ and laser strength $\varepsilon_0(t)$ are
considered as constant parameters. 
It is required that each of the solutions corresponds to only one Floquet state, such that
\begin{align}
\psi_{\pm}[\vec r,t';\omega(t),\varepsilon_0(t)] = e^{-\frac{i}{\hbar}\epsilon_\pm(t) t'}  \ \Phi_\pm[\vec r,t';\omega(t),\varepsilon_0(t)] .
\end{align}
%{\color{red} Zkontrolovat znamenko epsilon dle poznamek u Daniele Simsy.}

The instantaneous adiabatic states $\psi_{ad,\pm}$ are defined as an
approximation to the solutions of the time-dependent Schr\"odinger
equation, where the two time-scales are not divided,
\begin{align}
&i\hbar \frac{\partial}{\partial t} \psi_{ad}(\vec r,t)
\approx \cr &
\left \{ \hat H_0 + \hat V_{int-cw}[t;\omega(t),\varepsilon_0(t)]  \right\} \psi_{ad}(\vec r,t) ,
\end{align}
using the solutions including only the fast time coordinate $\psi_\pm$, and
assuming that the slow time coordinate does not cause any mixing of
the Floquet states such that,
\begin{align}
\psi_{ad,\pm}(\vec r,t) = e^{-\frac{i}{\hbar}\int\limits_{\tt}^t dt'\,\epsilon_\pm(t') }  \ \Phi_\pm[\vec r,t;\omega(t),\varepsilon_0(t)] .
\end{align}

\section{\label{AP:APTnum}Numerical verification of the convergence of adiabatic perturbation theory 
in the case of Gaussian chirped pulses}
\noindent
In this Section, we demonstrate the convergence of the
perturbation series for the amplitude $a_+$
given in Eqs.~\ref{EQperts} and \ref{V1} numerically.
We evaluated the perturbation series 
for the case of a linearly chirped Gaussian pulse
using the analytical definitions for quasienergy split (Eq.~\ref{EQdelred2}, see also Section~\ref{Sdelre} concerning
the sign of the quasienergy split)
 and non-adiabatic coupling (see Appendix~\ref{APNA0}, Eq.~\ref{NA}). 
 The values of the perturbation corrections
 after the
pulse is over (at $s\to\infty$) are denoted as
\begin{align}
	v_j \equiv  \bar v^{(j)}(s\to\infty) .
\end{align}
The quantities $v_j$ are dependent on the laser parameters $x$ (corresponding with
the peak pulse strength, Eq.~\ref{EQdefx}), $\balpha$ (the 
effective pulse chirp, Eq.~\ref{EQdefalpb}), 
and $\theta$ (the pulse area, Eq.~\ref{EQth});
\begin{figure}[!t]
\begin{center}
\includegraphics[width= 3 in
]{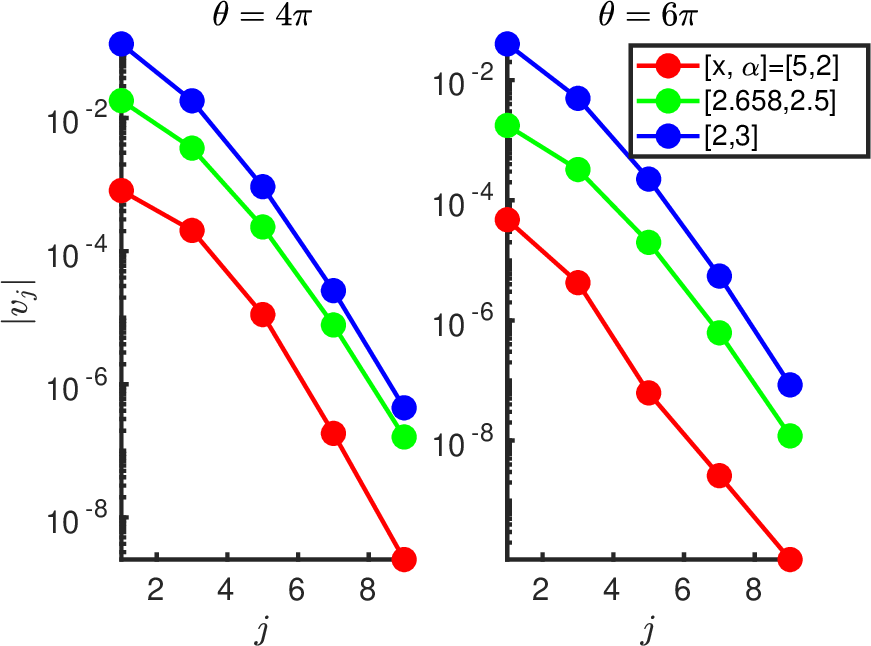}
\end{center}
\caption{
	Convergence of the perturbation series in the whole
	range of studied laser parameters is demonstrated.
	The amplitude $a_+$ of the field-free bound state after the pulse is over
	is calculated using an infinite perturbation series, Eq.~\ref{EQperts} where
	$s\to\infty$.
	Contributions of $|v_j|$, where $j$ is the perturbation
	order, have been obtained by a numerical evaluation of Eq.~\ref{V1}
	for different pulse parameters here defined by $x$, $\balpha$, and $\theta$ (see text).
	}
\mylabel{FigPerturb}
\end{figure}
We performed numerical calculations 
for three different sets of pulse parameters ($x$, $\balpha$, $\tau$), which are relevant to the studied
phenomenon, where the choice of the concrete values of $x$ and $\balpha$ taken
in the calculations has been $[x,\balpha]=\{[5,2],\ [2.658,2.5],\ [2,3]$.
The convergence is logarithmic with the perturbation order, see Fig.~\ref{FigPerturb},
which provides an evidence that the perturbation series is convergent.
The speed of convergence (the ratio between the subsequent orders) is raughly 
independent of the pulse area $\theta$ in the regime where
the studied phenomenon appears, i.e. $\theta$ is sufficently large ($\theta>4$).

\section{\label{APasymtp}Derivation of asymptotic series of transition points}
\noindent
We start our derivation from the approximate Eq.~\ref{EQ67}.
For the subsequent TP $s_{k+1}$ we can write the same equation
\begin{align}
	e^{-s_{k+1}^2} = -\(\frac{\balpha }{2}\)^2 s_{k+1}^2 .
	\mylabel{EQ69}
\end{align}
Now we define the difference
\begin{align}
	\xi = s_{k+1}^2 - s_k^2 
	\mylabel{EQ70}
\end{align}
and substitute to Eq.~\ref{EQ69} such that
\begin{align}
	e^{-s_{k}^2} e^{-\xi}= -\(\frac{\balpha}{2}\)^2 (s_{k}^2 + \xi).
	\mylabel{EQ71}
\end{align}
If we make the assumption that $\xi$ is a finite number in the asymptotic
limit $|s_k|\to\infty$, then $\xi$ can be neglected on the right hand side of Eq.~\ref{EQ71}:
\begin{align}
	e^{-s_{k}^2} e^{-\xi}= -\(\frac{\balpha}{2}\)^2 s_{k}^2 .
\end{align}
By substituting from Eq.~\ref{EQ67} we get
\begin{align}
	e^{-\xi} = 1
\end{align}
and from here
\begin{align}
	\xi = \pm 2i\pi .
\end{align}
Because $\xi$ does not depend on index $k$ for the asymptotic limit $|s_k|\to\infty$,
we see that $s_k^2$ take the form of the linear series in this limit such that,
\begin{align}
	s_k^2 = k\cdot 2i\pi + c,
	\mylabel{EQQ75}
\end{align}
where $c$ is an unknown complex constant.
Let us take a logarithm from Eq.~\ref{EQ67}:
\begin{align}
	s_k^2 = -\ln (-) - 2\ln\frac{\balpha}{2} - \ln s_k^2 + k'\cdot 2i\pi,
\end{align}
where we add an arbitrary multiplier of $k'\cdot 2i\pi$.
Now we substitute for the asymptotic series from Eq.~\ref{EQQ75}:
\begin{align}
	c
	=
	(k'-k)\cdot 2i\pi 
	- \ln (-) - 2\ln\frac{\balpha}{2} - \ln (k\cdot 2i\pi + c)
\end{align}
and we take the limit $k\to\infty$:
\begin{align}
	c
	=
	(k'-k)\cdot 2i\pi 
	- i\pi - 2\ln\frac{\balpha}{2} - \ln (k\cdot 2i\pi)
\end{align}
and separate the logarithm of imaginary unit:
\begin{align}
	&c = (k'-k)\cdot 2i\pi - i\pi - 2\ln\frac{\balpha}{2} - \ln (2k\pi) - \ln i \cr
	&\quad\quad = (k'-k)\cdot 2i\pi - 3i\pi/2 - 2\ln\frac{\balpha}{2} - \ln (2k\pi)  .
\end{align}
Because the choice of $k'$ is arbitrary one may write
\begin{align}
	c = i\pi/2 - \ln(2k\pi) - 2 \ln\(\frac{\balpha}{2}\).
\end{align}
From here we get
\begin{align}
	&s_k^2 = k\cdot 2i\pi + i\pi/2 - \ln(2k\pi)-2\ln\(\frac{\balpha}{2}\) ,\cr
	&\quad k\to\infty.
	%\mylabel{EQQ81}
\end{align}

\section{\myslabel{APNA}Various expressions for non-adiabatic coupling term}

\subsection{\label{APNA0}Non-adiabatic coupling element for linear Gaussian chirp}

\noindent
The non-adiabatic coupling element $\bar N(s)$ is defined in Eq.~\ref{EQNAf},
which can be also written as
 \begin{align}
 \bar N(s) = \frac{1}{2} \frac{d \bar\lambda(s)}{ds} 
 \(
 \frac{1}{1 + i \bar\lambda(s)}
 +
 \frac{1}{1 - i \bar\lambda(s)}
 \) .
  \end{align}
Using the definition of $\bar\lambda(s)$ within Eqs.~\ref{EQdelred2},
\begin{align}
& \bar\lambda(s) = 
e^{s^2/2} \cdot 
\(
\frac{\bar\alpha}{2}\, s  + \frac{i}{x}
\) ,
\revisited{EQdelred2}
\end{align}
we get
\begin{align}
	\frac{d\bar{\lambda}}{ds}  = \[\frac{\balpha}{2} (s^2+1) + \frac{i}{x}s\] e^{s^2/2} .
	\mylabel{aEQ85}
\end{align}
Using the definition of $\bar{\lambda}$ in Eq.~\ref{EQdelred2} we get also
\begin{align}
	\frac{1}{i\pm \frac{\Delta}{\Omega}} = \frac{1}{i\pm\(\frac{\balpha}{2} s + \frac{i}{x}\)e^{s^2/2}}.
\end{align}
From here,
\begin{align}
	&\bar N(s) = \frac{(s^2+1) + i\frac{2}{\balpha x}s}{4} 
	\cr
	&\times\quad
	\sum\limits_{z=\{+/-\}} 
	\frac{z}{\(is-\frac{2}{\balpha x}\)-z\frac{2}{\balpha}e^{-s^2/2}} .
	\mylabel{NA}
\end{align}

\subsection{\label{APNA1}First-order expansion of the non-adiabatic coupling term near the poles}

\noindent
We will start from the definition of the non-adiabatic coupling element for
the linearly chirped Gaussian pulse given in Eq.~\ref{NA}.
Let us denote $\xi = s - s_k$ a small distance from the EP $s_k$ in the complex plane of $s$
and substitute to the above equation:
\begin{align}
	&\bar N(s) = 
	\frac{s_k^2+2\xi \(s_k+\frac{i}{\balpha x}\)+\(1+\frac{2is_k}{\balpha x}\)}{4} 
	\cr & \times
	\sum\limits_{z=\{+/-\}} \frac{z}{i(s_k+\xi)-\frac{2}{\balpha x}-z\frac{2}{\balpha}
	e^{-s_k^2/2}
	e^{-s_k\xi}
	},
\end{align}
where we neglected the second-order terms $\xi^2$ in the exponential, as it is assumed $\xi\to 0$.
Additionally, we may neglect the term in the sum, which is not the pole. We may write,
\begin{align}
	&\bar N(s) = 
	\frac{s_k^2+2\xi \(s_k+\frac{i}{\balpha x}\)+\(1+\frac{2is_k}{\balpha x}\)}{4} 
	\cr &\times
	\frac{z_k}{i(s_k+\xi)-\frac{2}{\balpha x}-z_k\frac{2}{\balpha}
	e^{-s_k^2/2}
	e^{-s_k\xi}
	},
	\mylabel{aEQ90}
\end{align}
%\begin{align}
%	\bar N(s) = \frac{s_k^2+2\xi s_k+1}{4} \frac{z_k}{i(s_k+\xi)-z_k\frac{2}{\balpha}
%	e^{-s_k^2/2}
%	e^{-s_k\xi}
%	},
%	\mylabel{aEQ90}
%\end{align}
where $z_k$ is given by the condition,
\begin{align}
	is_k-z_k\frac{2}{\balpha} e^{-s_k^2/2} - \frac{2}{\balpha x} = 0 .
	\mylabel{aEQ91}
\end{align}
By substituting Eq.~\ref{aEQ91} to Eq.~\ref{aEQ90} we get
%\begin{align}
%	\bar N(s) = \frac{s_k^2+2\xi s_k+1}{4} \frac{z_k}{i(s_k+\xi)-i s_k e^{-s_k\xi} } .
%\end{align}
\begin{align}
	&\bar N(s) = 
	\frac{s_k^2+2\xi \(s_k+\frac{i}{\balpha x}\)+\(1+\frac{2is_k}{\balpha x}\)}{4} 
	\cr &\times
	\frac{z_k}{i(s_k+\xi)-\frac{2}{\balpha x}
	+e^{-s_k\xi}\(\frac{2}{\balpha x} - i s_k\)
	},
\end{align}
Now we simplify the denominator by taking the limit $\xi\to 0$ for the exponential
($e^{-s_k\xi}\approx 1-s_k\xi$):
\begin{align}
	&\bar N(s) = 
	\frac{s_k^2+2\xi \(s_k+\frac{i}{\balpha x}\)+\(1+\frac{2is_k}{\balpha x}\)}{4} 
	\cr &\times\quad
	\frac{z_k}{i\xi\(1+\frac{2is_k}{\balpha x}  + s_k^2\)} .
\end{align}
%\begin{align}
%	\bar N(s) = \frac{s_k^2+2\xi s_k+1}{4} \frac{z_k}{i\xi(1+ s_k^2) } .
%\end{align}
Taking the limit $\xi\to 0$ also in the nominator allows us to simplify the expression for $\bar N(s)$ 
down to
\begin{align}
	\lim\limits_{s\to s_k} 
	\bar N(s) = 
	\frac{z_k}{4i\xi} =
	\frac{z_k}{4i(s-s_k)} .
	\mylabel{EQ94}
\end{align}

\subsection{\label{NAsection}Asymptotic behavior of the non-adiabatic element}

\noindent
We start from the definition of the non-adiabatic coupling element given by Eq.~\ref{NA}.
The asymptotic behavior for $|s|\to\infty$ is obtained
if adding up the two terms in the sum such that
\begin{align}
	\bar N(s) = \frac{1}{\balpha}\frac
	{(s^2 + 1) + i\frac{2}{\balpha x} s}
	{\(is -\frac{2}{\balpha x}\)^2\, e^{s^2/2} - \(\frac{2}{\balpha}\)^2\,e^{-s^2/2}} .
\end{align}
If the real component
of $s$ is larger then the imaginary component
then the first term in the denominator prevails which yields,
\begin{align}
	& |\arg s| < \pi/4 \ or \ |\arg (-s)| < \pi/4: \cr
	& \bar N(|s|\to\infty) = 
	-\frac{1}{\balpha} \, e^{-s^2/2} ,
	\mylabel{EQ75}
\end{align}
while in the opposite case the second term in the denominator prevails which yields,
\begin{align}
	& |\arg s| > \pi/4 \ and \ |\arg (-s)| > \pi/4: \cr
	& \bar N(|s|\to\infty) = -\frac{\balpha}{4} \, s^2\, e^{s^2/2} .
	\mylabel{EQ76}
\end{align}
In either case the non-adiabatic coupling exponentially quadratically
decays to zero in all asymptotes. This behavior of the non-adiabatic coupling
elements in the complex plane of time is demonstrated using a numerical
evaluation in Fig.~\ref{FNA}.

\section{\myslabel{PuiseuxBeta1}Complex phase of local Puiseux coefficients for TPs based on linearly chirped Gaussian pulses}

\noindent
Let us analyze the complex phase for the local Puiseux
coefficient $\beta_k^{(1)}$,
\begin{align}
& \beta_{k}^{(1)} = 
\sqrt{\balpha \(\frac{\balpha }{2}s_k + \frac{i}{x}\) - 2s_k\, e^{-s_k^2}} 
\revisited{EQ106}
\end{align}
associated with the TP $s_k$ defined by
\begin{align}
& e^{-s_k^2} + \(\frac{\balpha}{2} s_k + \frac{i}{x}\)^2 = 0 .
\revisited{e1}
\end{align}

\subsection{TPs on the imaginary time axis}
\noindent
We will start for the case where the central TPs lie on the
imaginary axis and we will use Eq.~\ref{e1} to remove the
exponential from the definition of $\beta_k^{(1)}$ such that
\begin{align}
& \beta_{k}^{(1)} = 
\sqrt{\(\frac{\balpha}{2} s_k + \frac{i}{x}\)\, a} , \cr
&
a = \balpha + 2 s_k \(\frac{\balpha}{2} s_k + \frac{i}{x} \) .
\mylabel{EQG1}
\end{align}
$a$ is always real defined, having positive value for the
lower TP and negative for the upper TP on the imaginary axis.
Based on this fact the complex phases of the corresponding
coefficients are given by,
\begin{align}
& \arg \beta_{0i,low}^{(1)} = \frac{3\pi}{4} \pm \frac{\pi}{2}, \cr
& \arg \beta_{0i,upp}^{(1)} = \frac{\pi}{4} \pm \frac{\pi}{2} .
\mylabel{EQ147}
\end{align}
There are two possible values of the sign following from the square
root. We know, however, that the sign has to fit with the definition
of $(s-s_k)^{1/2}$ in Fig.~\ref{nakressym} and with the sign
of $\tilde \delta(s)$ on the real time axis, $s\in\Re$.
The lower TP $s_{0i,low}$ is very often very close to the real
time axis, so that we may use this point to determine the correct
sign.
We assume that  that the quasienergy split
on the real time axis near the center, namely at the point $s=0$,
can be written to a good approximation using the first order Puiseux
expansion such that
\begin{align}
\tilde \delta(s) \approx \beta_{0i,low}^{(1)} (s-s_k)^{1/2} .
\mylabel{EQ149}
\end{align}
The phase of the square root is determined based on the rules
explained in Fig.~\ref{nakressym} such that
\begin{align}
& \arg (s-s_k)\bigg|_{s=0} = \frac{3\pi}{2} \cr \rightarrow 
& \arg (s-s_k)^{1/2}\bigg|_{s=0} = \frac{3\pi}{4} .
\mylabel{EQ150}
\end{align}
The value of $\tilde \delta(s=0)$ is always purely real and positive,
from where we know that its phase is given by zero, $\arg \tilde \delta(0) = 0$.
Using Eqs.~\ref{EQ149} and \ref{EQ150} we determine the phase
of the coefficient $\beta_{0i,low}^{(1)} $ such that
\begin{align}
\arg \beta_{0i,low}^{(1)} = -\frac{3\pi}{4} ,
\mylabel{EQargbetl}
\end{align}
which is in accord with the upper definition given in Eqs.~\ref{EQ147}
derived independently for the general location of $s_{0i,low}$, 
not just near the real axis. Now, the sign uniquely derived as 
being $(+)$ for the specific case, can be generalized to any
$s_{0i,low}$.

Now we aim to determine the unique sign of $\beta_{0i,upp}^{(1)}$
so far  undetermined in the lower definition in Eq.~\ref{EQ147}.
In order to do this we will study the case where the TPs
$s_{0i,low}$ and $s_{0i,upp}$ are not far from their coalescence,
see Section~\ref{coal},
\begin{align}
s_{0i,low} \approx s_{0i,upp} .
\mylabel{EQ151}
\end{align}
Based on Eq.~\ref{EQ133}, we can write the two corresponding local Puiseux
expansion coefficients such that
\begin{align}
& \beta_{0i,low}^{(1)} =
\alpha_{0i,low} \alpha_{0i,upp}
(s_{0i,low} - s_{0i,upp})^{1/2} \cr
& \times \quad
\prod\limits_{l\ne (0i)} \alpha_l (s_{0i,low} - s_l)^{1/2} \quad , \cr
& \beta_{0i,upp}^{(1)} =
\alpha_{0i,low} \alpha_{0i,upp} 
(s_{0i,upp} - s_{0i,low})^{1/2} \cr
& \times \quad 
\prod\limits_{l\ne (0i)} \alpha_l (s_{0i,upp} - s_l)^{1/2} .
\mylabel{EQ152}
\end{align}
Not far from from coalescence, Eq.~\ref{EQ151},
the products concerning distant TPs can be considered equal,
\begin{align}
&f_0 = \prod\limits_{l\ne (0i)} \alpha_l (s_{0i,low} - s_l)^{1/2}
\cr
& \quad
\approx \prod\limits_{l\ne (0i)} \alpha_l (s_{0i,upp} - s_l)^{1/2} ,
\end{align}
and
\begin{align}
& \beta_{0i,low}^{(1)} =
\alpha_{0i,low} \alpha_{0i,upp}
(s_{0i,low} - s_{0i,upp})^{1/2} \ f_0 , \cr
& \beta_{0i,upp}^{(1)} =
\alpha_{0i,low} \alpha_{0i,upp} 
(s_{0i,upp} - s_{0i,low})^{1/2} \ f_0 .
\mylabel{EQ154}
\end{align}
From here it follows that $\beta_{0i,low}^{(1)}$ 
and $\beta_{0i,upp}^{(1)}$ differ only in their
phases near the coalescence.
Now we need to determine the phase of the square roots including
the complex vectors $(s_{0i,low} - s_{0i,upp})$ and
$(s_{0i,upp} - s_{0i,low})$, respectively.
It is possible to define the unique signs of the
square roots based on Fig.~\ref{nakressym},
which is justified by the fact that the
local Puiseux coefficients are defined based on Eq.~\ref{Puiseux0}
where $s$ has been replaced by the positions of the other TPs.
By applying the rules given in Fig.~\ref{nakressym} to the
square roots in Eq.~\ref{EQ154} we
obtain
\begin{align}
& \arg (s_{0i,low} - s_{0i,upp}) = -\frac{\pi}{2}  \cr
& \quad
\to \arg (s_{0i,low} - s_{0i,upp})^{1/2} = \frac{3\pi}{4} , \cr
& \arg (s_{0i,upp} - s_{0i,low}) = \frac{\pi}{2} \cr
& \quad
\to \arg (s_{0i,up} - s_{0i,low})^{1/2} = \frac{\pi}{4} ,
\end{align}
from where we get
\begin{align}
\beta_{0i,upp}^{(1)} = - i \ \beta_{0i,low}^{(1)} ,
\end{align}
which applies near the coalescence.
The actual argument of $\beta_{0i,upp}^{(1)}$ near the coalescence
is obtained using Eq.~\ref{EQargbetl} derived above,
\begin{align}
\arg \beta_{0i,upp}^{(1)} = - \frac{\pi}{4} .
\mylabel{EQargbetu}
\end{align} 
This equation applies not only for the case when the TPs
are near the coalescence, for which it has been derived.
This special case helps us to determine the sign 
for the general definition of the phase in the second line
of Eqs.~\ref{EQ147}, which implies that the phase
Eq.~\ref{EQargbetu} applies for the upper TP in general.

\subsection{TPs near the coalescence}
\noindent
As the TPs lie on the imaginary time axis, the complex phases
of their corresponding first order local Puiseux coefficients
are defined by Eqs.~\ref{EQargbetl} and \ref{EQargbetu}.
In the coalescence, the auxiliary variable $a$ defined
in Eq.~\ref{EQG1} is zero and very near to the coalescence, $a$
is infinitesimally small.

Let us define the TPs near the coalescence as
\begin{align}
s_k = s_{coal} + \delta_s \ ,
\mylabel{EQG13}
\end{align}
where $s_k$ may stand for any TP near the coalescence,
\begin{align}
s_k\in\{s_{0i,low}, s_{0i,upp}, s_0, \bar s_0 \}\ .
\end{align}
The auxiliary variable $a$ defined in Eq.~\ref{EQG1}
which measures how near is the coalescence of the TPs
is expressed using $\delta_s$ such that
\begin{align}
a = \ 2\, \delta_s \ \(\balp s_{coal} + \frac{i}{x} \) + \balp\, \delta_s^2 \ .
\end{align}
We know that the coalescence always occurs on the positive
imaginary time axis so that we can rewrite the above equation such that
\begin{align}
a = \ 2i\, \delta_s \ \(\balp |s_{coal}| + \frac{1}{x} \) + \balp\, \delta_s^2 \ .
\mylabel{EQG16}
\end{align} 
As long as the TPs occur on the imaginary time axis,
$\delta_s$ is an imaginary number and thus $a$ is real
defined. On the other side of the coalescence, $\delta_s$
is a real defined small number.
This implies that $a$ is an imaginary small number
very near the coalescence,
\begin{align}
&\arg a(s_0) = \frac{\pi}{2} ,
&\arg a(\bar s_0) = -\frac{\pi}{2} , \cr
& s_0\approx \bar s_0 \approx s_{coal} .
\end{align}  
where the sign is defined by the sign of the variation $\delta_s$.
The complex phase of the bracket below the square root in Eq.~\ref{EQG1}
is approximately given by $\pi/2$ as $\delta_s$ approximates to zero.
Putting this together with the complex phase of $a$ we obtain
the complex phases of $\beta_0^{(1)}$ and $\bar \beta_0^{(1)}$,
\begin{align}
&\arg \beta_0^{(1)} = \pm \frac{\pi}{2},
&\arg \bar \beta_0^{(1)} =  \frac{\pi}{2} \pm \frac{\pi}{2} .
\mylabel{EQG18}
\end{align}  

The sign is determined using the fact that the
quasienergy split is real and positive defined
at the point $s=0$ on the real time axis.
By studying the continuity of the quasienergy split on the imaginary
time axis, using Eq.~\ref{zdef} and Fig.~\ref{FIGcoal},
we find out that the quasienergy split
is also a positive define real defined number in
the point of coalescence, $s=s_{coal}$, assuming that
the coalescence occurs, and this would remain
true also where the two TPs are near the coalescence,
which is the case studied here.
We express the quasienergy split
at the point $s_{coal}$ using the 
general Puiseux expansion, Eq.~\ref{Puiseux0},
which includes the explicit
contributions of the two near TPs $s_0$ and $\bar s_0$:
\begin{align}
&\tilde\delta(s_{coal}) = \alpha_0 \balpha_0 \sqrt{(s_{coal}-s_0) (s_{coal}-\bar s_0)} \cr
& \quad\times\quad\prod\limits_{k\ne 0} \alpha_k (s_{coal}-s_k)^{1/2} .
\end{align} 
Now we substitute the known the complex phases,
where the phases of the square roots $(s_{coal}-s_0)^{1/2}$ and
$(s_{coal}-\bar s_0)^{1/2}$ are determined based on the
rules given in Fig.~\ref{nakressym} assuming that the
two complex vectors are perpendicular to the imaginary
axis prior to the application of the square roots.
We obtain,
\begin{align}
\arg \tilde\delta(s_{coal}) & = 
\arg \(\alpha_0 \balpha_0 
\prod\limits_{k\ne 0} \alpha_k (s_{coal}-s_k)^{1/2}\) 
+ \frac{\pi}{2} \cr
& = 0.
\mylabel{EQG20}
\end{align} 
Now we apply the definition of $\beta_k^{(1)}$ in Eqs.~\ref{EQ133}
to our case:
\begin{align}
& \beta_0^{(1)} = \alpha_0 \bar \alpha_0 (s_0-\bar s_0)^{1/2}
\prod\limits_{k\ne 0} \alpha_k (s_0 - s_k)^{1/2}, \cr
& \bar\beta_0^{(1)} = \bar\alpha_0  \alpha_0 (\bar s_0-s_0)^{1/2}
\prod\limits_{k\ne 0} \alpha_k (\bar s_0 - s_k)^{1/2} .
\end{align}
The products over the higher TPs in the above equations can
be simply approximated by using $s_0\approx \bar s_0 \approx s_{coal}$
such that
\begin{align}
& \beta_0^{(1)} = f_0\, (s_0-\bar s_0)^{1/2} , \cr
& \bar\beta_0^{(1)} = f_0\, (\bar s_0-s_0)^{1/2} ,\cr
& f_0 = \alpha_0 \bar \alpha_0\,\prod\limits_{k\ne 0} \alpha_k (s_{coal} - s_k)^{1/2}\ .
\end{align}
In Eq.~\ref{EQG20} we determined the complex phase of 
$f_0$ as $\arg f_0 = -\pi/2$.
The complex arguments of $(s_0 - \bar s_0)$ and
$(\bar s_0 - s_0)$ are 0 and $\pi$, respectively,
to which the square roots are applied based on Fig.~\ref{nakressym}.
From here we get
\begin{align}
& \arg \beta_0^{(1)} = -\frac{\pi}{2},
& \arg \bar \beta_0^{(1)} = 0 ,
\mylabel{EQG23}
\end{align}
which is in accord with Eqs.~\ref{EQG18}.

\subsection{Distant TPs}
\noindent
Let us analyze $\beta_{k}^{(1)}$ for the distant TPs.
By applying the asymptotic limit for the TPs, see Section~\ref{ASTP} and Eq.~\ref{EQ67}, we get the approximate value for $\beta_{k}^{(1)}$ 
given by,
\begin{align}
\beta_{k\to\infty}^{(1)} = \pm \sqrt{\frac{\balpha}{2}}\  s_k^{3/2}.
\mylabel{EQG24}
\end{align}
Let us determine the argument of this quantity. 
By taking the uppermost limit of $k\to\infty$ for $s_k$ using Eq.~\ref{EQQ82} we get
\begin{align}
s_k \approx \sqrt{2i\pi k} = |\sqrt{2\pi k}| \cdot e^{i\pi/4} \, ,
\mylabel{EQG25}
\end{align}
where we assume only the phase $\pi/4$ and not $-3\pi /4$
based on the fact that $s_k$ occurs in the upper imaginary
half plane.
Similarly we determine the limit of $\bar s_k$ as
\begin{align}
\bar s_k \approx |\sqrt{2\pi k}| \cdot e^{3i\pi/4} \, .
\mylabel{EQG26}
\end{align}
The argument of $\beta_{k}^{(1)}$ based on Eqs.~\ref{EQG24}
and \ref{EQG25} is given by 
\begin{align}
\arg \beta_{k\to\infty}^{(1)} = \frac{3\pi}{8} - n\pi,
\quad n\in\{0\cdots 1\} ,
\mylabel{EQG27}
\end{align}
and similarly the argument of $\bar \beta_{k}^{(1)}$
based on Eqs.~\ref{EQG24} and \ref{EQG26} is given by 
\begin{align}
\arg \beta_{k\to\infty}^{(1)} = \frac{\pi}{8} - n\pi,
\quad n\in\{0\cdots 1\} .
\mylabel{EQG28}
\end{align}

\begin{figure}[!h]
	\begin{tabular}{ l c}
		(a) $\quad$ & \includegraphics[width= 2.5 in ]{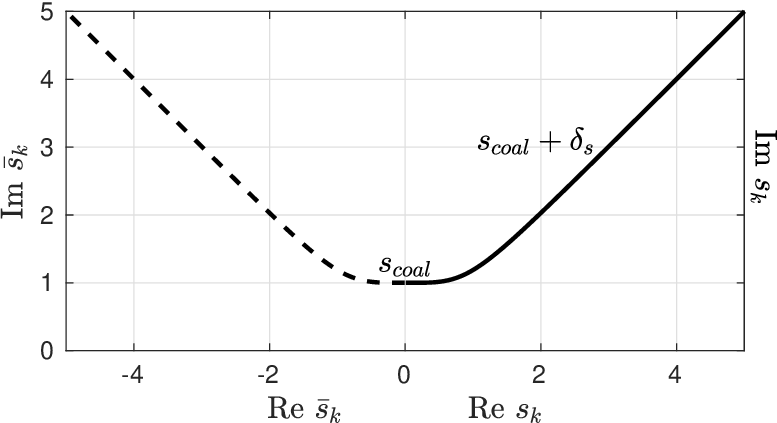} \\
		     & \\
		(b)  & \includegraphics[width= 2.5 in ]{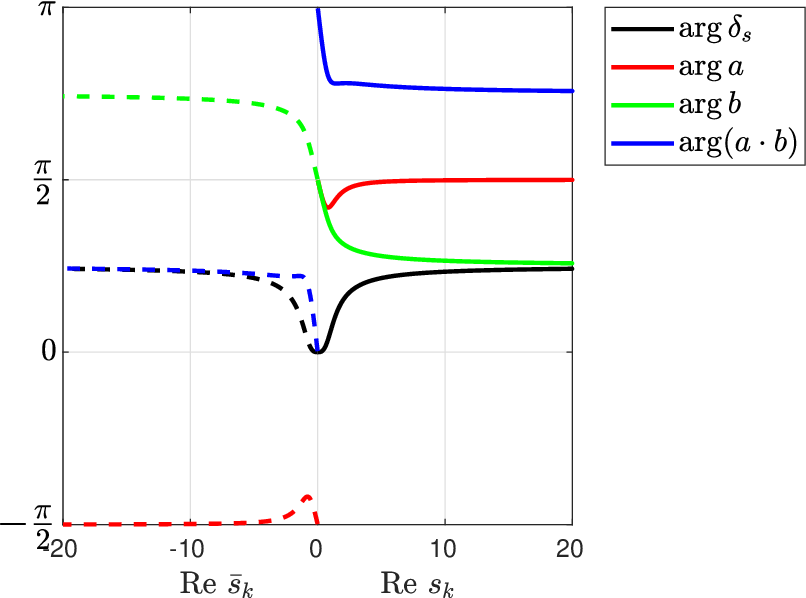} \\
		(c)  & \includegraphics[width= 2.5 in ]{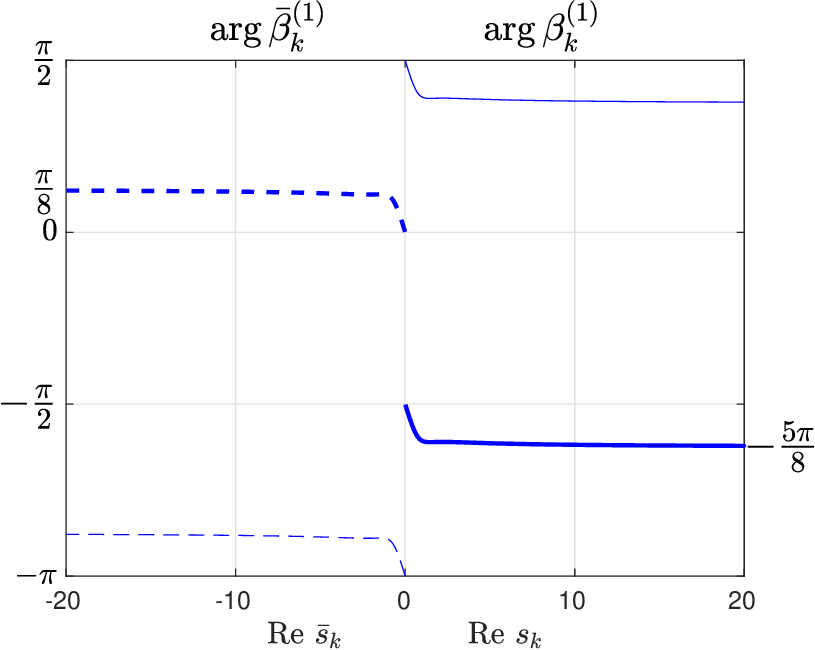} \\
	\end{tabular}
	\caption{
		This figure shows how one can determine uniquely the 
		complex phase
		of the local Puiseux coefficients $\beta_k^{(1)}$
		and $\bar \beta_k^{(1)}$
		for $k\to\infty$.
		(a) $s_k$ and $\bar s_k$ represent a specific
		pair of TPs, which we pursue as they move in the complex
		time-plane upon a continuous change of the laser
		parameters, starting at their coalescence ($k=0$)
		up to their joining the asymptotic
		limit ($k\to\infty$).
		(b) shows the complex phases of the variables
		$\delta_s$, $a$, and $b$, which are related to 
		the positions of $s_k$ and $\bar s_k$, respectively.
		(c) shows the complex phases of the local
		Puiseux coefficients $\beta_k^{(1)}$ and $\bar\beta_k^{(1)}$
		as related to the positions of the TPs $s_k$ and $\bar s_k$.
		As the Puiseux coefficients are related by the square root
		to the product $(a\cdot b)$, their correct signs 
		(denoted by the thick lines)
		are
		determined based on the known values of the complex phases
		of $\beta_0^{(1)}$ and $\bar\beta_0^{(1)}$
		near the coalescence of the TPs (which occurs for 
		$\re \, s_0 = 0$ and $\re \, \bar s_0 = 0$, respectively).
	}
	\mylabel{nakresphas}
\end{figure}
We will solve the uncertainty of the sign due to the square root
based on a heuristic reasoning using the continuity as
the TPs $s_0$ and $\bar s_0$ defined near the coalescence,
Eq.~\ref{EQG13}, move into the complex plane.
We assume that the positions of $s_0$ and $\bar s_0$
are continually changed in the complex plane upon 
the variation of the laser parameters $\balpha$ and $x$,
such that they depart from $s_{coal}$ and penetrate deep
to the complex plane. As this is happening, $\delta_s$
is increased, Eq.~\ref{EQG13}, first acquiring real values.
Later, as the laser parameters are varied continually,
$s_0$ and $\bar s_0$ penetrate yet deeper to the complex
plane, while another pair of the TPs appears 
near the central part of the complex time plane.
When this happens, the $k$ index of the monitored TPs 
is upgraded to $1$, such that $s_0 \to s_1$,  $\bar s_0 \to \bar s_1$.
$\delta_s$ is also no more a real number, as $s_k$ approaches
the asymptotic limit $k\to\infty$ defined above.
This process is shown in Fig.~\ref{nakresphas}a.

The point of doing this reasoning is that we know
the complex phase of $\beta_0^{(1)}$ and  $\bar \beta_0^{(1)}$
near the coalescence (Eq.~\ref{EQG23}). The idea
of continuity will navigate us to the right sign
as $k$ is changed for a concrete pair of the TPs from $k=0$
to $k\to\infty$.

Let us now determine the phase of $a$ defined in Eq.~\ref{EQG1}
and $b$ which we define for the present purpose as
\begin{align}
b = \frac{\balpha}{2} s_k + \frac{i}{x} ,
\end{align}
where clearly
\begin{align}
\beta_k^{(1)} = \sqrt{b \cdot a} .
\end{align}
Now, using the definition of $\delta_s$ in Eq.~\ref{EQG13}
and using \ref{EQG16}
\begin{align}
&a = 2 i \delta_s \cdot A + \balpha \delta_s^2 ,  \cr
&b = i B + \frac{\balpha}{2} \delta_s ,
\end{align}
where $A$ and $B$ are real positive defined numbers.
The evolution of the complex phase of $\delta_s$ corresponds
with the trajectory of $s_k$, see Fig.~\ref{nakressym}a and
~\ref{nakressym}b.
The evolution of the complex phases of $a$ and $b$ 
can be raughly guessed if we substitute positive real numbers for
$A$, $B$, and $\balpha$, as shown in Fig.~\ref{nakressym}b.
The complex phase of $\beta_k^{(1)}$ defined by the square
root allows for the different sign, which we show in
Fig.~\ref{nakressym}c and which is in agreement with
our previous considerations, see Eqs.~\ref{EQG27}
and \ref{EQG28}.
Now, using the idea of continutity, the 
correct signs of $\beta_k^{(1)}$ and $\bar\beta_k^{(1)}$
may be determined from the known values 
for $\beta_0^{(1)}$ and $\bar\beta_0^{(1)}$, i.e.
for $\re s_k = 0$, which are given in Eq.~\ref{EQG23}.
As implied by Fig.~\ref{nakressym}c, the
arguments of $\beta_k^{(1)}$ and $\bar\beta_k^{(1)}$
in the asymptotic limit are given by
\begin{align}
&\arg \beta_{k\to\infty}^{(1)} = -\frac{5\pi}{8},
&\arg \bar \beta_{k\to\infty}^{(1)} = \frac{\pi}{8}.
\mylabel{EQG35}
\end{align}
Note that Eqs.~\ref{EQG35} satisfy the symmetry relation
given in Eq.~\ref{EQSYMB}.

\section{Analytical derivation of asymptotic expressions for the equivalue lines}

\subsection{\label{APEQL1}Equivalue lines approaching the imaginary axis in the asymptotic limit}
\noindent
In the limit $\im\, s\to\infty$, $\tilde \delta(s)$ is given by the exponential term:
\begin{align}
\tilde \delta(s) \approx \pm  e^{-s^2/2} ,
\revisited{EQ9d11}
\end{align}
as follows from Eq.~\ref{EQdelred2}.
The asymptotic form of 
$\tilde \delta(s)$ enters the definition of the equivalue line (Eq.~\ref{BC2}) through its complex phase given by
\begin{align}
\arg \tilde \delta(s_{BC}) = -(\re s_{BC})(\im s_{BC}) + (\mp 1 + 1) \frac{\pi}{2} .
\mylabel{EQASph}
\end{align}
The left hand side of Eq.~\ref{BC2} is defined as $ds_{BC}/|ds_{BC}|$,
where we have set 
\begin{align}
\lambda = |d s_{BC}|
\mylabel{EQlam}
\end{align}
 (just as in the numerical
calculation in Section~\ref{EQLPHIN}).
Now we use the fact that for the studied asymptote, the
equivalue line is almost parallel with the imaginary time axis.
This fact is implemented by the following mathematical assumption:
\begin{align}
&\xi = \frac{d \re s_{BC}}{d \im s_{BC}}, &\xi\to 0.
\end{align}
Using this assumption, we simplify the increment $|d s_{BC}|$ such that
\begin{align}
&
|ds_{BC}| = |d\re s_{BC} + i\, d \im s_{BC}|\cr
&\quad
\approx d\im s_{BC} \cdot (1+\frac{1}{2}\xi^2).
\end{align}
$\xi$ is also used to express $d s_{BC}$ simply as,
\begin{align}
d s_{BC} = d\re s_{BC} + i\, d \im s_{BC}
= d\im s_{BC}\cdot (i + \xi).
\end{align}
Both expressions for $|d s_{BC}|$ and $d s_{BC}$
are used to simplify the ratio $ds_{BC}/|ds_{BC}|$
on the left hand side of Eq.~\ref{BC2} (with $\lambda = |d s_{BC}|$).
\begin{align}
\frac{d s_{BC}}{|d s_{BC}|} \approx i + \xi - \frac{1}{2}i\xi^2 \approx \exp{\(-i\xi + \frac{i\pi}{2}\)}.
\mylabel{EQAS1}
\end{align}

Now we substitute Eq.~\ref{EQAS1} for the left hand side 
and Eq.~\ref{EQASph} for the right hand side of Eq.~\ref{BC2},
respectively.
By applying the logarithm on the both sides we get,
\begin{align}
\xi = (\re s_{BC})(\im s_{BC}) \mp \pi/2 - n\pi .
\end{align}
The additive phase is not given by $2n\pi$ but
$n\pi$, respecting the fact that the r.h.s. in Eq.~\ref{BC2} includes both
possible signs.
By substituting $\xi\to 0$ in the above equation one gets:
\begin{align}
\re s_{BC} = \frac{(n\pm 1/2)\pi}{\im s_{BC}}  .
\revisited{EQ718}
\end{align}

\subsection{\label{APEQL2}Equivalue lines parallel to the real axis in the asymptotic limit}

\noindent
Let us investigate the asymptotic analytical
behavior of the ``green'' equivalue lines in Fig.~\ref{F3}, where
$|\re s|\gg|\im s|$. 
The energy split defined in Eq.~\ref{EQdelred2} can be simplified
such that
\begin{align}
\tilde \delta(s) \approx \pm \(\frac{\balpha}{2} \cdot s + \frac{i}{x} \) .
\revisited{EQG8}
\end{align}
The asymptotic parts of these
equivalue lines are nearly parallel to the real axis, which fact
can be expressed mathematically by defining $\xi$  as the ratio
\begin{align}
&\xi = \frac{d \im s_{BC}}{d \re s_{BC}}, &\xi\to 0
\end{align}
representing an infinitesimally small number.

First, we will use $\xi$ to evaluate the left hand side of Eq.~\ref{BC2} 
(using the definition of $\lambda=|ds_{BC}|$ in Eq.~\ref{EQlam}).
We proceed by finding the expressions for $|ds_{BC}|$,
\begin{align}
&
|ds_{BC}| = |d\re s_{BC} + i d \im s_{BC}|\cr
&\quad\approx d\re s_{BC} \cdot (1+\frac{1}{2}\xi^2),
\end{align}
and $ds_{BC}$,
\begin{align}
d s_{BC} = d\re s_{BC}\cdot (1 + i\xi),
\end{align}
from where we get the sought for asymptotic expression
for the left hand side of Eq.~\ref{BC2},
\begin{align}
\frac{d s_{BC}}{|d s_{BC}|} \approx 1  + i\xi - \frac{1}{2}\xi^2 \approx \exp{\(i\xi\)}.
\mylabel{EQG9}
\end{align}

Now we need to evaluate also the asymptotic form
on the right hand side of Eq.~\ref{BC2}, namely for the
argument of $\tilde \delta(s)$, which is defined
by the asymptotic form Eq.~\ref{EQG8}.
Using the fact that $|s|$ is very large, $|s|\to\infty$,
we can use the first order Taylor expansion such that
\begin{align}
\arg \tilde \delta(s) = \arg \(1 + \frac{2i}{\balpha x s}\) + n\pi + \arg s,
\end{align}
where $n$ maybe odd or even, according to the sign of $\tilde\delta(s)$.
Additionally, we may use the fact that the
 expression in the bracket is approximately given by an exponential in the limit
$|1/s|\to 0$:
\begin{align}
\arg \tilde \delta(s) = \arg \exp{\(\frac{2i}{\balpha x s}\)} + n\pi + \arg s.
\end{align}
By using $|\re s|\ll|\im s|$ we obtain,
\begin{align}
\arg \bar \delta(s) = \frac{2}{\balpha x \re s} + n\pi + \frac{\im s}{\re s} .
\end{align}

Based on Eq.~\ref{BC2}, we compare the phases of its l.h.s. (defined by $\xi$, see Eq.~\ref{EQG9})
and r.h.s. defined just above:
\begin{align}
\xi = -\frac{2}{\balpha x \re s} - n\pi - \frac{\im s}{\re s} .
\end{align}
%Now we substitute $\xi\to 0$ to get the sought for asymptotic
%relation for the equivalue lines, which reads
%\begin{align}
%\im s = -\frac{2}{\balpha x} - n \pi \re s .
%\end{align}
%$n=0$
%\begin{align}
%\im s = -\frac{2}{\balpha x} .
%\end{align}
Now we substitute for $\xi$ from Eq.~\ref{EQG9}:
\begin{align}
\frac{d\im s}{d\re s} = -\frac{2}{\balpha x \re s} - \frac{\im s}{\re s} .
\end{align}
From here we get the relation
\begin{align}
\frac{d\im s}{\frac{2}{\balpha x} + \im s} = -\frac{d\re s}{\re s} ,
\end{align}
which can be rewritten as
\begin{align}
d\ln\({\frac{2}{\balpha x} + \im s}\) = -d\ln{(\re s)} = d\ln\frac{1}{\re s} .
\end{align}
By performing the integration we obtain the final relation
\begin{align}
{\frac{2}{\balpha x} + \im s} = \frac{c}{\re s} ,\quad \re s>0,
\end{align}
or equivalently so
\begin{align}
{\im s} = \frac{c}{\re s} -\frac{2}{\balpha x} ,\quad \re s>0 ,
\revisited{EQ114}
\end{align}
where
$c$ is an unknown constant.

\subsection{\label{APEQL3}Crossing of the equivalue lines and the real time axis}

\noindent
Eq.~\ref{EQ114} shows that the equivalue lines 
approach the line which lies parallel below the real axis at 
$\im s = -2/(\balpha x)$.
The equivalue lines associated with different
TPs $s_k$ cross the real time axis at different
points $s_{0k}$, Fig.~\ref{FigContDDP}. The crossing points
are related to the unknown constant $c$ in Eq.~\ref{EQ114}
which will be denoted as $c_k$ to demonstrate its association
with the concrete TP. The relation of the crossing points
$s_{0k}$ and the constants $c_k$ is defined as
\begin{align}
s_{0k} =(\pm) \frac{\balpha\, x\, c_k}{2} .
\revisited{EQsrecr}
\end{align}

The constants $c_k$ can be determined from the asymptotic
expressions for exceptional points, Eqs.~\ref{EQQ81} and \ref{EQQ82} for the
equivalue lines, where the corresponding TP $s_k$ is located well in the asymptotic limit,
$|s_k|\gg 0$. Eq.~\ref{EQ114} for the equivalue line is rewritten such that
\begin{align}
c_k = \frac{1}{2}{\im s^2} +\frac{2}{\balpha x}\re s .
\end{align}
Now by substituting $s=s_k$ and using Eqs.~\ref{EQQ81} and \ref{EQQ82} to define $s_k$ we
get,
\begin{align}
& c_k = (k\pi + \frac{\pi}{4}) 
+ \[ (k\pi + \frac{\pi}{4})^2 + \frac{\ln^2(k\balpha\pi)}{4} \]^\frac{1}{4} \cr
&\times\quad
(\cos\phi_k - \sin\phi_k), \cr
& \phi_k = \arg\[1 + \frac{1}{8k} + i \frac{\ln(k\balpha\pi)}{4\pi k}\] .
\end{align}
By taking the asymptotic limit $k\to\infty$ from this expression we get
\begin{align}
& c_k \approx (k\pi + \frac{\pi}{4}) 
+ (k\pi + \frac{\pi}{4})^{1/2} \[1 + \frac{\ln^2(k\balpha\pi)}{(2k\pi)^2} \]^\frac{1}{4}\cr
&\times\quad
(\cos\phi_k - \sin\phi_k), \cr
& \quad\quad \approx (k\pi + \frac{\pi}{4}) 
+ (k\pi + \frac{\pi}{4})^{1/2} \[1 + \(\frac{\ln(k\balpha\pi)}{4k\pi}\)^2 \]\cr
&\times\quad
(\cos\phi_k - \sin\phi_k) .
\end{align}
Further we approximate
\begin{align}
&\phi_k \approx \frac{\ln(k\balpha\pi)}{4\pi k}, \cr
&\cos\phi_k\approx 1 - \frac{1}{2}\(\frac{\ln(k\balpha\pi)}{4\pi k}\)^2, \cr
&\sin\phi_k\approx \phi_k \approx \frac{\ln(k\balpha\pi)}{4\pi k},
\end{align}
such that
\begin{align}
& c_k \approx (k\pi + \frac{\pi}{4}) 
+ (k\pi + \frac{\pi}{4})^{1/2} \[1 + \(\frac{\ln(k\balpha\pi)}{4k\pi}\)^2 \]\cr
&\times\quad
\( 1 - \frac{\ln(k\balpha\pi)}{4\pi k}  - \frac{1}{2}\(\frac{\ln(k\balpha\pi)}{4\pi k}\)^2\).
\end{align}
By keeping only the leading order terms we get,
\begin{align}
& c_k \approx (k\pi + \frac{\pi}{4}) 
+ (k\pi + \frac{\pi}{4})^{1/2} 
\( 1 - \frac{\ln(k\balpha\pi)}{4\pi k} \).
\revisited{EQ940}
\end{align}
The assymptotic expression for the equivalue lines approaching the real axis in the
asymptotic limit given by Eqs.~\ref{EQ114} and \ref{EQ940} are illustrated
in Fig.~\ref{FigBCasym}.

\section{\label{connect}Connecting contour between two branchcuts in the asymptotic limit}

\noindent
This Appendix Section concerns asymptotic parts of the new contour
integration defined in this Paper. It is shown that the asymptotic
area has a zero contribution to the obtained integral.

\subsection{Linear connector between two neighbouring branchcuts in the asymptotic limit $\im s\to\infty$}
\noindent
%The absolute value of the exponential term is switched between two different finite values
%when passing between two branchcuts in the
%{\it asymptote} $s\to i \infty$, as we know from the fact that the next equivalue line is
%associated with a different EP. Now we will evaluate $\gamma(s)$ between the two branchcuts.
%
The contours connecting between two branchcuts in the asymptote $\im\,s\to\infty$
are illustrated by the short abscissas in Fig.~\ref{FigCont} on the 
upper boarder of the graph ($\im\,s=10$). 
These abscissas connect between the branchcuts, say of the TPs $s_k$ and $s_{k+1}$.
Let us define the connecting line as
\begin{align}
&\im s = const \to \infty,
\quad \re s = \frac{(\xi+1/2)\pi}{\im s}, \cr
&\quad \xi\in\Re, \ \xi\in(k,k+1),\ k\ge 0,
\mylabel{EQ9d1}
\end{align}
where the definition of $\re s$ is based on the asymptotic behavior of the
equivalue lines (which also represent the branchcuts), Eq.~\ref{EQ718}.

\subsection{Non-analytical behavior along the connecting contour}
\noindent
Now, we want to draw the attention of the reader to the fact that the
two points connected by the abscissa are parts of two equivalue lines which both correspond to
the same TP, $s_{k+1}$. This is because the branchcut of the TP $s_k$, which is also the equivalue line
from the TP  $s_k$, is merged with another equivalue line, which comes from TP  $s_{k+1}$ (and is
not a branchcut), as $\im \,s\to\infty$. This is clearly seen in Fig.~\ref{FigCont}.

Let us focus on the calculation of $\gamma(s)$ along the asymptotic connecting contour.
One may calculate $\gamma(s)$ along the connecting contour
such that
\begin{align}
&\gamma(s) = \gamma(s_k) 
+ \int\limits_{\re s_{BC,k}}^{\re s_{EL,k+1}}  Q(s') \, d\re s' \cr
&\quad + \int\limits_{\re s_{EL,k+1}}^{\re s}  Q(s') \, d\re s'
\cr
&\quad\re\, s_{BC,k}< \re\,s<\re\,s_{BC,k+1}, \cr
&\quad \im s = \im\, s_{BC,k} = \im\,s_{BC,k+1} \to \infty,
\mylabel{EQ9d2a}
\end{align}
where $\re\, s_{BC,k}$ defines the initial value, which starts at
the branchcut corresponding to the TP $s_k$. 
$\re\, s_{EL,k+1}$ is a point that is only infinitesimally shifted from the branchcut,
\begin{align}
\re\, s_{EL,k+1} = \re\, s_{BC,k} + \delta, \quad \delta\to 0.
\end{align}
$\re\, s_{EL,k+1}$ is a part of the equivalue line, which is associated with the
neighbouring TP $s_{k+1}$, where the asymptotic behaviors of this equivalue line  and
the branchut are the same. Clearly, the value of $\gamma(s)$ jumps
from $\gamma(s_k)$ to $\gamma(s_{k+1})$ on this infinitesimally small distance therefore
we may simply rewrite Eq.~\ref{EQ9d2a} such that
\begin{align}
&\gamma(s) = 
\gamma(s_{k+1}) 
+ \int\limits_{\re\, s_{EL,k+1}}^{\re\, s}  Q(s') \, d\re\, s' \cr
&\quad =
\gamma(s_{k+1}) 
+ \int\limits_{\re\, s_{BC,k}}^{\re\, s}  Q(s') \, d\re\, s' ,
\mylabel{EQ9d2}
\end{align}
where the point on the equivalue line ($\re\, s_{EL,k+1}$)  has been
replaced by the point on the branchcut ($\re\, s_{BC,k}$).

\subsection{Penetration of equivalue lines from the lower complex half plane of time}
\noindent
As a matter of fact,
when the laser strength parameter $x$ reaches a certain lower bound for a given chirp $\balpha$,
then some equivalue lines that start at the TPs on the lower half-plane end up in the asymptotic
limit $\im\,s\to\infty$. This phenomenon is illustrated in Fig.~\ref{FigPenet}.
This fact alters the previous conclusions only in the fact that the value of 
$\gamma(s_k)$ jumps several times at the asymptotic point $s_{BC,k}$
until it reaches the final value of $\gamma(s_{k+1})$, however, the result given by
Eq.~\ref{EQ9d2} is not altered.
\begin{figure}[h!]
	\begin{center}
		\includegraphics[width=3in ]{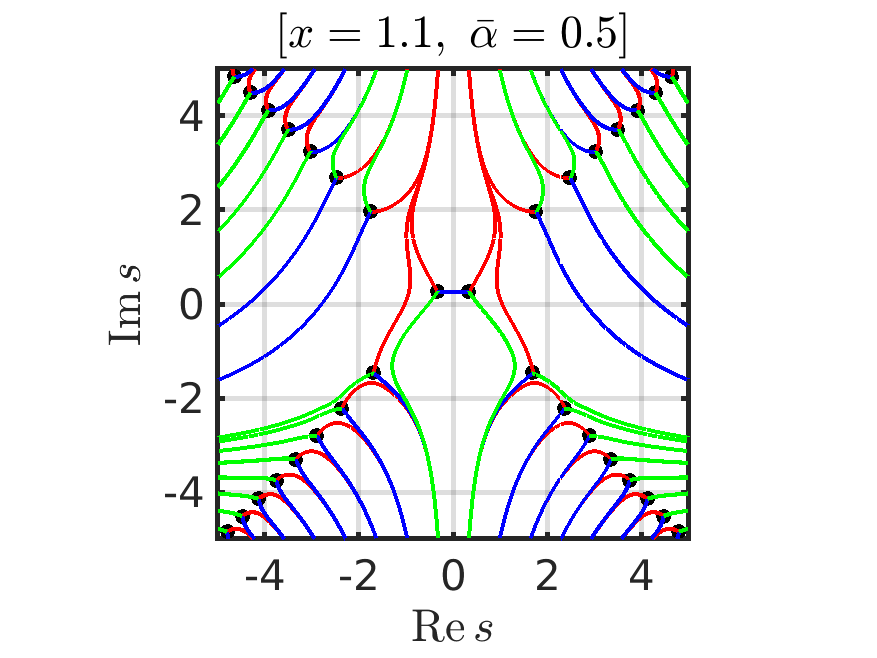}
		\includegraphics[width=3in ]{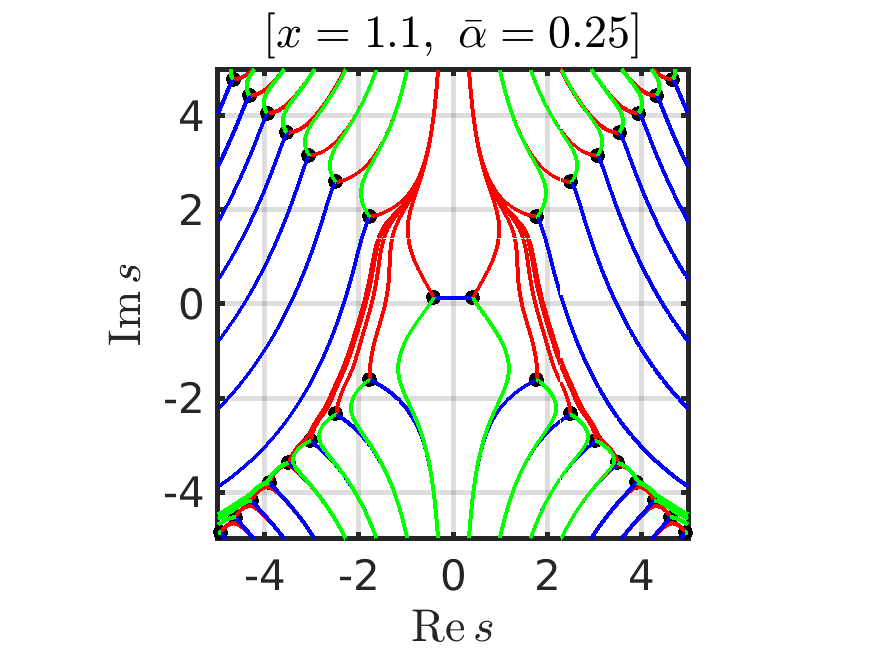}
	\end{center}
	\caption{When the laser strength parameter $x$ reaches a certain lower bound for a given chirp $\balpha$,
		then some equivalue lines that start at EPs from the lower half of the complex plane end up in the asymptotic
		limit $\im\,s\to\infty$, i.e. in the integration half-plane.}
	\mylabel{FigPenet}
\end{figure}

\subsection{Exponent in the asymptotic limit}
\noindent
Based on Eq.~\ref{EQ9d1}, the point $s_{BC,k}$ corresponds to $\xi=k$,
\begin{align}
& \re\, s_{BC,k} = \frac{(k+1/2)\pi}{\im s}.
%	\mylabel{EQ14d3}
\end{align}
$Q(s)$ is defined in Eq.~\ref{Palp} as a reduced imaginary part of 
the quasienergy split.
Note that the energy split $\tilde \delta(s)$, which is defined in Eq.~\ref{EQdelred2},
reads as
\begin{align}
&\tilde \delta(s)^2 =
e^{-[(\re s)^2 - (\im s)^2]} e^{-2i\im s\re s} 
\cr &\quad 
+\(\frac{\balpha }{2}(\re s + i\im s)+\frac{i}{x}\)^2
\mylabel{EQ11d6}
\end{align}
if the real and imaginary parts of $s$ are separated.
Now, Eq.~\ref{EQ11d6} is simplified for the case of the asymptotic limit, which 
is characterized by $\re\, s \ll \im\, s$ and $\im\,s\gg 1$, from where
\begin{align}
&\tilde \delta(s)^2=\cr &\quad
e^{(\im s)^2} e^{-2i\im s\re s} 
- \(\frac{\balpha}{2}\)^2 (\im s)^2
\cr&\quad
\approx
e^{(\im s)^2} e^{-2i\im s\re s}  ,
\end{align}
from where
\begin{align}
&\tilde \delta(s) \approx z
e^{(\im s)^2/2} e^{-i\im s\re s} \ ,
\mylabel{EQ11d8}
\end{align}
where $z$ is an unknown sign.
At this point it is useful to substitute for $\re\, s$ using the contour
parameter $\xi$ as defined in Eq.~\ref{EQ9d1}:
\begin{align}
&\tilde \delta(\xi) \approx -i\,z
e^{(\im s)^2/2} e^{-i\pi\xi} \, .
\mylabel{EQ14d11}
\end{align}
$Q(s)$ defined by Eq.~\ref{Palp} is derived from Eq.~\ref{EQ14d11} such that
\begin{align}
&Q(s) \approx -z\, x\, e^{(\im s)^2/2} \, \cos(\pi\xi) \, .
\mylabel{EQ14d13}
\end{align}
Now, we will substitute $Q(s)$ to Eq.~\ref{EQ9d2}, where we
change the integration variable $\re\, s'$ for $\xi$ as defined in Eq.~\ref{EQ9d1},
\begin{align}
&\gamma(s) = \gamma(s_{k+1}) + \frac{\pi}{\im s}
\int\limits_{k}^{\xi}  Q(\xi)\, d\xi ,
\mylabel{EQ14d14}
\end{align}
such that
\begin{align}
&\gamma(s) = \gamma(s_{k+1}) 
- z\cdot \frac{\pi\, xe^{(\im s)^2/2}}{\im s}\sin(\pi\xi) ,\cr
& \quad k=floor(\xi) ,
\mylabel{EQ14d15}
\end{align}
where $\xi$ is defined in Eq.~\ref{EQ9d1}.
Clearly, as $\xi$ acquires the limiting values on the integration contour, i.e.
$k$ or $k+1$, then $\gamma(s)$ is finite given by $\gamma(s_{k+1})$.
However, in between these two points, the absolute value of $\gamma(s)$ is infinite
as $\im\,s\to\infty$.
If additionally the sign of $\gamma(s)$ is positive for $s_k < s <s_{k+1}$
then the integrand in Eq.~\ref{v1} is infinitesimally
small along the connecting contour in the asymptote, by which
the use of the proposed integration contour would be justified. In the opposite case, the
integrand would be infinite and the integration contour could not be used.

\subsection{Sign alteration in the energy split when switching between equivalue lines}
\noindent
We can rewrite Eq.~\ref{EQ14d15} such that
\begin{align}
&\gamma(s) = \gamma(s_{k+1}) +
[z\cdot (-)^{k+1}]\, \frac{\pi\, xe^{(\im s)^2/2}}{\im s}|\sin(\pi\xi)| ,\cr
& \quad k=floor(\xi) .
\mylabel{EQ14d18}
\end{align}
Clearly, $\gamma(s)$ would be positive defined if
\begin{align}
z=(-)^{k+1} .
\end{align}
As long as $z$ represents the sign of the energy split $\tilde \delta(s)$ in the asymptotic limit,
see Eqs.~\ref{EQ11d6}--\ref{EQ11d8}, this indicates
that the sign of the energy split would be changed when passing from one connecting line
to another. 
This requirement makes sense if one recalls the fact that the sign is changed
across the branchcuts.

Yet, only the sign alteration is not sufficient to assure that $z=(-)^{k+1}$, and not $z=(-)^k$,
which would lead to a negative defined $\gamma(s)$. 
Namely, for $k=0$, $z<0$ for $\re\,s>\re s_{BC,0}$, and $z>0$ for 
$\re s_{BC,0}>s>0$, while $\im\,s\to\infty$.
It follows from Eq.~\ref{EQ11d8} that if this is satisfied then also
$\im\, \tilde\delta(s)<0$ (or equivalently $Q(s)<0$) as $\re s\to +0$, while $\im\,s\to\infty$.
This happens if the same condition is satisfied also on the real axis:
\begin{align}
Q(s) < 0 ,\quad  \re\,s\to +0,\quad \im\,s=0.
\end{align}
This condition is satisfied for bound to resonance transitions as follows
from the definition in Eqs.~\ref{Gb}. By this we proved that the contribution
of the connecting lines between the branchcuts in the asymptotic limit
$s\to i\infty$ is given by zero.

\section{\label{THINF}Large pulse area limit for the branchcut contributions}

\noindent
Let us specify the laser parameters for which the limit defined by Eq.~\ref{EQ11d30}
is relevant by substituting for $\beta_k^{(1)}$ from Eq.~\ref{EQ106}:
\begin{align}
\frac{x\Gamma\tau}{3\hbar} 
\left |
\balpha \(\frac{\balpha }{2}s_k + \frac{i}{x}\) - 2s_k\, e^{-s_k^2}
\right |^{1/2}
\to \infty,
\end{align}
which can be rewritten using the definition of the {\it pulse area $\theta$} (Eq.~\ref{EQthinf})
such that
\begin{align}
\frac{\theta}{3\hbar} \cdot \sqrt\frac{2}{\pi}\cdot
\left |
\balpha \(\frac{\balpha }{2}s_k + \frac{i}{x}\) - 2s_k\, e^{-s_k^2}
\right |^{1/2}
\to \infty.
\end{align}
The limit is exact for the very large pulse are $\theta\to\infty$,
while it is at least approximately correct whenever the left hand side exceeds $\pi/2$, i.e.
$\theta$ satisfies the condition
\begin{align}
\theta > \frac{3\hbar\pi}{2}
\sqrt\frac{\pi}{2}\cdot
\left |
\balpha \(\frac{\balpha }{2}s_k + \frac{i}{x}\) - 2s_k\,  e^{-s_k^2}
\right |^{-1/2} .
\end{align}

\section{\label{Topology}Continuity upon crossing the separatrix of odd and even layouts}

\noindent
We derived two different formulas for the survival probability applicable
when encircling EP in the frequency--laser strength plane. The applicability
of one or the other formula depends on the layout of the central pair of TPs
in the complex adiabatic-time plane, which can be either ``odd'' or ``even''.
There is a division (separatrix) in the laser parameter plane, which designates the areas
of the odd and even layouts, respectively, see Fig.~\ref{Fxa}.
The separatrix is defined by the coalescence of the two TPs in the complex adiabatic-time
plane. As the two TPs coalesce, a special second order TP arises for which the first order 
term in the Puiseux series is equal to zero.

The crossing of the separatrix is accompanied by the change of the formulas,
which apply for
the survival probability (of the initial bound state) $p_1$  such that 
\begin{align*}
	& p_{1,odd} = \frac{\pi^2}{9} \exp\[-\frac{2\,\theta}{\hbar\sqrt{2\pi}} \frac{\bar\gamma(s_{0i}) }{x}\], \revisited{EQ14d44} \\
	& p_{1,even} =  \cr & \frac{4 \pi^2}{9} 
	\exp\[-\frac{2\,\theta}{\hbar\sqrt{2\pi}} \frac{\bar\gamma(s_{0i}) }{x}\]
	\cos^2\[\frac{\theta}{\hbar\sqrt{2\pi}} \frac{\phi(s_{0}) }{x}\] \, .
	\cr
	&\revisited{EQ14d21b} 
\end{align*}
Above we focused on the fact that
these formulas predict the switch of the monotonic and oscillatory behavior and
we showed physically achievable examples where this switch could be observable.
Now we shall discuss a change of the survival probability $p_1$ upon crossing
the separatrix, i.e. in the infinitesimal neighborhood of the crossing.

The function $\phi(s_0)=0$ at the coalescence, which is shown in Fig.~\ref{Figphix0}a.
(This fact can be proved starting from the definition Eq.~\ref{gamma}: At the coalescence,
$s_0$ is purely imaginary, therefore the integration path is along the imaginary axis.
At the same time, $\tilde\delta(s)$ is real-defined, which follows from the symmetric
relation Eq.~\ref{EQ8d10} and the fact that $\tilde\delta(s)$ is continuous in the
infinitesimal neighborhood of the integration path. From here, the integral over
$\tilde\delta(s)$ along this path is purely imaginary, thus $\phi(s_0)=0$.)
By substitution of $\phi(s_0)$ we write the survival probability at the separatrix:
\begin{align}
	& p_{1,odd}^{separatrix} = \frac{\pi^2}{9} \exp\[-\frac{2\,\theta}{\hbar\sqrt{2\pi}} \frac{\bar\gamma(s_{0i}) }{x}\] \ne
	\cr
	&\quad p_{1,even}^{separatrix} = \frac{4 \pi^2}{9} 
	\exp\[-\frac{2\,\theta}{\hbar\sqrt{2\pi}} \frac{\bar\gamma(s_{0i}) }{x}\] .
\end{align}
The left and right hand sides of the equation are different due to the different prefactor,
as the rest of the functions is continuous at the crossing of the separatrix.
Namely, the first-order perturbation theory predicts a topological phenomenon
at the crossing of the separatrix in the laser pulse parameter plane
for the asymptotic limit $\theta\to\infty$.

%\subsection{Numerical fitting of the survival probability to the asymptotic formulas}

%\noindent
Keeping in mind that the formulas Eqs.~\ref{EQ14d44} and \ref{EQ14d21b}
have been derived for the asymptotic limit $\theta\to\infty$, 
the convergence to this limit near the separatrix should be investigated.
Let us use the asymptotic formulas Eqs.~\ref{EQ14d44} and \ref{EQ14d21b} as
a basis for a parameter fitting for finite pulse area $\theta<\infty$.
However for simplicity, instead of fitting the probability, we will fit the amplitude $v_1$
\begin{align}
	& v_1^{odd}= a(\theta) \exp\[-\frac{\theta}{\hbar\sqrt{2\pi}} \frac{\gamma(\theta) }{x}\], \cr
	& v_1^{even}= a(\theta)
	\exp\[-\frac{\theta}{\hbar\sqrt{2\pi}} \frac{\gamma(\theta) }{x}\]
	\cos\[\frac{\theta}{\hbar\sqrt{2\pi}} \frac{\phi(\theta) }{x}\] \, .\cr
	\mylabel{EQ15d2}
\end{align}
We will use the different fitting formulas on the different sides of the separatrix.
It is expected that the fitted parameters $a(\theta)$, $\gamma(\theta)$, and
$\phi(\theta)$ converge to the asymptotic limits as $\theta\to\infty$,
\begin{align}
	& a^{odd}(\theta\to\infty) = \frac{\pi}{3},
	& a^{even}(\theta\to\infty) = \frac{2\pi}{3}, \cr
	& \gamma^{odd}(\theta\to\infty) = \gamma(s_{0i}),
	& \gamma^{even}(\theta\to\infty) = \gamma(s_{0}), \cr
	& & \phi^{even}(\theta\to\infty) = \phi(s_{0}).
\end{align}

We use a numerical integration along the real axis to obtain $v_1$.
As a matter of fact, numerical fitting procedures are not stable if the exponential
parameter brings the value of $|v_1|$ below the computational precision.
For these reasons we investigate a separatrix crossing which occurs
not too far from $x=1$, where $\gamma(s_0)$ and $\gamma(s_{0i})$ are almost zero,
in particular, the separatrix crossing at $\balpha=0.5$ and $x=1.0305$.
We show the results of the fitting in Fig.~\ref{FigPref}.

Fig.~\ref{FigPref}a shows that the prefactor converges to different values
for large pulse areas $\theta$ at the two opposite sides of the separatrix,
which approves our previous conclusions based on analytical calculations.
However, the convergence is slowed down in a close neigborhood of the
separatrix. If we look at the exponent (Fig.~\ref{FigPref}b), we 
conclude that a change of the amplitude due to the exponent change when
crossing the separatrix, compensates the jump in the prefactor several
times at so large values of $\theta$. This explains the fact that
the abrupt change of the
prefactor is not demonstrated as an abrupt change of the survival 
probability $p_1$ when crossing the separatrix.
\begin{figure}[!h]
\begin{center}
	(a) \includegraphics[width= 3 in ]{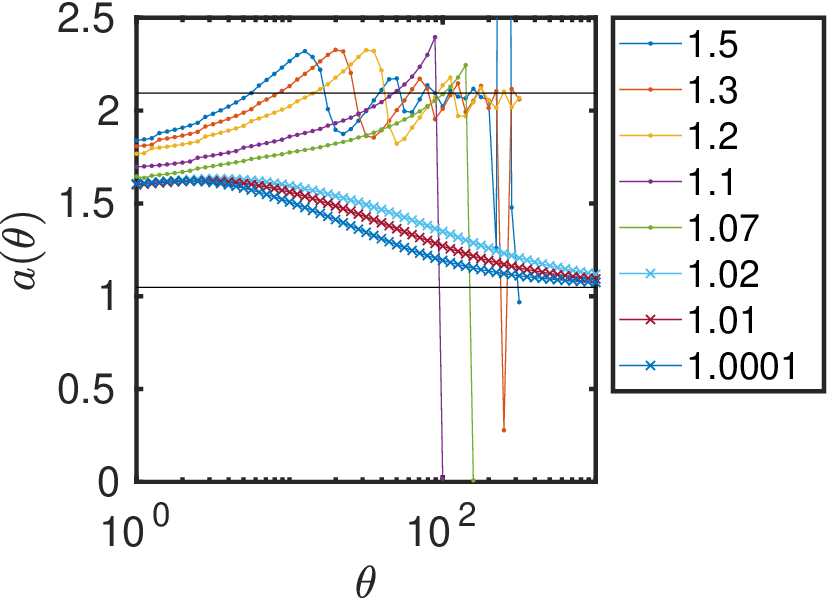}
	(b) \includegraphics[width= 3 in ]{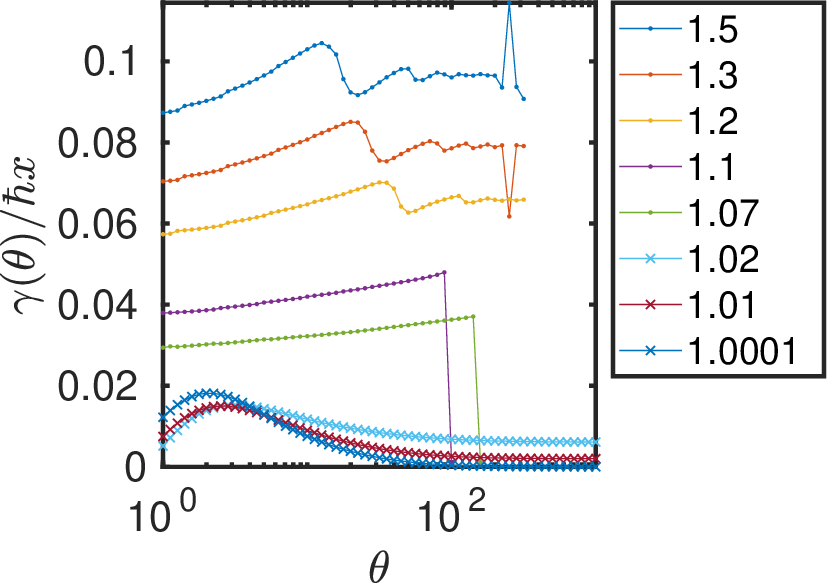}
\end{center}
\caption{
	Results of numerical fitting of the amplitude $v_1$
	calculated using first order perturbation theory to the
	asymptotic formulas given by Eqs.~\ref{EQ15d2}:
	(a) prefactor; (b) exponent.
	The laser parameter $x$ is changed from 1.0001 to 1.5, which
	$\balpha=0.5$ for all calculations. The separatrix would be crossed
	at $x=1.0305$. 
	}
\mylabel{FigPref}
\end{figure}

%\printbibliography

%\bibliography{jourtab,etg,etges,foumouo,heioniz,helium,helium2,moiseyevnew,zamastil,kapralova,burgers,theory,wigner,hedoubly,laserdyn,laserdyn2,laserdyn3,laserdyn4,laserdyn5,xuvlasers,alon,civis,field,helium3,helium4,excep,excep2,rap,pulsearea,aux}

\begin{thebibliography}{10}
\expandafter\ifx\csname url\endcsname\relax
  \def\url#1{\texttt{#1}}\fi
\expandafter\ifx\csname urlprefix\endcsname\relax\def\urlprefix{URL }\fi
\expandafter\ifx\csname href\endcsname\relax
  \def\href#1#2{#2} \def\path#1{#1}\fi

\bibitem{Rotter:2009}
I.~Rotter, {A Non-Hermitian Hamilton Operator and the Physics of Open Quantum
  Systems}, J. Phys. A-Math. Theor. {42} (2009) 153001.
\newblock \href {https://doi.org/{10.1088/1751-8113/42/15/153001}}
  {\path{doi:{10.1088/1751-8113/42/15/153001}}}.

\bibitem{Heiss:2012}
W.~D. Heiss, {The Physics of Exceptional Points}, J. Phys. A-Math. Theor. {45}
  (2012) 444016.
\newblock \href {https://doi.org/{10.1088/1751-8113/45/44/444016}}
  {\path{doi:{10.1088/1751-8113/45/44/444016}}}.

\bibitem{Rotter:2015}
I.~Rotter, J.~P. Bird, {A Review of Progress in the Physics of Open Quantum
  Systems: Theory and Experiment}, Rep. Prog. Phys. {78} (2015) 114001.
\newblock \href {https://doi.org/{10.1088/0034-4885/78/11/114001}}
  {\path{doi:{10.1088/0034-4885/78/11/114001}}}.

\bibitem{Klaiman:2008}
S.~Klaiman, U.~Guenther, N.~Moiseyev, {Visualization of Branch Points in
  PT-Symmetric Waveguides}, Phys. Rev. Lett. {101} (2008) 080402.
\newblock \href {https://doi.org/{10.1103/PhysRevLett.101.080402}}
  {\path{doi:{10.1103/PhysRevLett.101.080402}}}.

\bibitem{Doppler:2016}
J.~Doppler, A.~A. Mailybaev, J.~Bohm, U.~Kuhl, A.~Girschik, F.~Libisch, T.~J.
  Milburn, P.~Rabl, N.~Moiseyev, S.~Rotter, {Dynamically Encircling an
  Exceptional Point for Asymmetric Mode Switching}, Nature. {537} (2016) 76.
\newblock \href {https://doi.org/{10.1038/nature18605}}
  {\path{doi:{10.1038/nature18605}}}.

\bibitem{Miri:2019}
M.-A. Miri, A.~Alu, {Exceptional Points in Optics and Photonics}, Science.
  {363} (2019) 7709.

\bibitem{Liertzer:2012}
M.~Liertzer, L.~Ge, A.~D. Stone, H.~E. Tuereci, S.~Rotter, {Pump-Induced
  Exceptional Points in Lasers}, Phys. Rev. Lett. {108} (2012) 173901.

\bibitem{Peng:2014}
B.~Peng, S.~K. Ozdemir, S.~Rotter, H.~Yilmax, M.~Liertzer, F.~Monifi, C.~M.
  Bender, F.~Nori, L.~Yang, {Loss-Induced Suppression and Revival of Lasing},
  Science. {346} (2014) 328.

\bibitem{Feng:2014}
L.~Feng, Z.~J. Wong, R.~M. Ma, Y.~Wang, X.~Zhang, {Single-Mode Laser by
  Parity-Time Symmetry Breaking}, Science. {346} (2014) 972.
\newblock \href {https://doi.org/{10.1126/science.1258479}}
  {\path{doi:{10.1126/science.1258479}}}.

\bibitem{Ozdemir:2019}
S.~K. Ozdemir, S.~Rotter, F.~Nori, L.~Yang, {Parity-Time Symmetry and
  Exceptional Points in Photonics}, Nature. Mater. {18} (2019) 783.

\bibitem{Kapralova-Zdanska:2014}
P.~R. Kapralova-Zdanska, N.~Moiseyev, {Helium in Chirped Laser Fields as a
  Time-Asymmetric Atomic Switch}, J. Chem. Phys. {141} (2014) 014307.
\newblock \href {https://doi.org/{10.1063/1.4885136}}
  {\path{doi:{10.1063/1.4885136}}}.

\bibitem{Peng:2016}
B.~Peng, W.~Cao, C.~Qu, J.~Wen, L.~Jiang, Y.~Xiao, {Anti-parity-time symmetry
  with flying atoms}, Nat. Phys. {12} (2016) 1139.

\bibitem{Oberreiter:2018}
L.~Oberreiter, J.~Burkhardt, J.~Main, G.~Wunner, {Population transfer at
  exceptional points in the spectra of the hydrogen atom in parallel electric
  and magnetic fields}, Phys. Rev. A {98} (2018) 013417.

\bibitem{Li:2019}
J.~Li, A.~K. Harter, J.~Liu, L.~de~Melo, Y.~N. Joglekar, L.~Luo, {Observation
  of Parity-Time Symmetry Breaking Transitions in a Dissipative Floquet System
  of Ultracold Atoms}, Nat. Commun. {10} (2019) 855.

\bibitem{Estrada:1986}
H.~Estrada, L.~S. Cederbaum, W.~Domcke, {Vibronic Coupling of Short-lived
  Electronic States}, J. Chem. Phys. {84} (1986) 152.
\newblock \href {https://doi.org/{10.1063/1.450165}}
  {\path{doi:{10.1063/1.450165}}}.

\bibitem{Lefebvre:2009}
R.~Lefebvre, O.~Atabek, M.~Sindelka, N.~Moiseyev, {Resonance Coalescence in
  Molecular Photodissociation}, Phys. Rev. Lett. {103} (2009) 123003.
\newblock \href {https://doi.org/{10.1103/PhysRevLett.103.123003}}
  {\path{doi:{10.1103/PhysRevLett.103.123003}}}.

\bibitem{Cederbaum:2011}
L.~S. Cederbaum, Y.-C. Chiang, P.~V. Demekhin, N.~Moiseyev, {Resonant Auger
  Decay of Molecules in Intense X-ray Laser Fields: Light-Induced Strong
  Nonadiabatic Effects}, Phys. Rev. Lett. {106} (2011) 123001.

\bibitem{Leclerc:2017}
A.~Leclerc, D.~Viennot, G.~Jolicard, R.~Lefebvre, O.~Atabek, {Exotic States in
  the Strong-Field Control of H$_2^+$ Dissociation Dynamics: From Exceptional
  Points to Zero-Width Resonances}, J. Phys. B-At. Mol. Opt. Phys. {50} (2017)
  234002.

\bibitem{Benda:2018}
Z.~Benda, T.-C. Jagau, {Locating Exceptional Points on Multidimensional
  Complex-Valued Potential Energy Surfaces}, J. Phys. Chem. Lett. {9} (2018)
  6978.

\bibitem{Uzdin:2011}
R.~Uzdin, A.~Mailybaev, N.~Moiseyev, {On the Observability and Asymmetry of
  Adiabatic State Flips Generated by Exceptional Points}, J. Phys. A-Math.
  Theor. {44}~({43}) (2011) 435302.
\newblock \href {https://doi.org/{10.1088/1751-8113/44/43/435302}}
  {\path{doi:{10.1088/1751-8113/44/43/435302}}}.

\bibitem{Gilary:2013}
I.~Gilary, A.~A. Mailybaev, N.~Moiseyev, {Time-Asymmetric
  Quantum-State-Exchange Mechanism}, Phys. Rev. A {88} (2013) 010102.
\newblock \href {https://doi.org/{10.1103/PhysRevA.88.010102}}
  {\path{doi:{10.1103/PhysRevA.88.010102}}}.

\bibitem{Xu:2016}
H.~Xu, D.~Mason, L.~Y. Jiang, J.~G.~E. Harris, {Topological Energy Transfer in
  an Optomechanical System with Exceptional Points}, Nature. {537} (2016) 80.
\newblock \href {https://doi.org/{10.1038/nature18604}}
  {\path{doi:{10.1038/nature18604}}}.

\bibitem{Zhong:2018}
Q.~Zhong, M.~Khajavikhan, D.~N. Christodoulides, R.~El-Ganainy, {Winding around
  non-Hermitian singularities}, Nat. Commun. {9} (2018) 4808.

\bibitem{Fernandez:2020}
L.~J. Fernandez-Alcazar, H.~Li, F.~Ellis, A.~Alu, T.~Kottos, {Robust Scattered
  Fields from Adiabatically Driven Targets around Exceptional Points}, Phys.
  Rev. Lett. {124} (2020) 133905.

\bibitem{Feilhauer:2020}
J.~Feilhauer, A.~Schumer, J.~Doppler, A.~A. Mailybaev, J.~Bohm, U.~Kuhl,
  N.~Moiseyev, S.~Rotter, {Encircling exceptional points as a non-Hermitian
  extension of rapid adiabatic passage}, Phys. Rev. {102} (2020) 040201.

\bibitem{LETTER}
P.~R. Kapralova, M.~Sindelka, N.~Moiseyev, {Coalescence of Two Branch Points in
  Complex Time Marks the End of Rapid Adiabatic Passage and the Start of Rabi
  Oscillations}, J. Phys. A-Math. Theor. (Submitted to the special issue on
  Claritons and the Asymptotics of Ideas: the Physics of Michael Berry.).

\bibitem{Dykhne:1962}
A.~M. Dykhne, {Adiabatic Perturbation of Discrete Spectrum States}, Soviet
  Physics JETP {14} (1962) 941.

\bibitem{Davis:1975}
J.~P. Davis, P.~Pechukas, {Nonadiabatic Transitions Induced by a Time-Dependent
  Hamiltonian in the Semiclassical/Adiabatic Limit: The Two-State Case}, J.
  Chem. Phys. {64} (1976) 3129.

\bibitem{Dridi:2010}
G.~Dridi, S.~Guerin, H.~R. Jauslin, D.~Viennot, G.~Jolicard, {Adiabatic
  Approximation for Quantum Dissipative Systems: Formulation, Topology, and
  Superadiabatic Tracking}, Phys. Rev. A {82} (2010) 022109.

\bibitem{Child:1978}
M.~S. Child, {Semiclassical Effects in Heavy-Particle Theory}, Adv. in At. Mol.
  Phys. {14} (1979) 225.

\bibitem{McCall:1967}
S.~L. McCall, E.~L. Hahn, {Self-induced transparency by pulsed coherent light},
  Phys. Rev. Lett. {18} (1967) 908.

\bibitem{McCall:1969}
S.~L. McCall, E.~L. Hahn, {Self-induced transparency}, Phys. Rev. {183} (1969)
  183.

\bibitem{AllenEberly}
L.~Allen, J.~H. Eberly, {Optical Resonance and Two-Level Atoms}, Dover
  publications, inc., New York, 1987.

\bibitem{Vitanov:2001}
N.~V. Vitanov, T.~Halfmann, B.~W. Shore, K.~Bergmann, {Laser-Induced Population
  Transfer by Adiabatic Passage Techniques}, Annu. Rev. Phys. Chem. {52} (2001)
  763.
\newblock \href {https://doi.org/{10.1146/annurev.physchem.52.1.763}}
  {\path{doi:{10.1146/annurev.physchem.52.1.763}}}.

\bibitem{Schilling:2006}
R.~Schilling, M.~Vogelsberger, D.~A. Garanin, {Non-adiabatic Transitions for a
  Decaying Two-level System: Geometrical and Dynamical Contributions}, J. Phys.
  A-Math. Gen. {39} (2006) 13727.

\bibitem{Dridi:2012}
G.~Dridi, S.~Guerin, {Adiabatic Passage for a Lossy Two-level Quantum System by
  a Complex Time Method}, J. Phys. A-Math. Gen. {45} (2012) 185303.

\bibitem{Moiseyev}
N.~Moiseyev, {Non-hermitian quantum mechanics}, Cambridge University Press, New
  York, 2011.

\bibitem{Zener:1932}
C.~Zener, {Non-adiabatic Crossing of Energy Levels}, Proceedings. Of. The.
  Royal. Society. Of. London. Series. A-Mathematical. And. Physical. Sciences.
  {137} (1932) 696.

\bibitem{Wittig:2005}
C.~Wittig, {The Landau-Zener Formula}, J. Phys. Chem. B {109} (2005) 8428.

\bibitem{Vutha:2010}
A.~C. Vutha, {A simple approach to the Landau-Zener formula}, Eur. J. Phys.
  {31} (2010) 389.
\newblock \href {https://doi.org/{10.1088/0143-0807/31/2/016}}
  {\path{doi:{10.1088/0143-0807/31/2/016}}}.

\bibitem{Shevchenko:2010}
S.~N. Shevchenko, S.~Ashhab, S.~Nori, {Landau-Zener-St\"uckelberg
  interferometry}, Phys. Rep. {492} (2010) 1.

\bibitem{Thorson:1971}
W.~R. Thorson, J.~B. Delos, S.~A. Boorstein, {Studies of the potential-curve
  crossing problem. I. Analysis of Stueckelberg's method}, Phys. Rev. A {4}
  (1971) 1052.

\bibitem{Delos:1972}
J.~B. Delos, W.~R. Thorson, {Studies of the potential-curve crossing problem.
  II. General theory and a model for close crossings}, Phys. Rev. A {6} (1972)
  728.

\bibitem{Ota:2018}
T.~Ota, K.~Hitachi, K.~Muraki, {Landau-Zener-Stueckelberg interference in
  coherent charge oscillations of a one-electron double quantum dot}, Sci. Rep.
  {8} (2018) 5491.

\bibitem{Tannor}
D.~J. Tannor, {Introduction to quantum mechanics -- a time-dependent
  perspective}, University Science Books, 2007.

\bibitem{Vitanov:1998}
N.~V. Vitanov, K.-A. Suominen, {Nonlinear Level-Crossing Models}, Phys. Rev. A
  {59} (1999) 4580.

\bibitem{Yan:2010}
Y.~Yan, B.~Wu, {Integral definition of transition time in the Landau-Zener
  model}, Phys. Rev. A {81} (2010) 022126.
\newblock \href {https://doi.org/{10.1103/PhysRevA.81.022126}}
  {\path{doi:{10.1103/PhysRevA.81.022126}}}.

\bibitem{Pick:2019}
A.~Pick, P.~R. Kapralova-Zdanska, N.~Moiseyev, {Ab-initio Theory of
  Photoionization via Resonances}, J. Chem. Phys. {150} (2019) 204111.

\end{thebibliography}

\end{document}